\documentclass[
 aps, pra,
 amsmath,amssymb,
 11pt,
 final,
tightenlines,
 twoside,
 twocolumn,
 nofloats,
nofootinbib,
 superscriptaddress,
showkeys,
showkeywords,
 ]
{revtex4-2}

\usepackage[T2A]{fontenc}
\usepackage[utf8x]{inputenc}
\usepackage[russian,english]{babel}
\usepackage{graphicx}
\usepackage{dcolumn}
\usepackage{bm}
\usepackage{longtable}

\input{maik.rty}

\setcitestyle{authoryear,round}
\def\squareforqed{\hbox{\rlap{$\sqcap$}$\sqcup$}}

\def\sq{\ifmmode\squareforqed\else{\unskip\nobreak\hfil
\penalty50\hskip1em\null\nobreak\hfil\squareforqed
\parfillskip=0pt\finalhyphendemerits=0\endgraf}\fi}

\def\utw{\smash{\rlap{\lower5pt\hbox{$\sim$}}}}

\def\udtw{\smash{\rlap{\lower6pt\hbox{$\approx$}}}}

\def\diameter{{\ifmmode\mathchoice
{\ooalign{\hfil\hbox{$\displaystyle/$}\hfil\crcr
{\hbox{$\displaystyle\mathchar"20D$}}}}
{\ooalign{\hfil\hbox{$\textstyle/$}\hfil\crcr
{\hbox{$\textstyle\mathchar"20D$}}}}
{\ooalign{\hfil\hbox{$\scriptstyle/$}\hfil\crcr
{\hbox{$\scriptstyle\mathchar"20D$}}}}
{\ooalign{\hfil\hbox{$\scriptscriptstyle/$}\hfil\crcr
{\hbox{$\scriptscriptstyle\mathchar"20D$}}}}
\else{\ooalign{\hfil/\hfil\crcr\mathhexbox20D}}%
\fi}}

\begin{document}

\selectlanguage{english}


\title{The Influence of the Bar on the Dynamics of Globular Clusters in the
Central Region of the Milky Way. Frequency Analysis of Orbits
According to GaiaEDR3 Data}

\author{\firstname{A.~T.}~\surname{Bajkova}}
 \email{bajkova@gaoran.ru}
 \affiliation{Central (Pulkovo) Astronomical Observatory RAS, St. Petersburg, 196140 Russia}

\author{\firstname{A.~A.}~\surname{Smirnov}}
 \affiliation{Central (Pulkovo) Astronomical Observatory RAS, St. Petersburg, 196140 Russia}

\author{\firstname{V.~V.}~\surname{Bobylev}}
 \affiliation{Central (Pulkovo) Astronomical Observatory RAS, St. Petersburg, 196140 Russia}

\begin{abstract}
Abstract—This work is devoted to studying the influence of the bar on the orbital dynamics of globular clusters. The orbits of 45 globular clusters in the central galactic region with a radius of 3.5 kpc were analyzed using spectral dynamics methods in order to identify objects captured by the bar. To form the 6D phase space required for orbit integration, the most accurate astrometric data to date from the Gaia satellite (EDR3), as well as new refined average distances to globular clusters, were used. Since the parameters of the Milky Way bar are known with very great uncertainty, the orbits were constructed and their frequency analysis was carried out with varying the mass, length and angular velocity of rotation of the bar in a wide range of values with a fairly small step. The integration of orbits was carried out at 2.5 billion
years ago. As a result, bar-supporting globular clusters were identified for each set of bar parameters. For the first time, an analytical expression has been obtained for the dependence of the dominant frequency $f_X$ on the angular velocity of rotation of the bar. In addition, the probabilities of capturing globular clusters by the bar were determined when the bar parameters were varied in certain ranges of values according to a random distribution law. A list of 14 globular clusters with the most significant capture probabilities is given, with five GCs—NGC6266, NGC6569, Terzan 5, NGC6522, NGC6540 - showing the probability capture by bar
$\geq 0.2$. A conclusion is made about the regularity of the orbits of globular clusters based
on the calculation of approximations of the maximum characteristic Lyapunov exponents.

\bigskip


\end{abstract}

\maketitle

\section{INTRODUCTION}

Currently, about 170 galactic globular clusters(GCs) are known. According to theoretical estimates, their number in the Galaxy can reach 200(Ogorodnikov, 1965). More than 150 GCs have complete astrometric (positional) and kinematic measurements (proper motions and radial velocities) to
form the 6D phase space necessary for constructing orbits. Studying the orbital motion of GCs is of great importance for studying the evolution of the Galaxy,since they are the oldest objects, whose age reaches 13 billion years.

In most studies, the motion of galactic globular clusters is considered in an axisymmetric stationary potential. However, the real potential of our Galaxy is neither axisymmetric nor stationary. And first of all, this is due to the fact that in the center of the Galaxy
there is a rotating elongated bar.

The previous stage of our studies of the influence of the bar on the orbital characteristics of the GCs in the central region of the Milky Way galaxy was devoted to the analysis of such orbital parameters as apocentric and pericentric distances, eccentricity and maximum distance from the galactic plane (Bajkova et al., 2023). The goal of this work is to identify globular clusters captured by the bar, depending on the parameters of the bar: mass, length and angular
velocity of rotation, which are currently known with great uncertainty.

To solve the problem, the spectral dynamics method proposed in Binney and Spergel (1982) was
used. Themethod is that each orbit can be associated with characteristic oscillation frequencies $f_X, f_Y, f_Z$ and $f_R$ along the $X, Y, Z$  coordinates and cylindrical distance $R=\sqrt{X^2+Y^2}$, respectively. To do this, the amplitude spectra of the time series $x(t), y(t), z(t)$ and $R(t)$ are calculated and the frequencies corresponding to the largest values of the spectrum are found. These are the characteristic, or dominant, frequencies. For the task of determining the orbits captured by the bar, the frequencies $f_X$ and $f_R$ are calculated. If the frequency ratio is $f_R/f_X=2$, then the orbit is considered to be captured by the bar. In practice, the ratio $f_R/f_X=2\pm 0.1$ (Parul et al., 2020) is used.

The criterion we chose was also discussed in detail in Wang et al. (2016). These authors showed that such a criterion determines orbits that are in 2:1 resonance with the bar, that is, captured by the bar.

Below is a description of the used model of the gravitational potential of the Galaxy, which includes a three-component axisymmetric part (bulge, disk, halo) with a built-in central elongated bar in the form of a triaxial ellipsoid. The model adopted in this work is, of course, simplified and does not take into account many interesting effects caused by the interaction of
various subsystems of the Galaxy. For example, one of such effects is the occurrence of bar librations due to the influence of the force field of a triaxial halo-feedback effect (Kondratyev et al., 2022). And bar librations can affect the value of dominant frequencies. However, such subtle issues lie beyond the scope of the research carried out, but are of interest
for subsequent work.

An alternative approach to solving the problem of orbit classification using spectral dynamics methods, based on the Milky Way potential model obtained as a result of N-body calculations, is considered in the work by Smirnov et al. (2023). An important difference between this model and the one used in this work is that it is self-consistent and the bar is formed naturally due to the instability of the stellar disk relative to bar-like disturbances. This direction
of studying the influence of the bar is very interesting and promising and requires separate further research. In this work, we limit ourselves, as mentioned above, to the model when the bar component of the potential is simply “superimposed” on top of the disk, as is customary in most works devoted to the influence of the bar on the orbital dynamics of galactic objects.

The article is structured as follows. Section 2 gives a brief description of the accepted potential
models -- axisymmetric potential and potential with a bar. Section 3 provides references to the astrometric data used. In Section 4, 45 globular clusters were selected from the central part of the Galaxy. Section 5 provides a detailed description of the procedure for calculating dominant orbital frequencies. In Section 6, the properties of dominant frequencies are considered, an analytical expression is obtained for the dependence of the dominant frequency $f_X$ on the
angular velocity of rotation of the bar; an illustration of frequency analysis is given using the example of the globular cluster NGC6266. Section 7 is devoted directly to the frequency analysis of the orbits of all 45 globular clusters, constructed as a result of varying the bar parameters over a wide range of values, as well as estimating the probability of their capture. The problem of orbit regularity is considered based on the calculation of characteristic Lyapunov exponents.
The Conclusion lists the main results of the work.

\section{MODEL OF GALACTIC POTENTIAL}

\subsection{Axisymmetric potential}

The axisymmetric gravitational potential of the Galaxy is represented as the sum of three components~-- the central spherical bulge $\Phi_b(r(R,Z))$, the disk
$\Phi_d(R,Z)$ and the massive spherical dark matter halo
$\Phi_h(r(R,Z))$:
\begin{equation}
\begin{array}{lll}
\Phi(R,Z)=\Phi_b(r(R,Z))+\Phi_d(R,Z)+\Phi_h(r(R,Z)).
\label{pot}
\end{array}
\end{equation}
Here we use a cylindrical coordinate system ($R,\psi,Z$) with the origin at the center of the Galaxy. In a rectangular coordinate system $(X,Y,Z)$ with the origin at the center of the Galaxy, the distance to the star (spherical radius) will be equal to $r^2=X^2+Y^2+Z^2=R^2+Z^2$, while the $X$ axis is directed from the Sun to the Galactic center, the $Y$ axis is perpendicular
to the $X$ axis codirectional with the rotation of the Galaxy, the $Z$ axis is perpendicular to the galactic plane $X-Y$ towards the north galactic pole. Gravitational potential is expressed in units of 100 km$^2$ s$^{-2}$, distance -- in kpc, mass -- in units of galactic mass
$M_G=2.325\times 10^7 M_\odot$, corresponding to the gravitational constant $G=1$.

The bulge $\Phi_b(r(R,Z))$ and disk $\Phi_d(R,Z)$ potentials are represented in the form used by Miyamoto and Nagai (1975):
 \begin{equation}
  \Phi_b(r)=-\frac{M_b}{(r^2+b_b^2)^{1/2}},
  \label{bulge}
 \end{equation}
 \begin{equation}
 \Phi_d(R,Z)=-\frac{M_d}{\Biggl[R^2+\Bigl(a_d+\sqrt{Z^2+b_d^2}\Bigr)^2\Biggr]^{1/2}},
 \label{disk}
\end{equation}
where $M_b, M_d$ are masses, $b_b, a_d, b_d$ are the scale parameters of the components in kpc. The halo component is represented according to the work of Navarro et al. (1997):
 \begin{equation}
  \Phi_h(r)=-\frac{M_h}{r} \ln {\Biggl(1+\frac{r}{a_h}\Biggr)}.
 \label{halo-III}
 \end{equation}
Table ~\ref{t:model-III} shows the values of the parameters of the galactic potential model (\ref{bulge})--(\ref{halo-III}), which were found in the work of Bajkova and Bobylev (2016) as a result of fitting to the rotation curve of the Galaxy by Bhattacharjee et al. (2014), constructed from objects located at distances up to $R\sim200$ kpc. Note that when constructing this Galactic rotation curve, the following values of local parameters were used: $R_\odot=8.3$ kpc and $V_\odot=244$ km s$^{-1}$. In Bajkova and Bobylev (2016) model (\ref{bulge})--(\ref{halo-III}) is designated as model III. The adopted model is the best among the six galactic potential models considered in Bajkova and Bobylev (2017), since it provides the smallest discrepancy between the data of Bhattacharjee et al. (2014) and the model rotation curve.

\subsection{Bar model}

The model of a triaxial ellipsoid of Palou$\breve{s}$ et al.
(1993) was chosen as the potential of the central bar:
\begin{equation}
  \Phi_{bar} = -\frac{M_{bar}}{(q_b^2+X^2+[Ya/b]^2+[Za/c]^2)^{1/2}},
\label{bar}
\end{equation}
where $X=R\cos\vartheta, Y=R\sin\vartheta$, $a, b, c$ are three semi-axes of the bar, $q_b$ is the bar scale parameter; $\vartheta=\theta-\Omega_{b}t-\theta_{b}$, $tg(\theta)=Y/X$, $\Omega_{b}$ is the angular velocity of bar rotation, $t$ is the integration time, $\theta_{b}$ is the angle of orientation of the bar relative to the galactic axes $X,Y$, measured from the line connecting the Sun and the center of the Galaxy ($X$ axis) to the major axis of the bar in the direction of rotation of the Galaxy.

Modeling of the galactic potential with a bar was carried out by varying the following parameters
of the bar model. The bar mass was taken from the range of values: $M_{bar}=[430 - 110]\times M_G$ (in this case, the bar mass was subtracted from the bulge mass), the bar rotation speed values from interval: $\Omega_{b}=[10 - 60]$ km s$^{-1}$ kpc$^{-1}$, bar length from the interval: $q_b=[5 - 2]$ kpc. Looking ahead, we note that in the problem of calculating the probability of GC capture by the bar (Section 7), the ratio of the ellipsoid axes $a/b=[2 - 5]$ was also varied (see also Table~\ref{t:model-III}).

Based on information in numerous literature, the following bar parameters were used as basic:
$M_{bar}=430\times M_G=10^{10} M\odot$, which corresponds to 95\% of the bulge mass;
$\Omega_{b}=40$ km s$^{-1}$ kpc$^{-1}$; $q_b=5$ kpc; $\theta_{b}=25^o$; the ratio of the major axis of the ellipsoid to the minor axis $a/b=2.38$ in the projection $X-Y$, the flatness of the bar in the vertical direction $a/c=3.03$. The last four parameters are adopted, in particular, by Palou$\breve{s}$ et al. (1993).

 {\begin{table}[t]                                    
 \caption[]
 {\small\baselineskip=1.0ex
Values of parameters of the galactic potential model,$M_G=2.325\times 10^7 M_\odot$
}
 \label{t:model-III}
 \begin{center}\begin{tabular}{|c|r|}\hline
 $M_b$ &   443 M$_G$ \\
 $M_d$ &  2798 M$_G$ \\
 $M_h$ & 12474 M$_G$ \\
 $b_b$ & 0.2672 kpc  \\
 $a_d$ &   4.40 kpc  \\
 $b_d$ & 0.3084 kpc  \\
 $a_h$ &    7.7 kpc  \\
\hline\hline
 $M_{bar}$ & 110 -- 430 M$_G$ \\
 $\Omega_b$ & 10 -- 60 km s$^{-1}$ kpc$^{-1}$ \\
 $q_b$     &  2 -- 5 kpc  \\
 $\theta_{b}$ &  $25^o$   \\\hline
 $a/b$ & 2 -- 5  \\
 $a/c$ & 3.03  \\
   \hline
 \end{tabular}\end{center}\end{table}}

To integrate the equations of motion, we used the fourth-order Runge–Kutta algorithm. The value of
the peculiar speed of the Sun relative to the local standard of rest was taken equal to
$(u_\odot,v_\odot,w_\odot)=(11.1,12.2,7.3)\pm(0.7,0.5,0.4)$ km s$^{-1}$
according to work of Sch\"onrich et al. (2010). The elevation of the Sun above the Galactic plane is taken to be 17 pc in accordance with the work of Bobylev and Bajkova (2016).

\section{DATA}

Data on the proper motions of globular clusters are taken from the new catalog of Vasiliev and Baumgardt (2021), compiled from observations of Gaia EDR3 (Brown et al., 2021). Radial velocities are taken from Vasiliev (2019). The new average distances to globular clusters are taken from Baumgardt and Vasiliev (2021). A comparative analysis of new data on proper motions and distances with previous versions of catalogs is given, for example, in the work of Bajkova and Bobylev (2022). Here we only note that the accuracy of measuring new proper motions has increased on average by a factor of two compared to the measurements of Gaia DR2. As an analysis of the radial velocities presented in the catalog by Vasiliev (2019) shows, their errors are not significant and for the vast majority of GCs in our sample are less than 1\% of the values of the velocities themselves.
Proper motions and distances are burdened by more significant errors. All uncertainties in the data (about positions, distances, proper motions, radial velocities), as well as the uncertainty of the vector of the peculiar velocity of the Sun according to Sch\"onrich et al. (2010), were taken into account by the Monte Carlo method (1000 implementations) when calculating the uncertainties of the 6D-space components $(x_o,y_o,z_o,u_o,v_o,w_o)$ (see notation in Bajkova and Bobylev (2022)), later used to calculate orbits. The equations of motion of a test particle in the gravitational potential of the Galaxy are also presented in detail in the work of Bajkova and Bobylev (2022).

\section{SELECTION OF GLOBULAR CLUSTERS}

The catalog of GCs at our disposal, Bajkova and Bobylev (2022), contains 152 objects. The selection
of globular clusters belonging to the bulge/bar region from this set was made in accordance with the purely geometric criterion considered in the work of Massari et al. (2019) and also used by us in the work of Bajkova et al. (2020). It is very simple and consists in selecting GCs whose apocentric orbital distance does not exceed 3.5 kpc. Orbits are calculated in an axisymmetric potential.

Note that the use of classification results obtained in the works of Massari et al. (2019) and Bajkova et al. (2020) based on proper motions from the GaiaDR2 catalog and distances from Harris (2010), is not entirely correct in this work, since we are dealing with new data. Changes in data (especially in distances) can greatly affect the orbital characteristics of GCs. Therefore, we carried out a new classification. Application of a constraint on the apocentric distance
of the GC ($apo\leq 3.5$ kpc) made it possible to identify 39 objects, most of which coincided with the results of the previous classification, but there were also differences.

\begin{figure*}
{\begin{center}
 \includegraphics[width=0.3\textwidth,angle=-90]{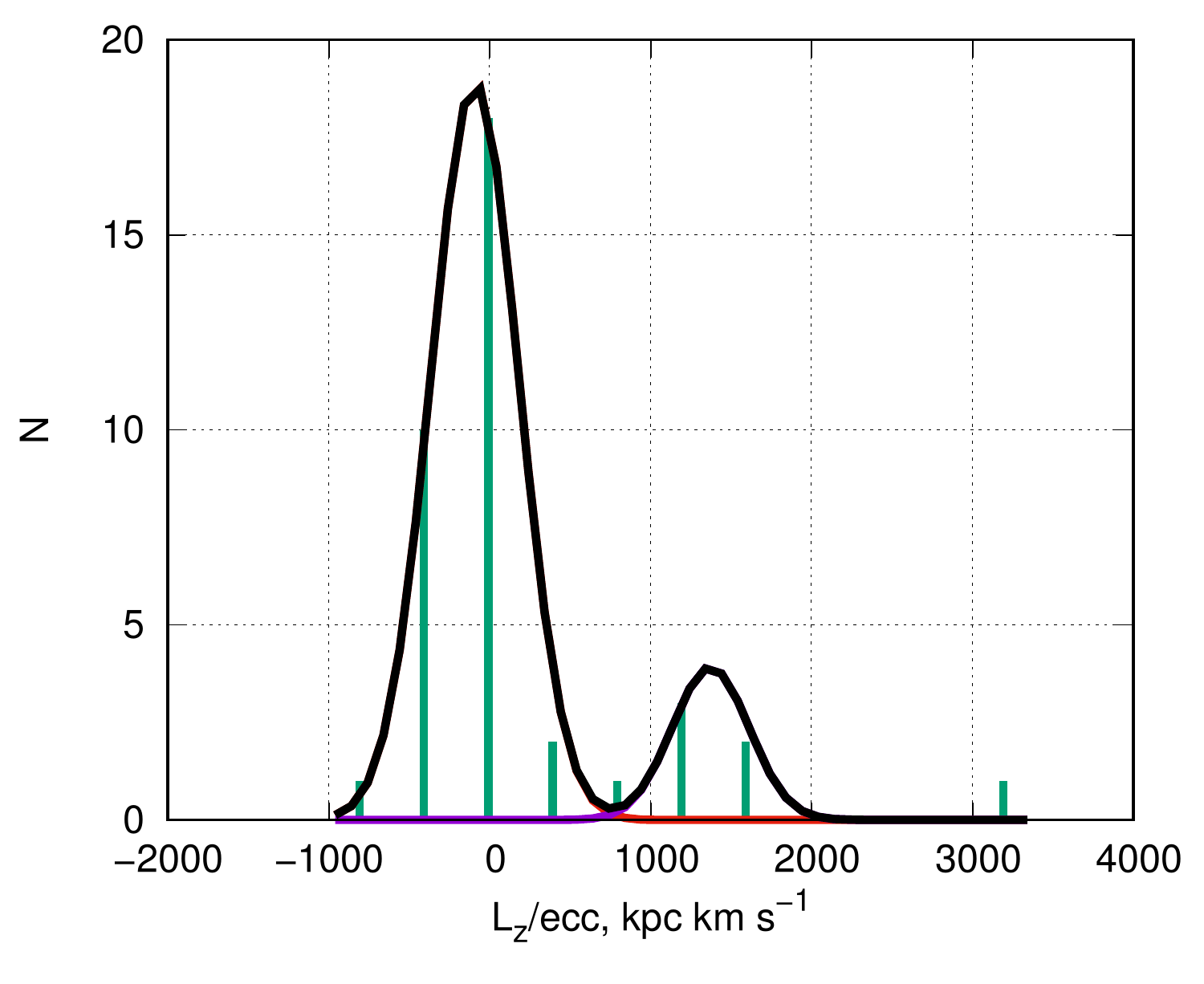}
\includegraphics[width=0.3\textwidth,angle=-90]{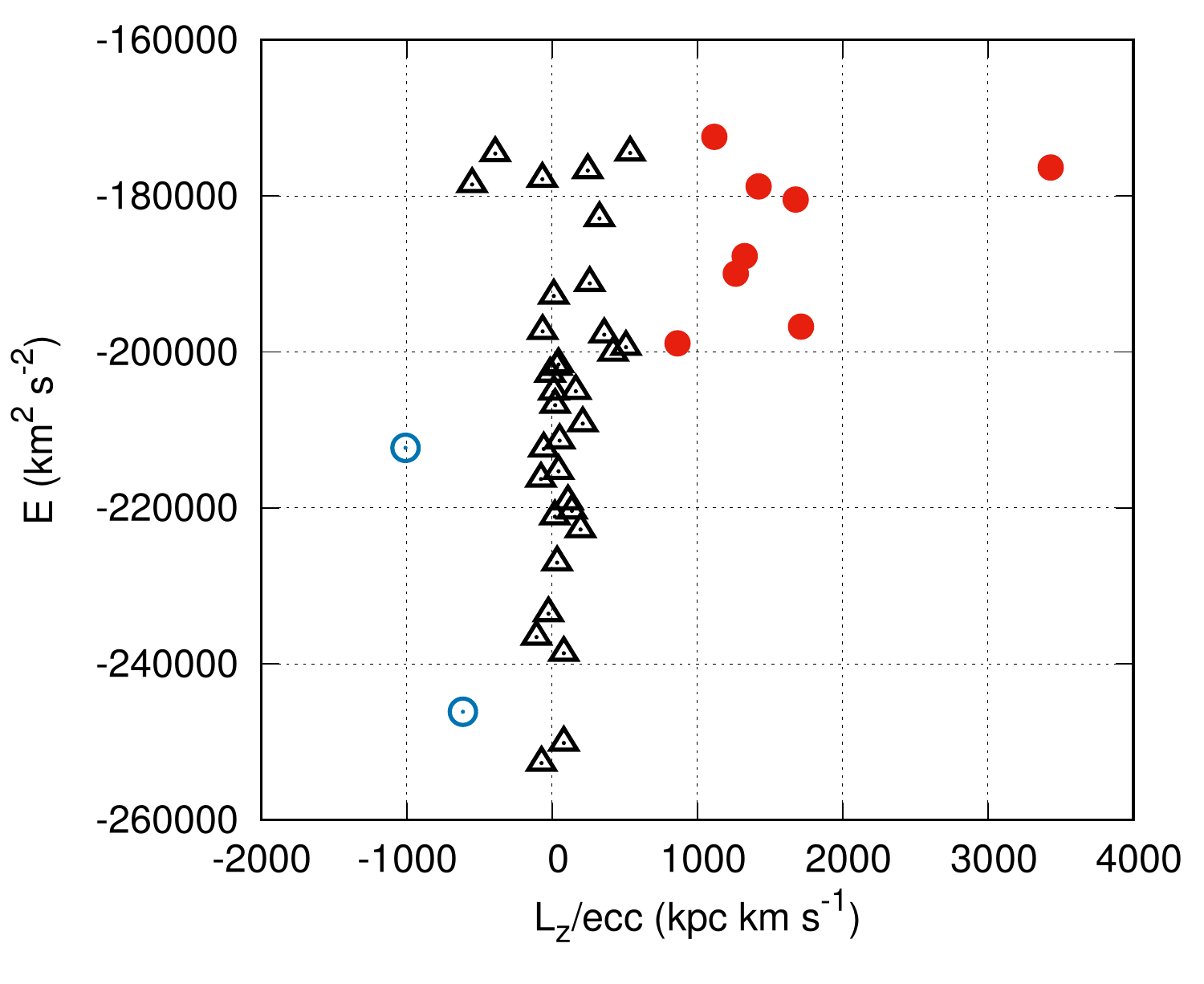}\

\hskip 0.5cm{a)}      \hskip 6.5cm{b)}\

\medskip

\includegraphics[width=0.3\textwidth,angle=-90]{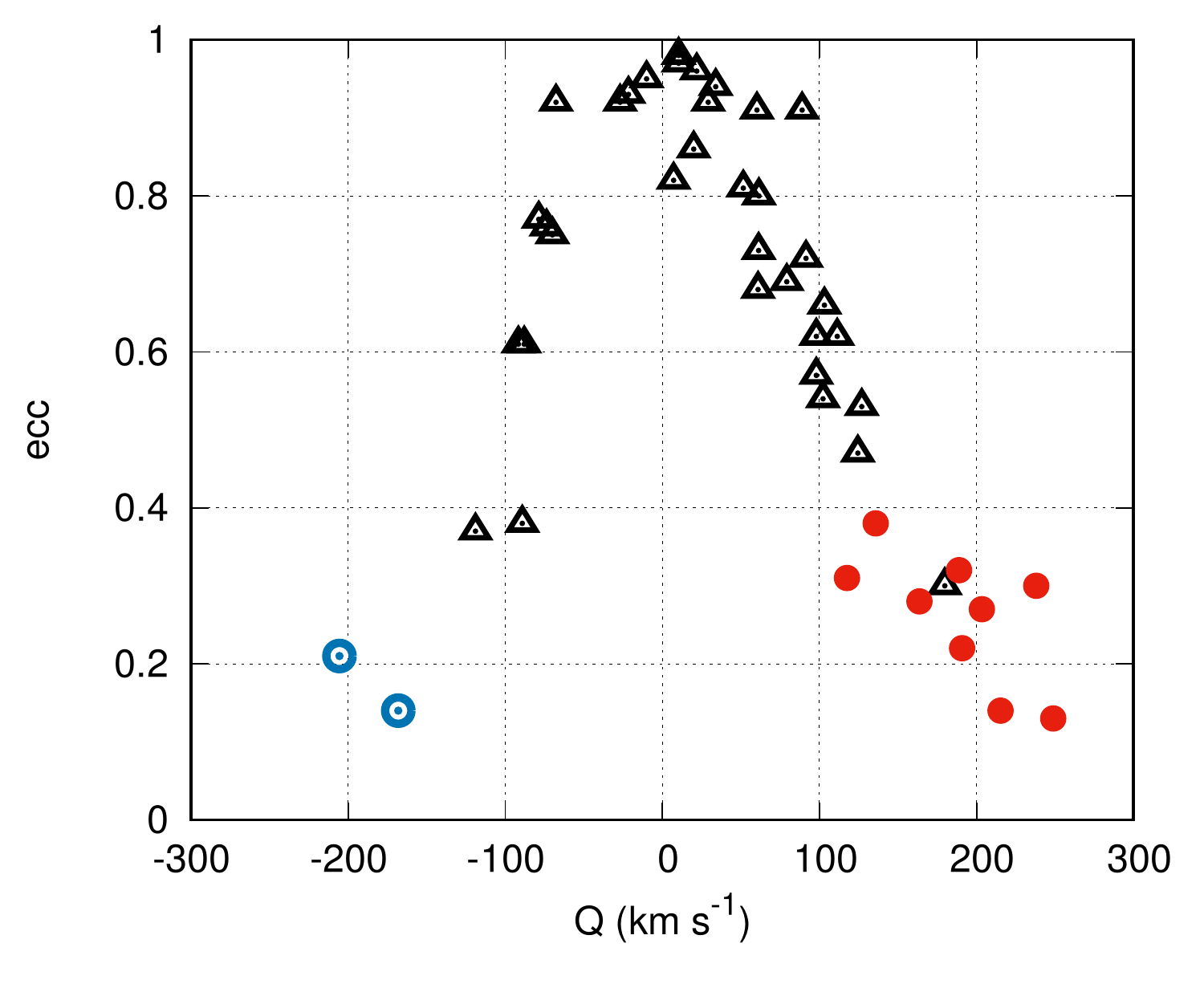}
\includegraphics[width=0.3\textwidth,angle=-90]{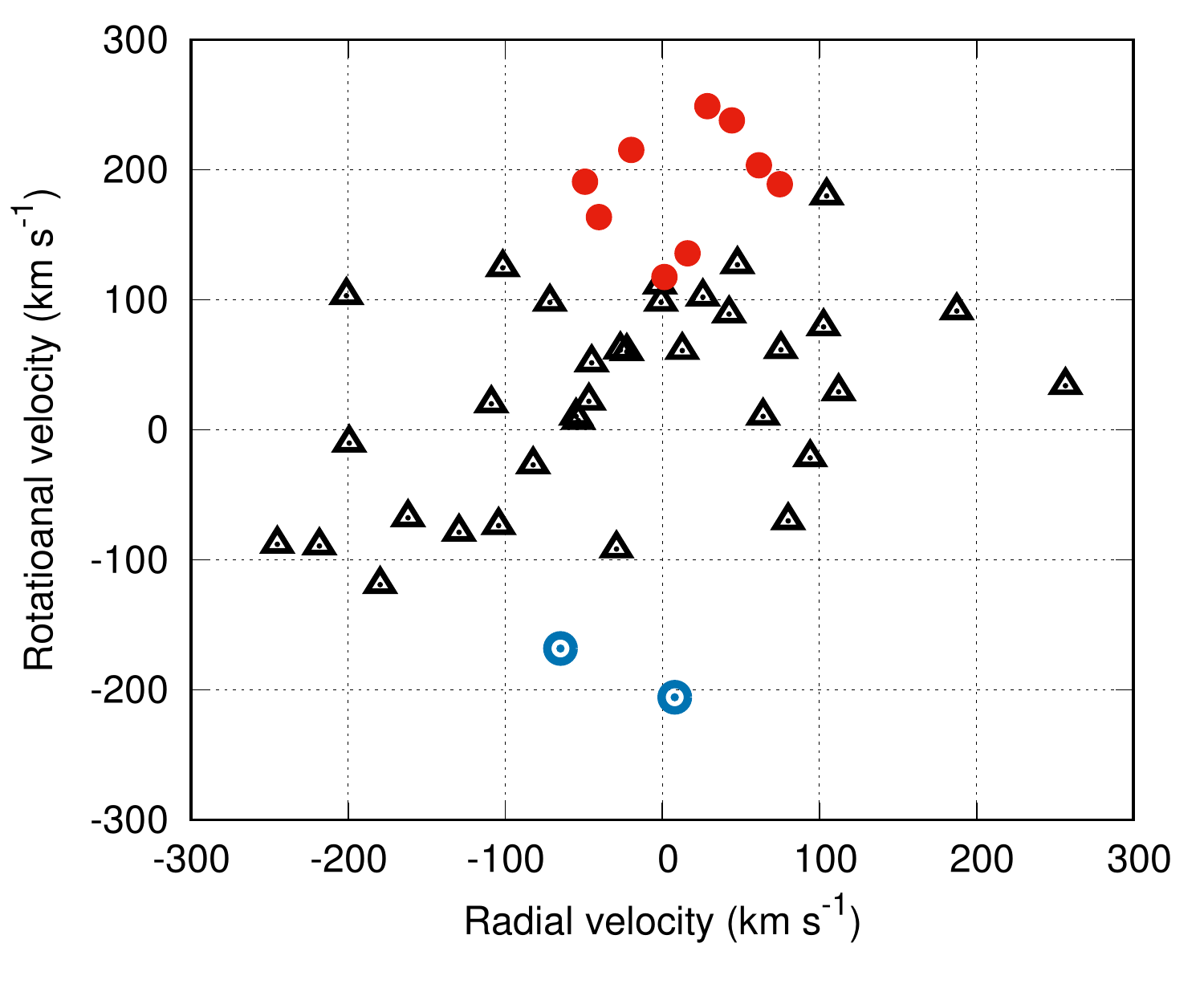}\

\hskip 0.5cm{c)}      \hskip 6.5cm{d)}\

\medskip

\caption{\small Identification of GCs belonging to the bulge/bar using the probabilistic criterion proposed in the work of Bajkova et al. (2020). Panel (a) shows a histogram of the GC distribution according to the $L_Z/ecc$ parameter. The bimodality of the distribution is clearly visible. The left Gaussian determines the probability that the GC belongs to the bulge, the right Gaussian determines the probability that the GC belongs to the disk. Panel (b) shows the $"L_Z/ecc$ -- total energy $E"$ diagram for the resulting sample of 45 GCs. Panel (c) shows the $"$circular velocity -- eccentricity$"$ diagram. Panel (d) shows the $"$radial velocity -- circular velocity$"$ diagram. Black triangles indicate GCs belonging to the bulge, red closed circles -- GCs belonging to the disk, blue open circles -- GCs with retrograde orbits, also belonging to the disk.}
\label{fL}
\end{center}}
\end{figure*}

Next, from the resulting set of GCs, we identified GCs belonging to the disk component of the Galaxy. To do this, we applied a probabilistic method for dividing GCs into subsystems of the Galaxy proposed in the work of Bajkova et al. (2020). It is based on the bimodality of the distribution of the parameter $L_Z/ecc$, where $L_Z$ is the vertical ($Z$) component of the angular momentum, ecc is the orbital eccentricity. An illustration of themethod is shown in Fig. ~\ref{fL}. As a result of applying this method, it was possible to identify 9 GCs belonging to the disk. The remaining 30 belong directly to the bulge. To the resulting sample of 39 objects, we added six more GCs belonging to the bulge region according to the classification of Massari et al. (2019), but not included in our sample due to a slight excess of the apocentric distance of the orbits
by 3.5 kpc.

Thus, we formed a sample of 45 GCs in the central region of the Galaxy, of which 34 belong to the bulge, 9 -- to the disk (ESO456-SC78, Terzan 3, Pismis 26, NGC6256, NGC 6304, NGC6569, NGC6540,
NGC6539, NGC6553), and 2 more objects (NGC6325, Djorg 2), having significant negative
rotational velocities, presumably also belong to the disk. A complete list of 45 objects with names and serial numbers is given below in Table~\ref{t:f}. The first 34 objects belong to the bulge, and the last 11 belong to the disk. The first 30 bulge objects are GCs identified as a result of applying the probabilistic separation method, the remaining four are GCs according to the classification of Massari et al. (2019). Designations of orbital parameters and formulas for their calculation can be found, in particular, in the work of Bajkova and Bobylev (2022), which also provides a Table with a complete set of parameters calculated in the axisymmetric potential we adopted. Fig.~\ref{fL} also shows the diagrams $"L_Z/ecc$ -- total energy $E"$, $"$circular speed -- eccentricity$"$ and $"$radial speed -- circular velocity$"$, from which the division of the GCs into bulge and disk subsystems is clearly visible.

\begin{figure*}
{\begin{center}
     \includegraphics[width=0.3\textwidth,angle=-90]{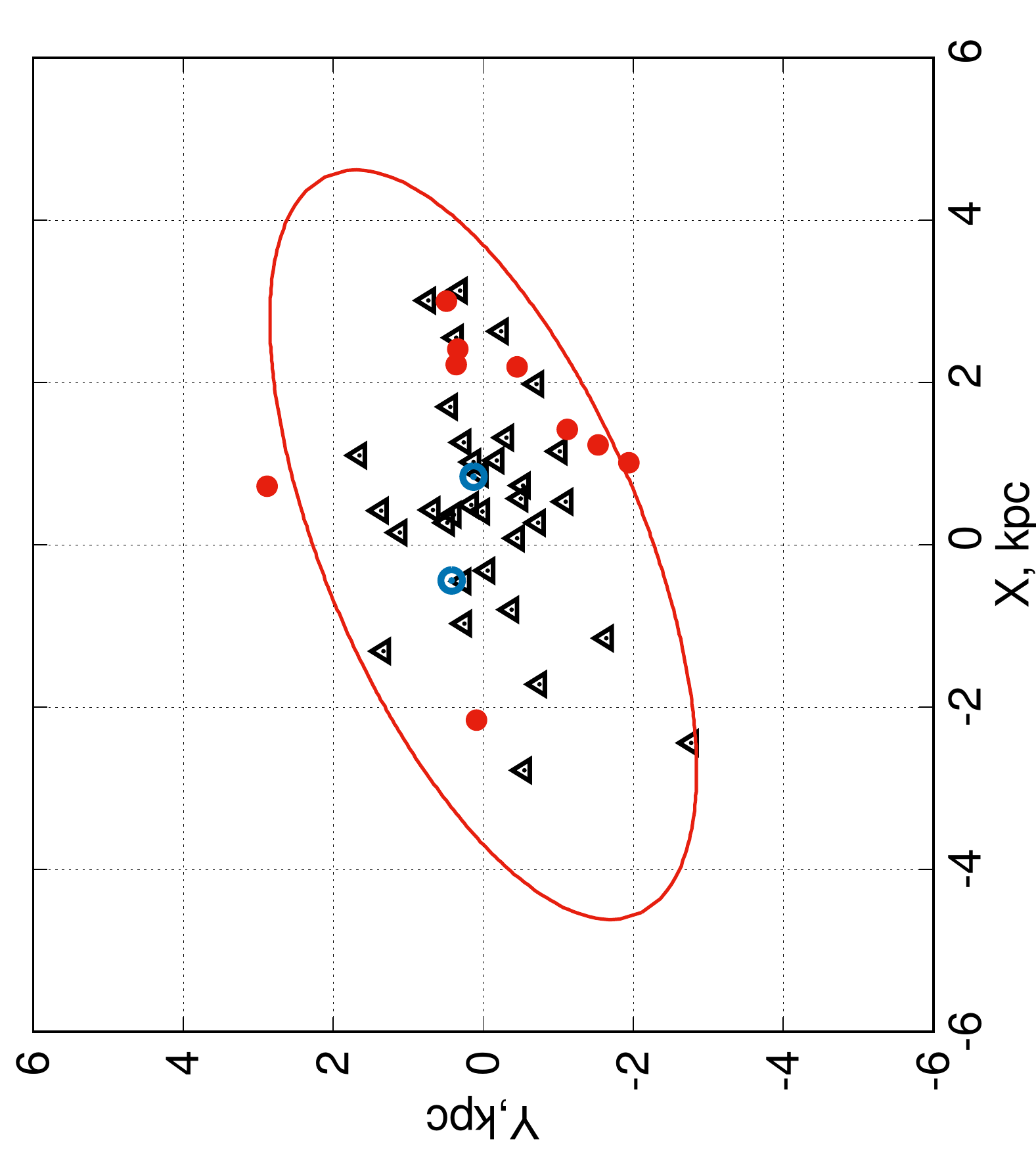}
     \includegraphics[width=0.3\textwidth,angle=-90]{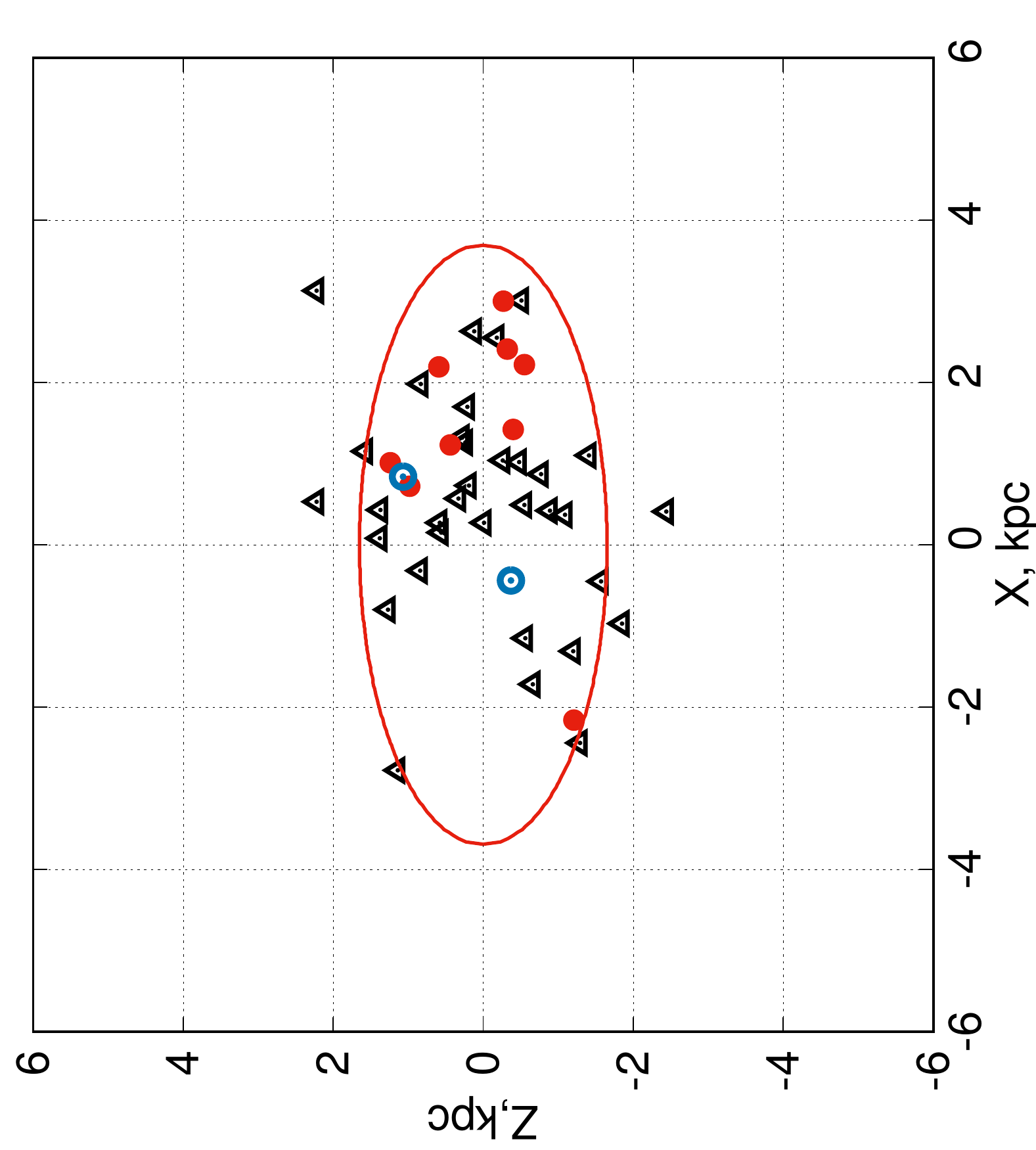}\

     \medskip

\hskip 0.5cm{a)}      \hskip 5.25cm{b)}\

\medskip

  \caption{\small Distribution of selected 45 GCs in the $X-Y$ (a) and $X-Z$ (b) projections of the galactic coordinate system. The designations of objects are the same as in the previous figure. The red line shows a cross section of the bar (triaxial ellipsoid) with basic parameters.}
\label{fB}
\end{center}}
\end{figure*}

Figure~\ref{fB} shows the distribution of the selected 45 GCs in the $X-Y$ and $X-Z$ projections of the galactic coordinate system. The figure also shows sections of the triaxial ellipsoid that describes the bar, with the basic parameters listed above.

 {\begin{table*}[t]                                    
 \caption[]
 {\small\baselineskip=1.0ex
List of clusters whose orbits were studied in this work.
  }
 \label{t:f}
 \begin{center}\begin{tabular}{|r|l||r|l||r|l||r|l||r|l|}\hline
 №  & ID       & №  & ID       & №  &  ID      & №  & ID         & №  &  ID      \\\hline
 1  &NGC 6144  &10  & BH 229   & 19 & NGC 6453 & 28 & NGC 6642   & 37 & NGC 6540 \\\hline
 2  &ESO 452-SC11&11  & Liller 1 & 20 & Terzan 9 & 29 & NGC 6717   & 38 & NGC 6325 \\\hline
 3  &NGC 6266  &12  & NGC 6380 & 21 & NGC 6522 & 30 & NGC 6723   & 39 & Djorg 2  \\\hline
 4  &NGC 6273  &13  & Terzan 1 & 22 & NGC 6528 & 31 & Terzan 3   & 40 & NGC 6171 \\\hline
 5  &NGC 6293  &14  & NGC 6401 & 23 & NGC 6558 & 32 & NGC 6256   & 41 & NGC 6316 \\\hline
 6  &NGC 6342  &15  & Palomar 6& 24 & NGC 6624 & 33 & NGC 6304   & 42 & NGC 6388 \\\hline
 7  &NGC 6355  &16  & Terzan 5 & 25 & NGC 6626 & 34 & Pismis 26  & 43 & NGC 6539 \\\hline
 8  &Terzan 2  &17  & NGC 6440 & 26 & NGC 6638 & 35 & NGC 6569   & 44 & NGC 6553 \\\hline
 9  &Terzan 4  &18  & Terzan 6 & 27 & NGC 6637 & 36 & ESO 456-SC78 & 45 & NGC 6652 \\\hline
 \end{tabular}\end{center}\end{table*}}

\section{CALCULATION OF DOMINANT FREQUENCIES}

\begin{figure*}
{\begin{center}

\includegraphics[width=0.4\textwidth,angle=-90]{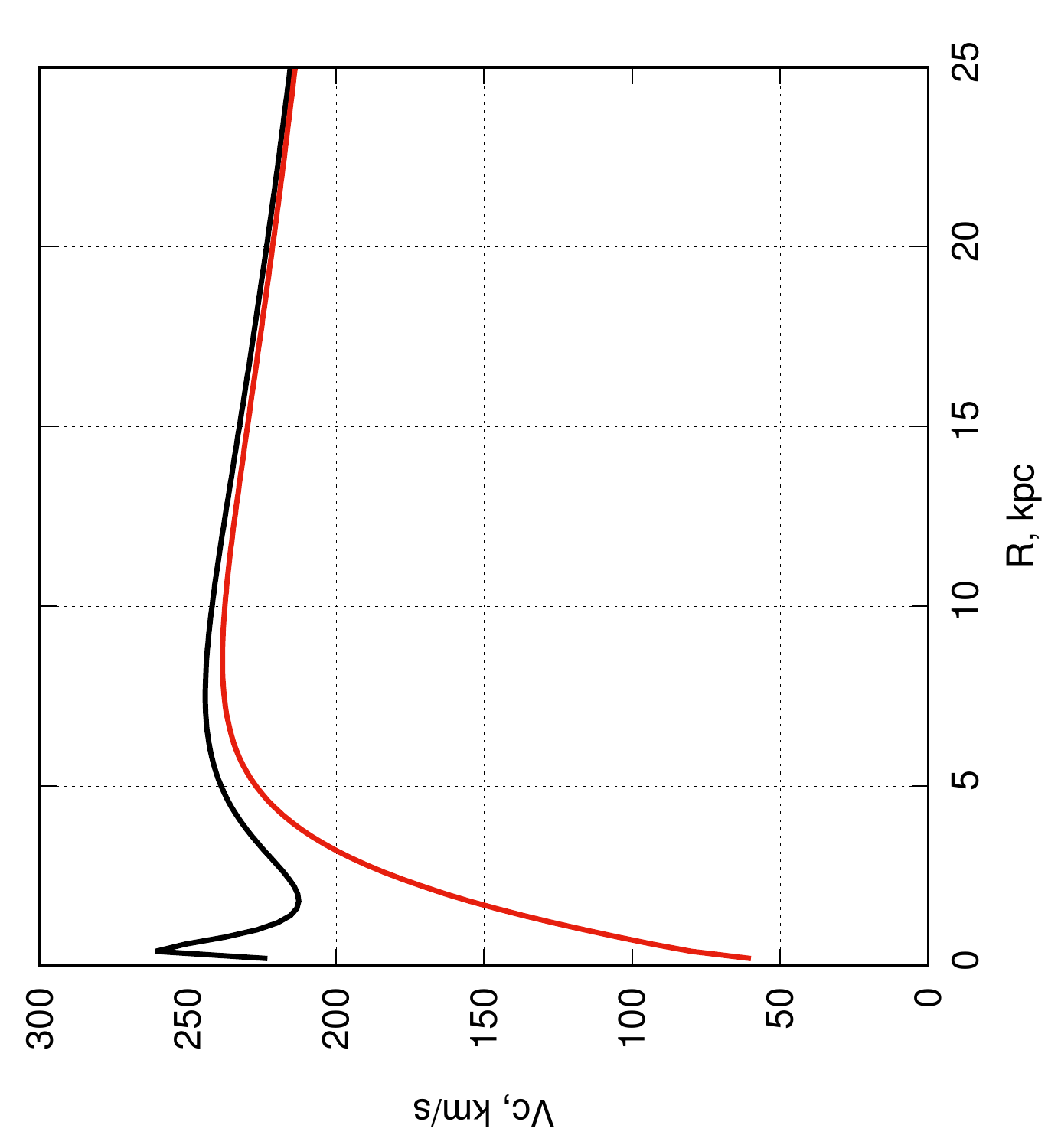}\
\caption{\small The rotation curve of the Galaxy with an axisymmetric potential without a bar (black line) and a non-axisymmetric potential including a bar (red line).}
\label{fcomp}
\end{center}}
\end{figure*}

For comparison, Fig.~\ref{fcomp} shows the rotation curve of the Galaxy for an axisymmetric potential without a bar (black line) and the rotation curve of a non=axisymmetric potential with a bar, averaged over azimuth (red line). Note that when adding a bar to the model, the bar mass $M_{bar}$ was subtracted from the bulge mass $M_b$. The rotation curve of the non-axisymmetric potential was obtained with the basic bar parameters listed in Section 2.2. The figure shows that the introduction of a bar with a mass almost equal to the mass of the bulge significantly affected the shape of the rotation curve in the central region of the Galaxy, which inevitably affected the orbital dynamics of the selected globular clusters.

Let’s move on to our task.

Spectral dynamics of orbits in practice is based on calculating the modulus of the discrete Fourier
transform (DFT) of uniform time series $x_n, y_n, z_n$ and $R_n, n=0,...,N-1$ ($N$ is the row length), by coordinates $X, Y, Z$  and cylindrical distance $R$.

So, for example, for the series $x_n, n=0,...,N-1$,
the formula for the DFT module will look like this:
\begin{equation}
\label{Four}
X_k=|\frac{1}{N} \sum_{n=0}^{N-1} x_n \exp{(-\jmath\frac{2\pi\times n\times k}{N})}|,
\end{equation}
where $ j=\sqrt{-1}~~,k=0,..., N-1$.

Similarly, you can write the formula for other sequences. In this case, the length of the series is
chosen equal to $N=2^\alpha$, where $\alpha$ is an integer greater than 0, so that the Fast Fourier Transform (FFT) algorithm can be used to calculate the DFT. The required length of the series is achieved by padding the actual series with zeros.

In our case, the length of the actual sequences is 2500, since we integrate orbits 2.5 billion years ago with an integration interval of 1 million years. Before calculating the DFT, we first center the series of coordinates (i.e., we get rid of the constant component), then we supplement the resulting sequences $x_n, y_n, z_n$  and $R_n$ with zero samples at $n>2500$ until the length of the entire analyzed sequence is $N=8192=2^{13}$. Note that adding zeros to the original sequence is also useful from the point of view of increasing the accuracy of the coordinates of the spectral components, which is important in our case, since our task is to determine the maximum
(dominant) frequencies of the amplitude spectrum. Since the interval between sequence samples in time is $\Delta_t=0.001$ billion years (Gyr), then the analyzed frequency range, which is a periodic function, is $F=1/\Delta_t=1000$ Gyr$^{-1}$. The frequency discrete is $\Delta_F= F/N \approx 0.122$ Gyr$^{-1}$. In the future, for convenience, we will indicate on the graphs not the physical frequencies, but the numbers $k$ of samples of the discrete Fourier transform (\ref{Four}). The transition from $k$ to physical frequency can be made using the formula: $f=k\times\Delta_F\approx k\times 0.122$. As mentioned above, the dominant, or characteristic, frequency of the analyzed series is the one in which the power spectrum (\ref{Four}) reaches its maximum value.

\section{PROPERTIES OF DOMINANT FREQUENCIES}

\begin{figure*}
{\begin{center}
\includegraphics[width=0.275\textwidth,angle=-90]{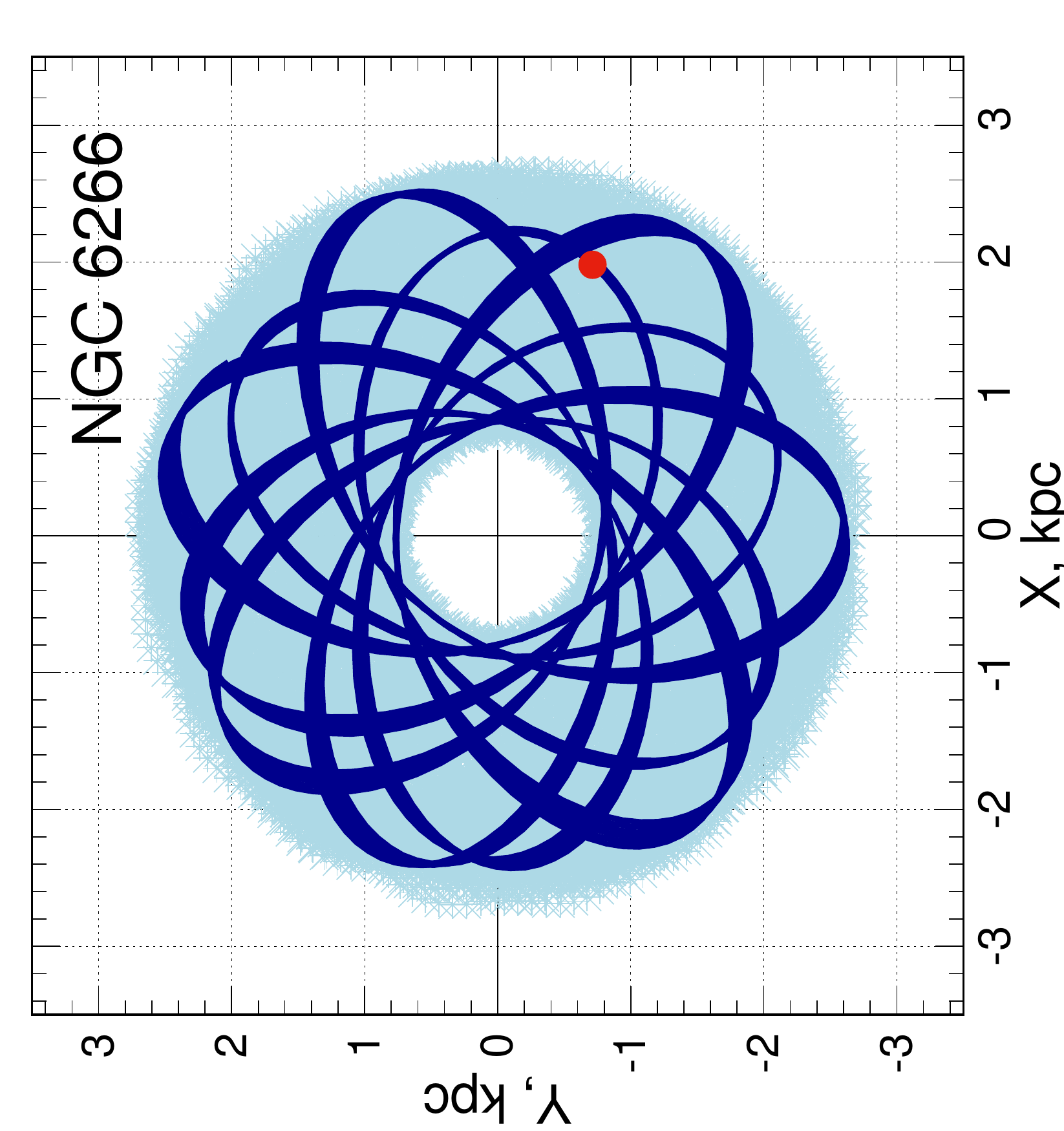}
\includegraphics[width=0.275\textwidth,angle=-90]{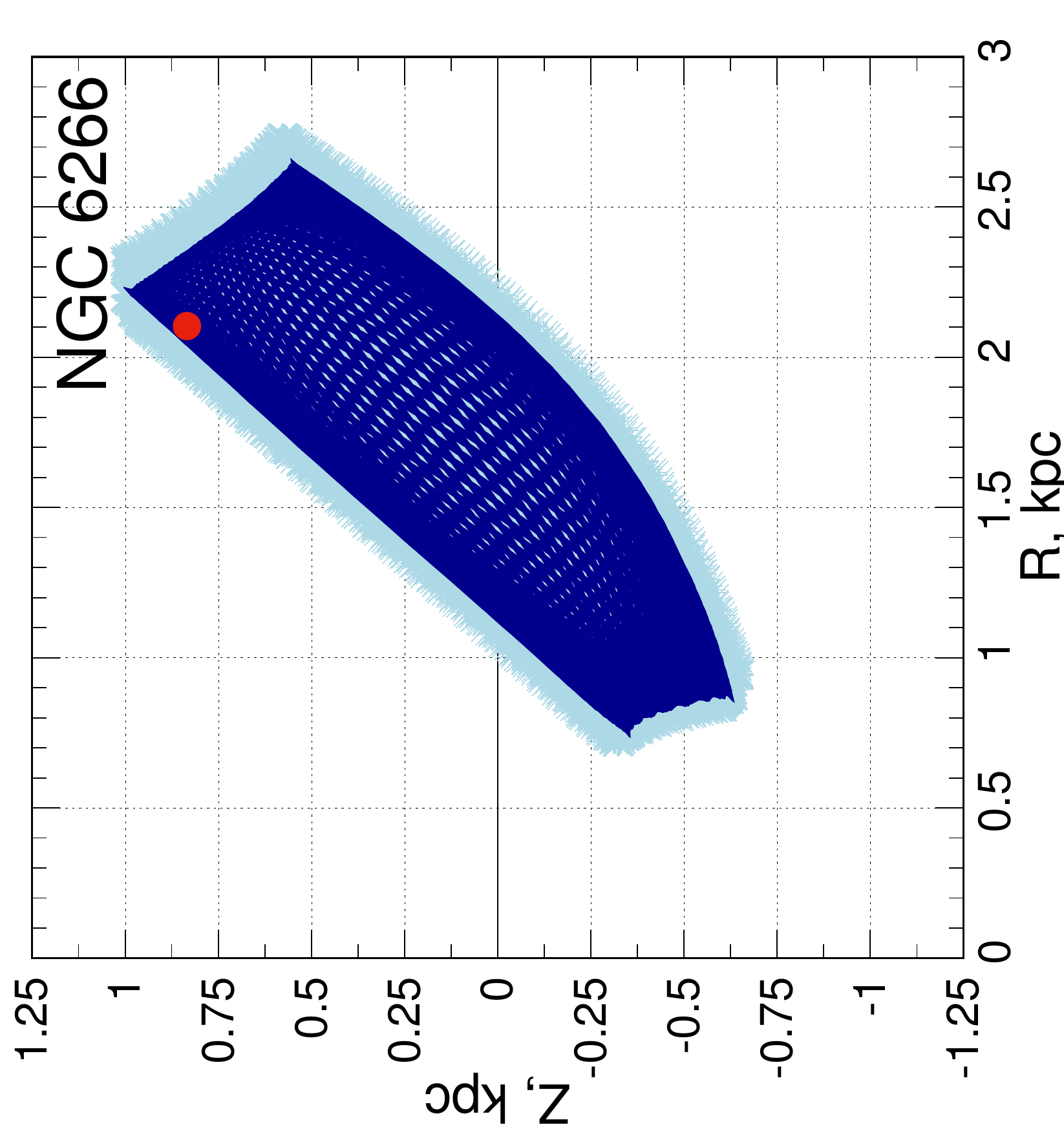}
\includegraphics[width=0.275\textwidth,angle=-90]{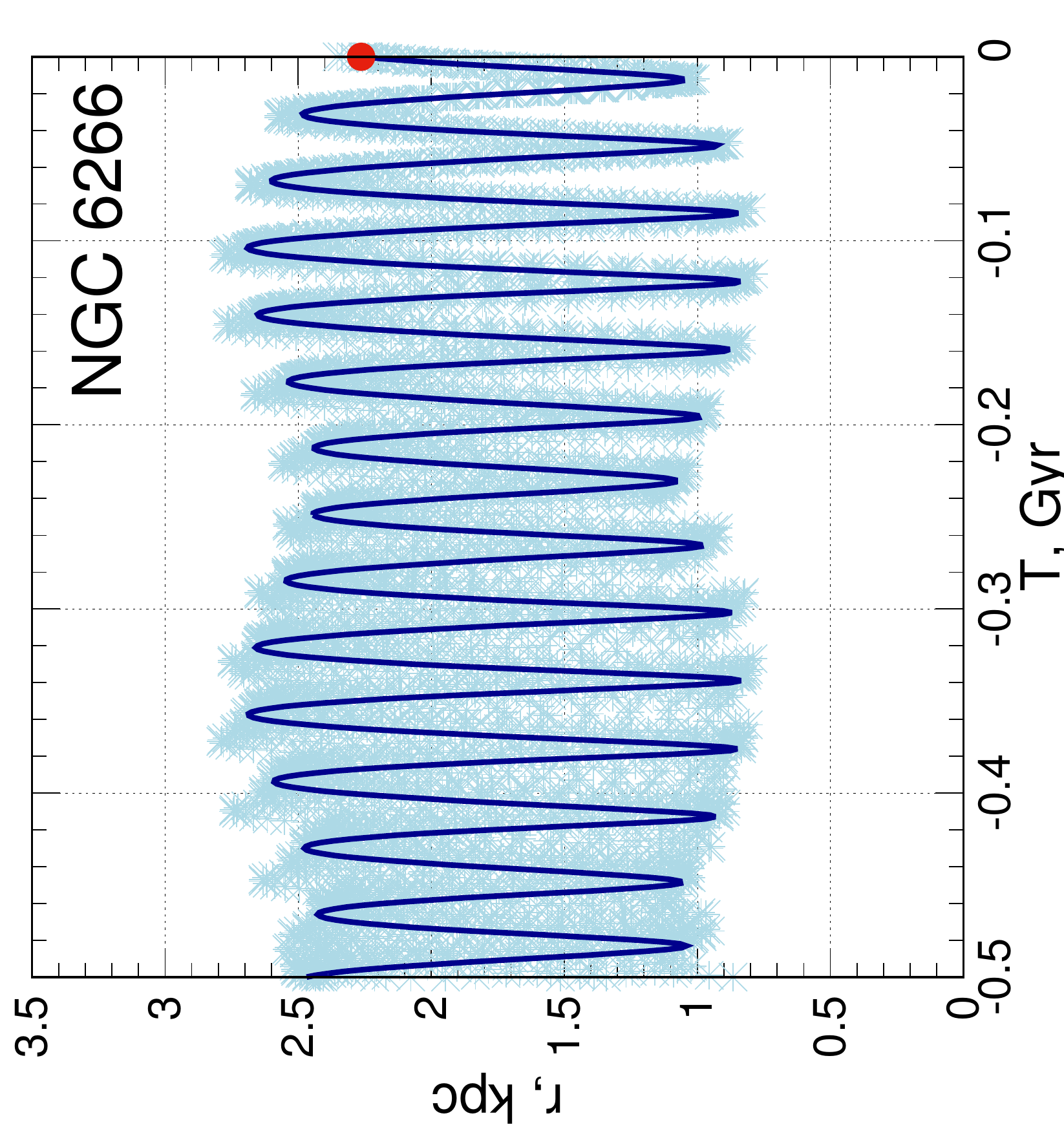}\

\medskip

\includegraphics[width=0.275\textwidth,angle=-90]{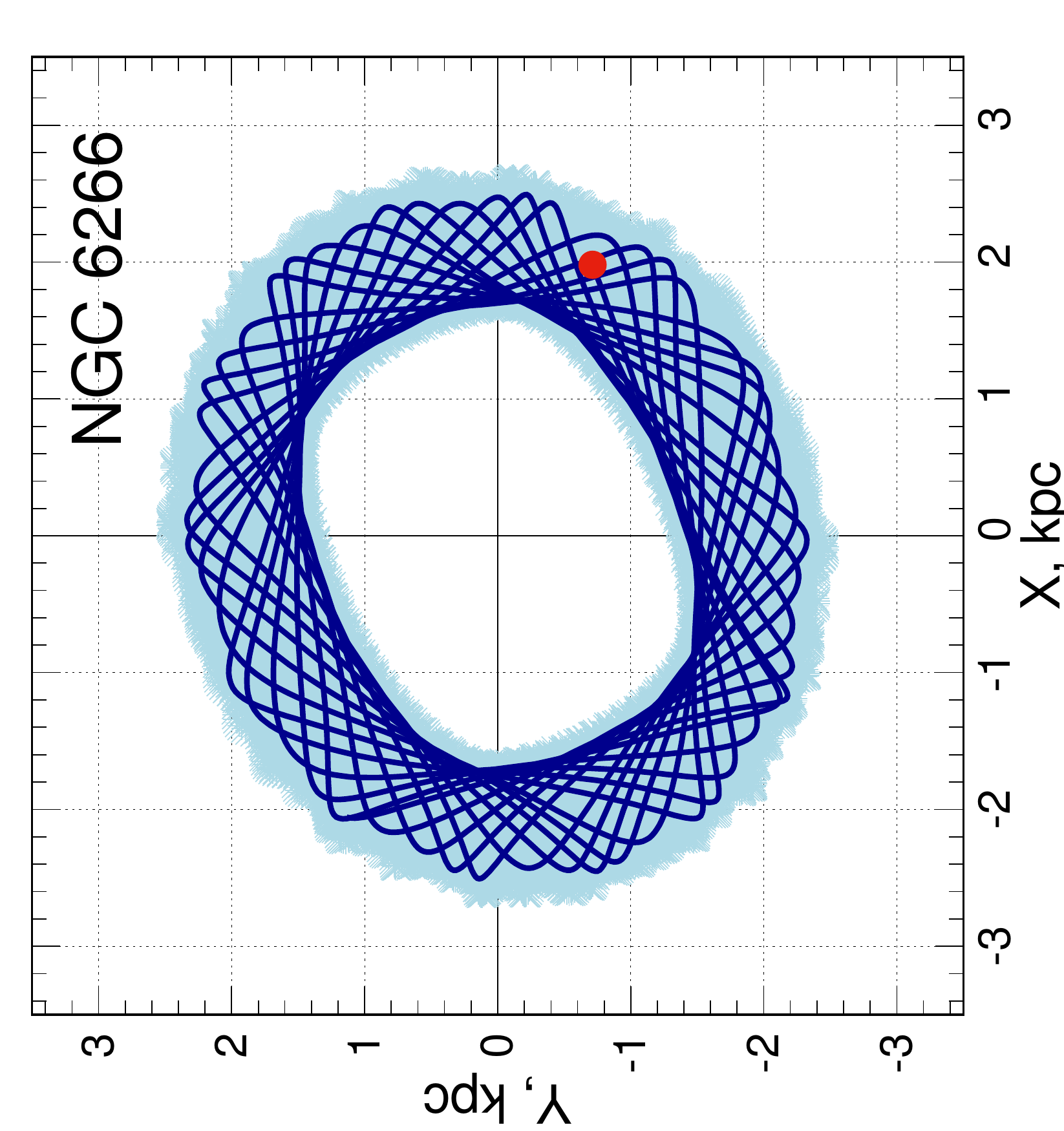}
\includegraphics[width=0.275\textwidth,angle=-90]{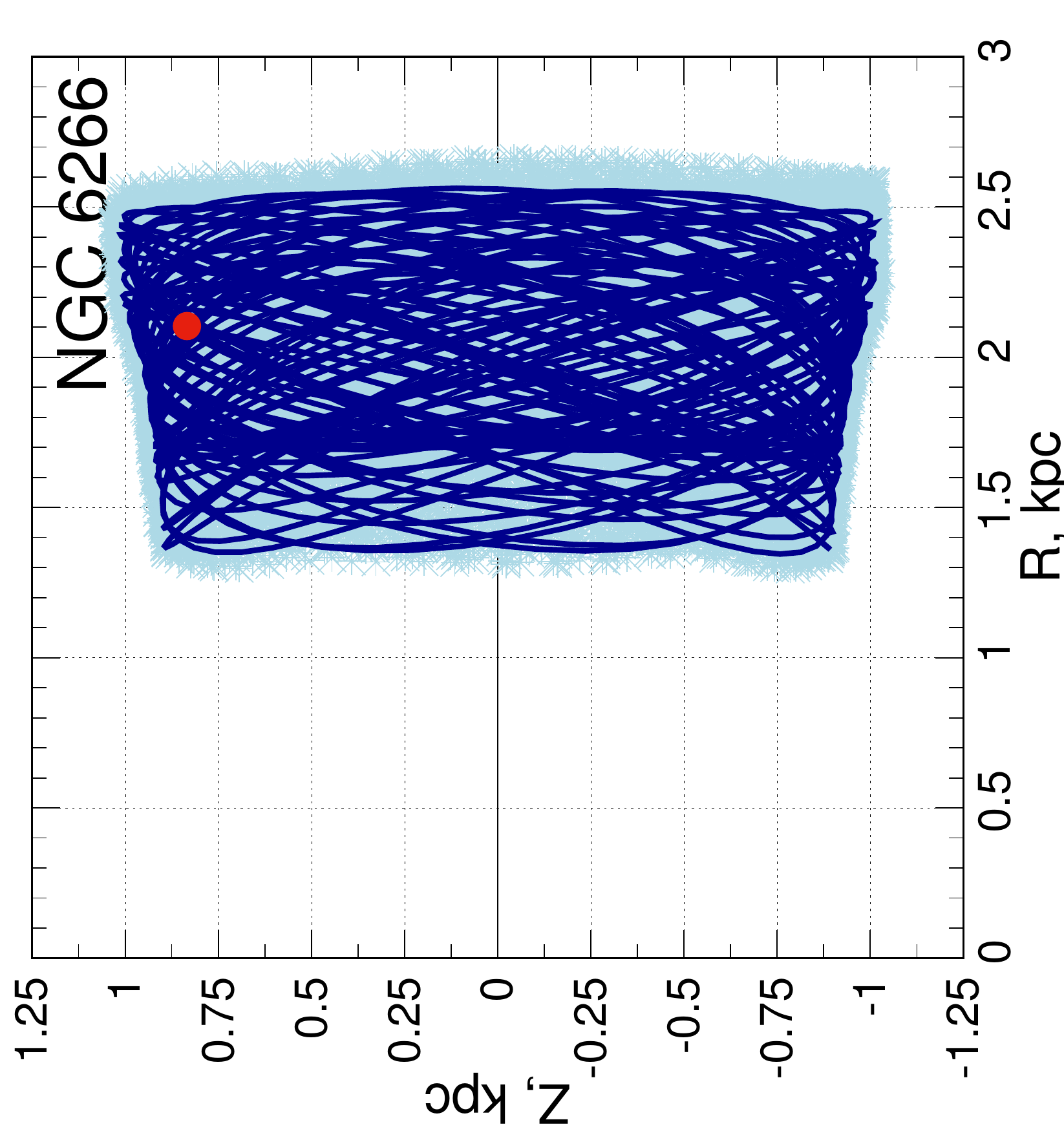}
\includegraphics[width=0.275\textwidth,angle=-90]{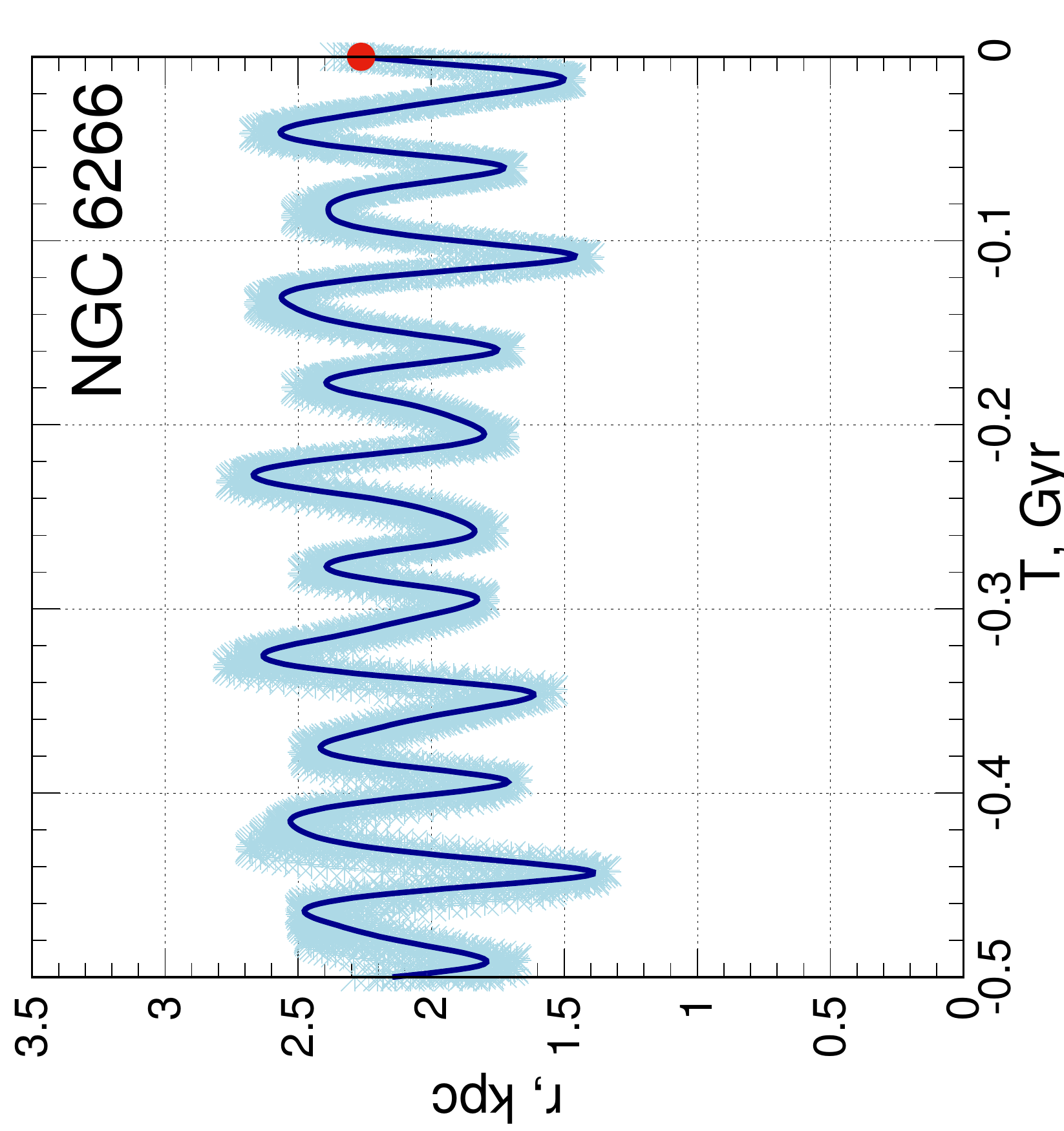}\
\caption{\small Orbits of the globular cluster NGC6266 in an axisymmetric potential (top panels) and in a non-axisymmetric potential with the following bar parameters: $M_{bar}=430\times M_G$, $q_b=5$ kpc, $\Omega_b = 40$ km s$^{-1}$ kpc$^{-1}$ (bottom panels). The $X-Y$  projections (left panels), $R-Z$  projections (middle panels) and the dependence of the orbital radius vector $r=\sqrt{X^2+Y^2+Z^2}$ versus time (right panels). The orbits corresponding to the nominal initial values of the phase space are shown in dark blue. The orbits obtained from the Monte Carlo simulation are shown in blue. In the case of a non-axisymmetric potential, the orbits are calculated in a rotating bar system.}
\label{fcomp2}
\end{center}}
\end{figure*}

\begin{figure*}
{\begin{center}
\includegraphics[width=0.175\textwidth,angle=-90]{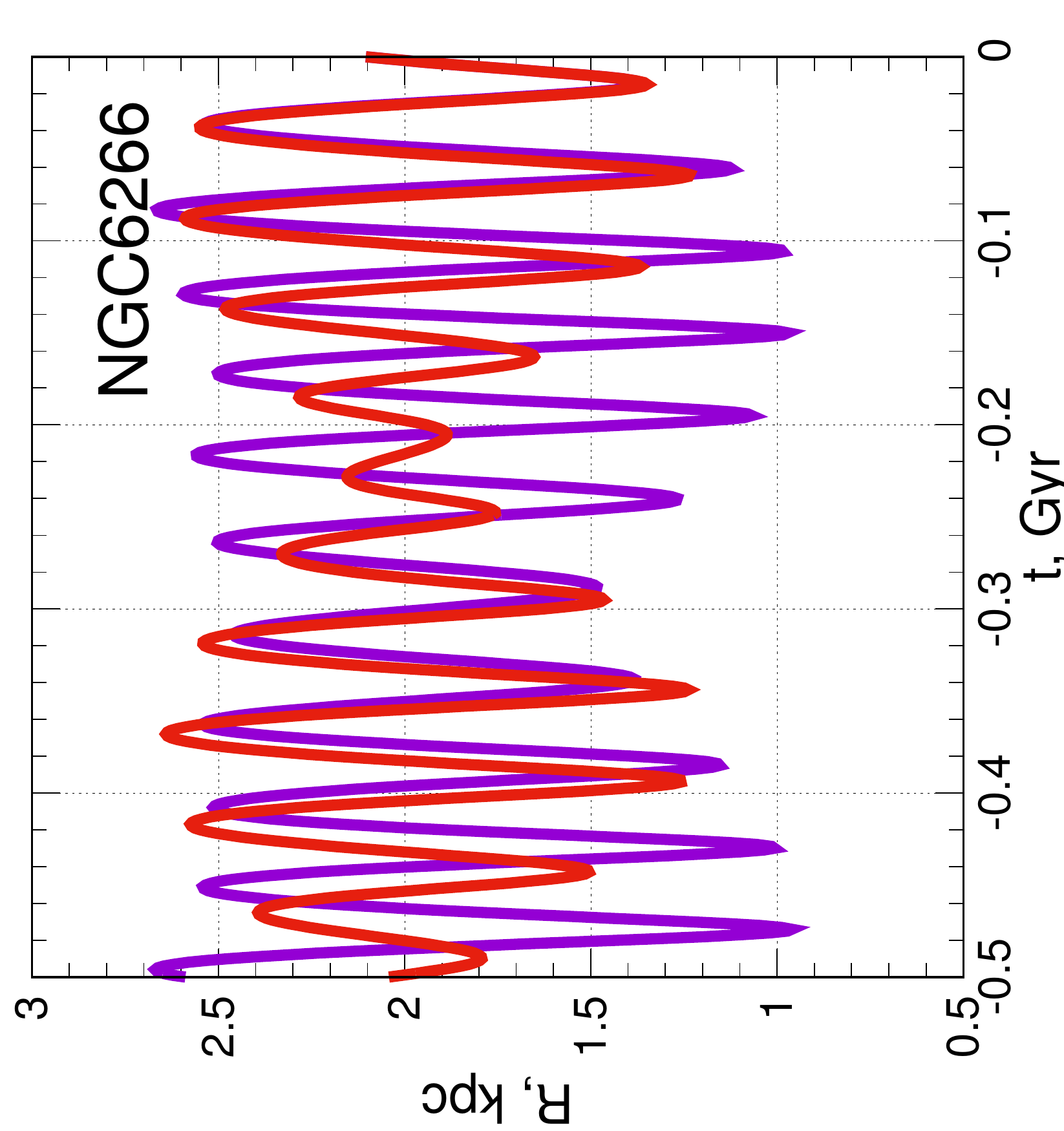}
\includegraphics[width=0.175\textwidth,angle=-90]{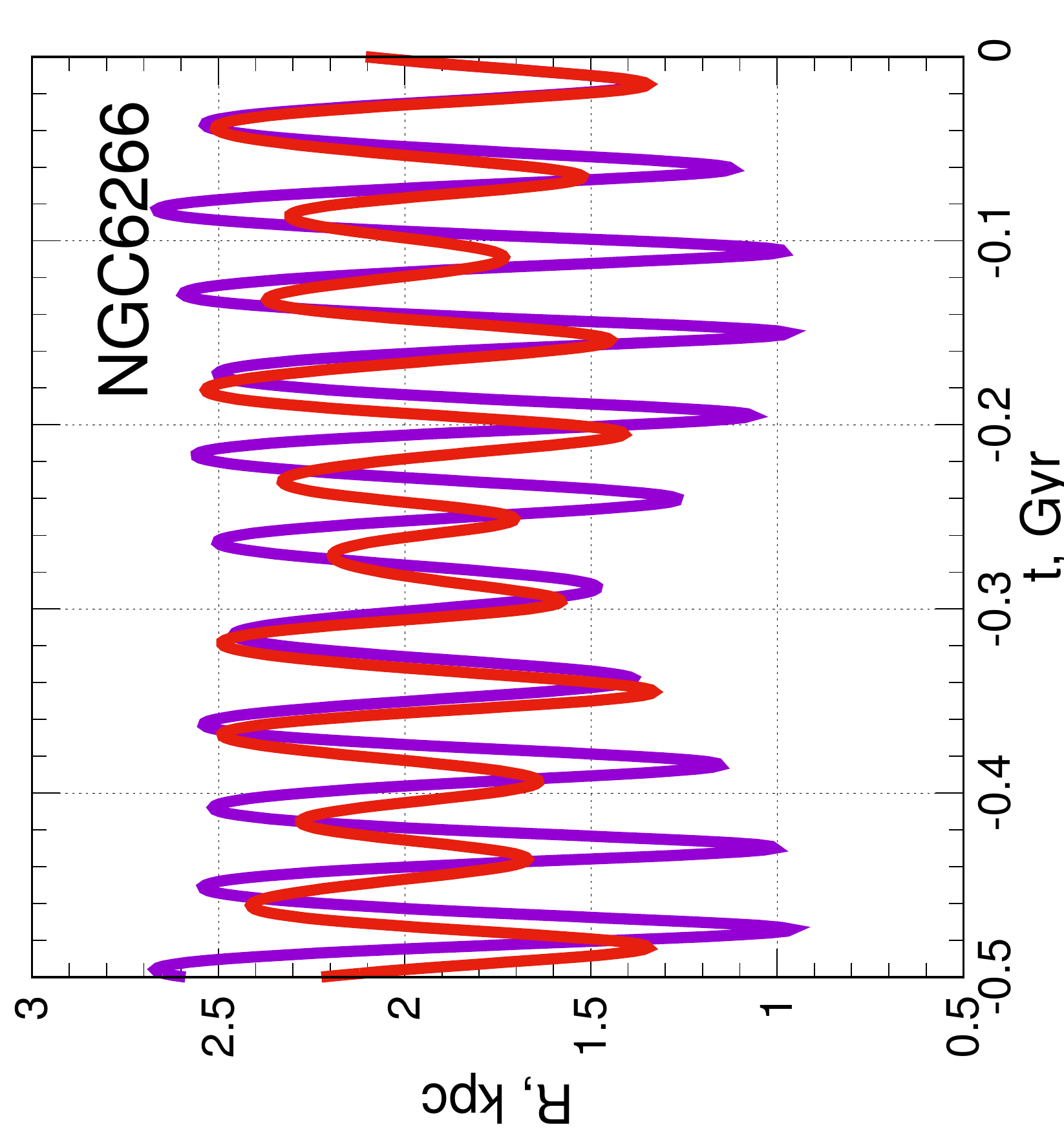}
\includegraphics[width=0.175\textwidth,angle=-90]{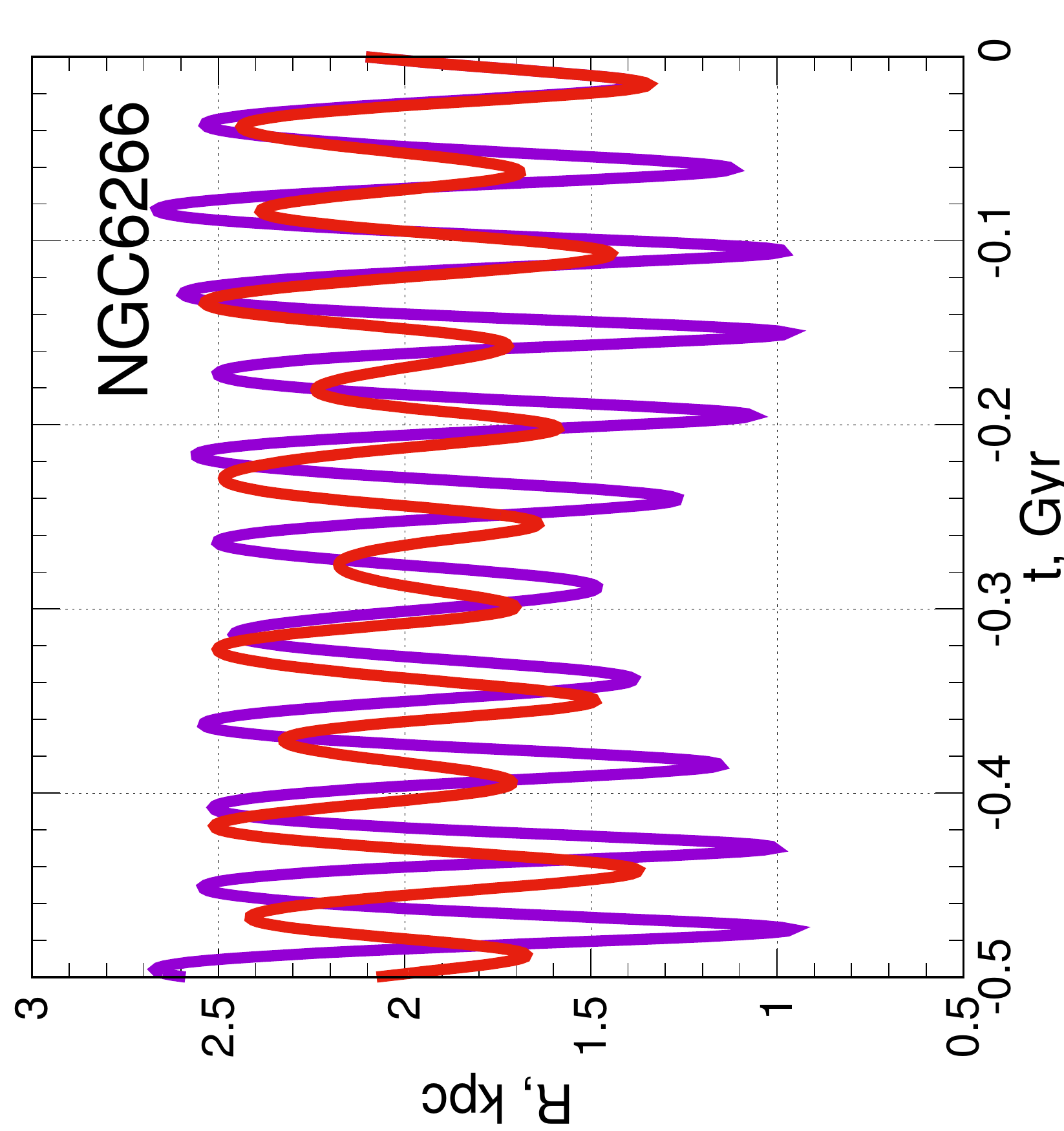}
\includegraphics[width=0.175\textwidth,angle=-90]{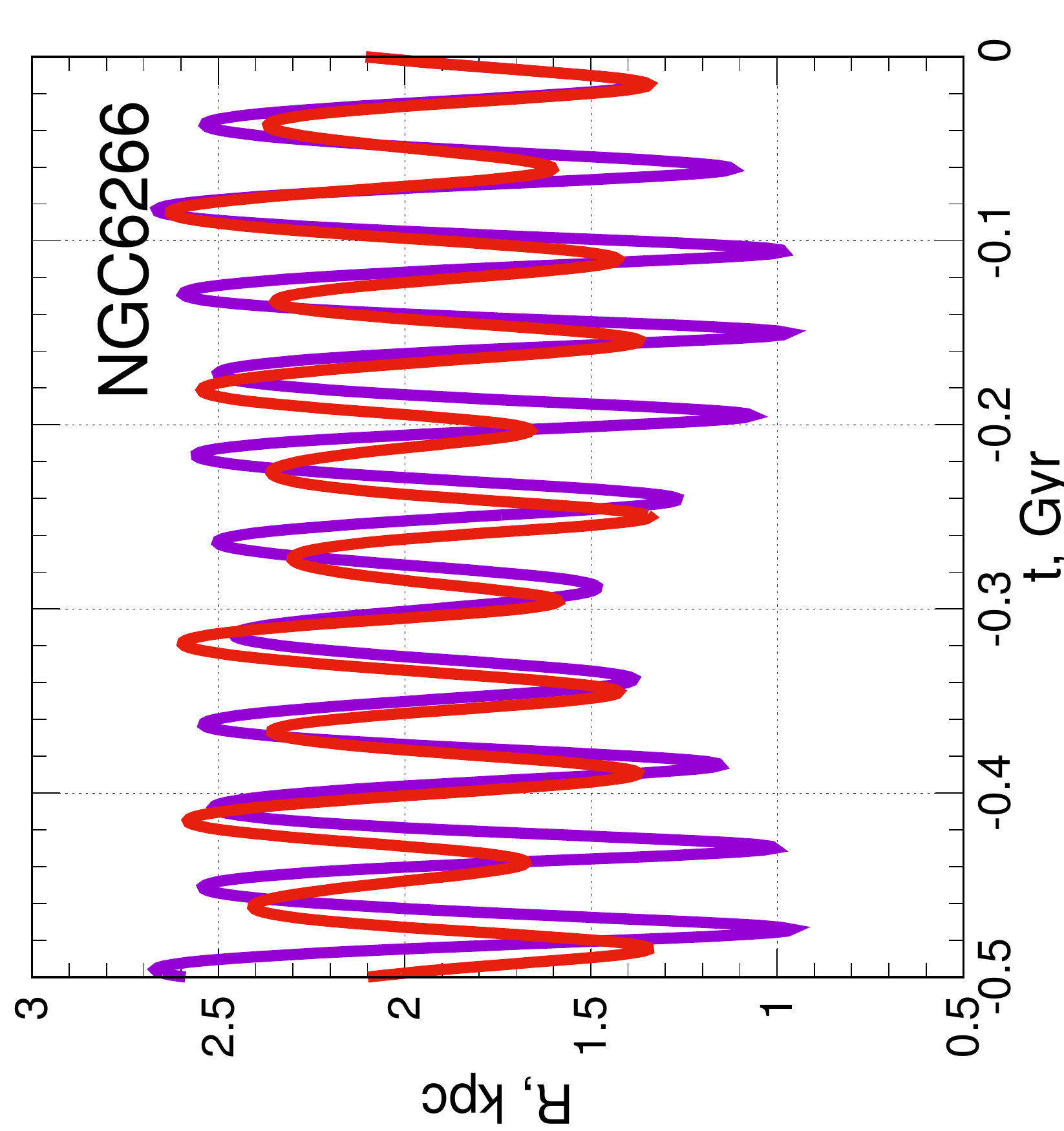}
\includegraphics[width=0.175\textwidth,angle=-90]{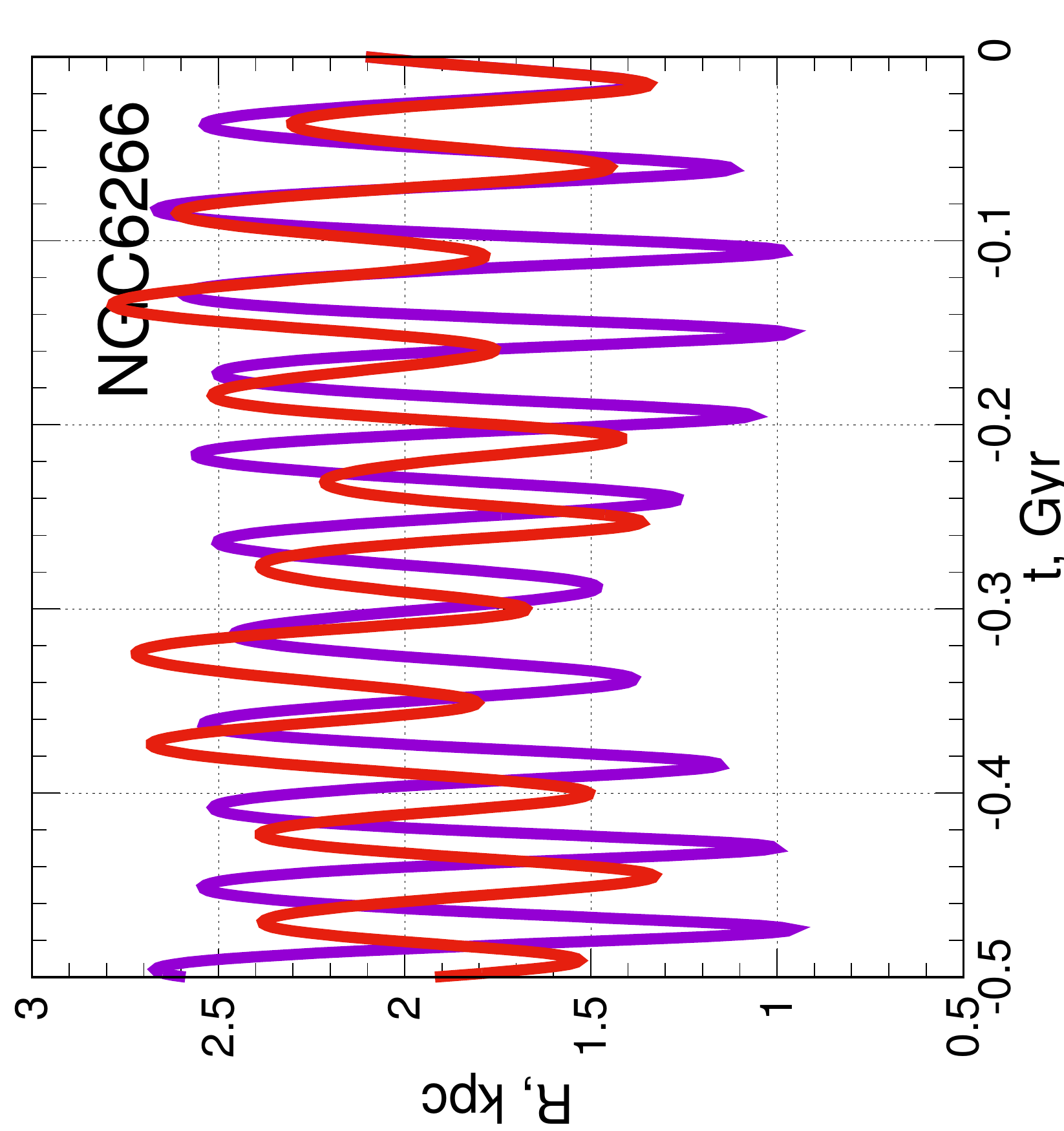}\

\medskip

\includegraphics[width=0.175\textwidth,angle=-90]{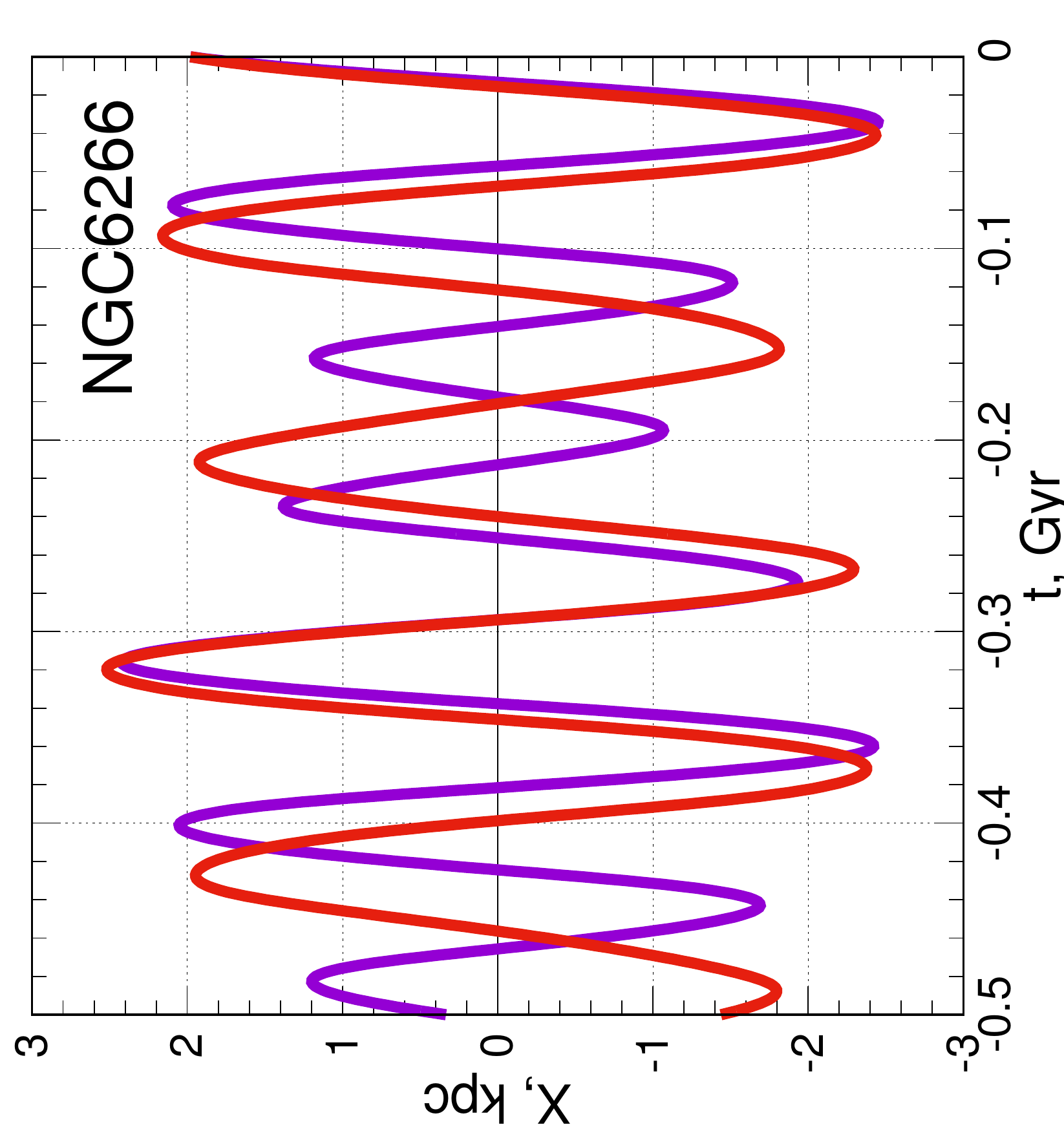}
\includegraphics[width=0.175\textwidth,angle=-90]{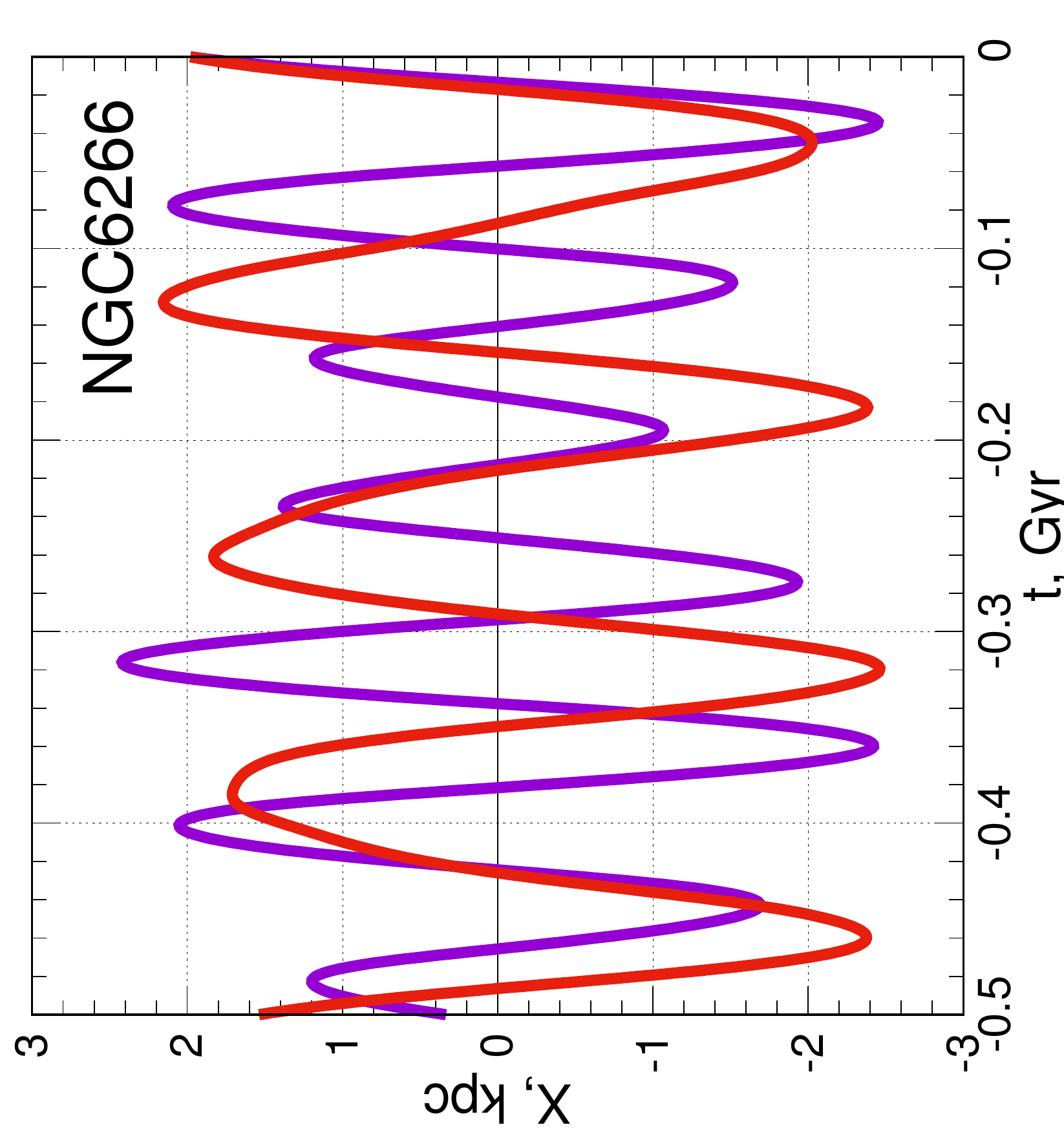}
\includegraphics[width=0.175\textwidth,angle=-90]{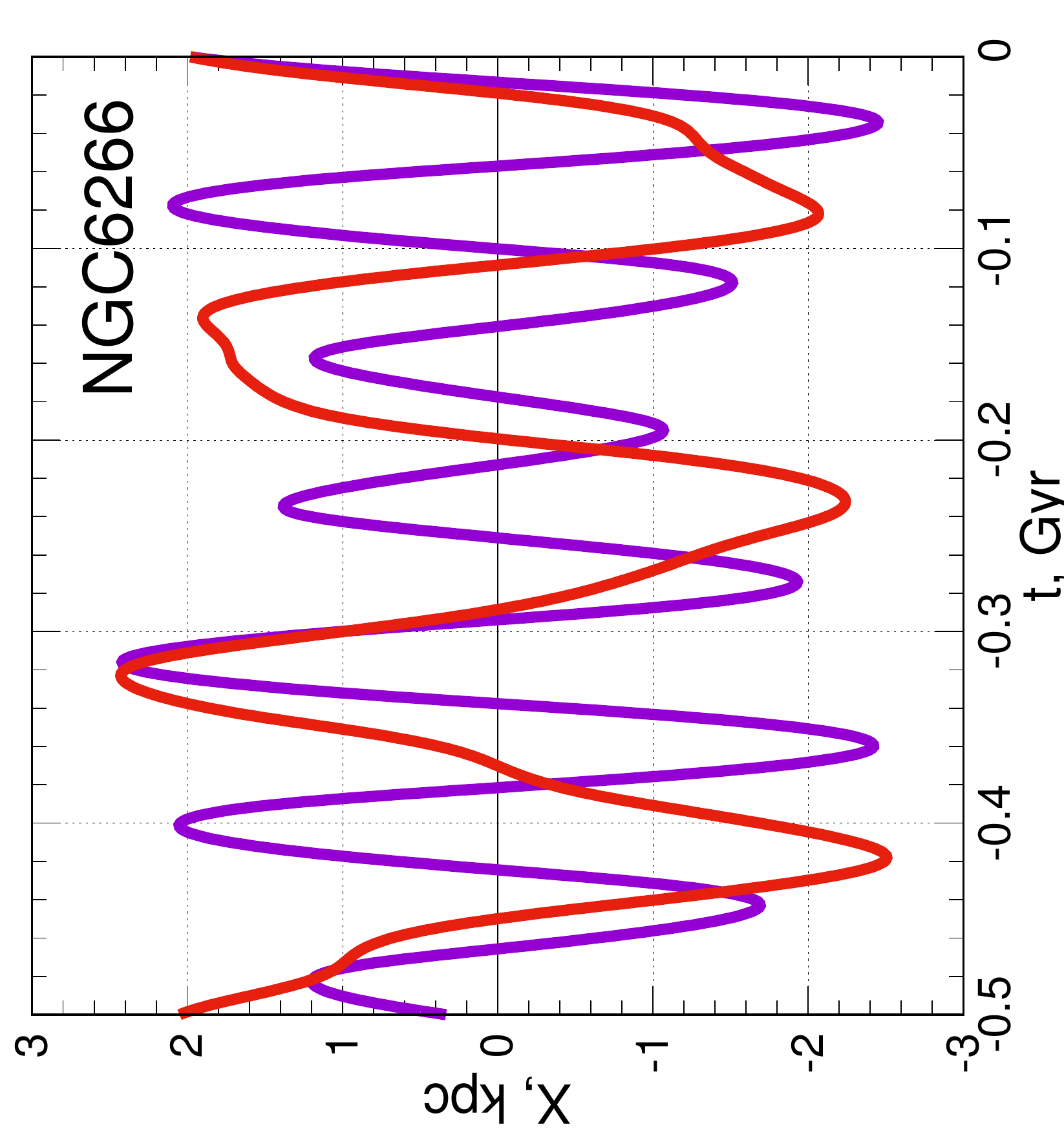}
\includegraphics[width=0.175\textwidth,angle=-90]{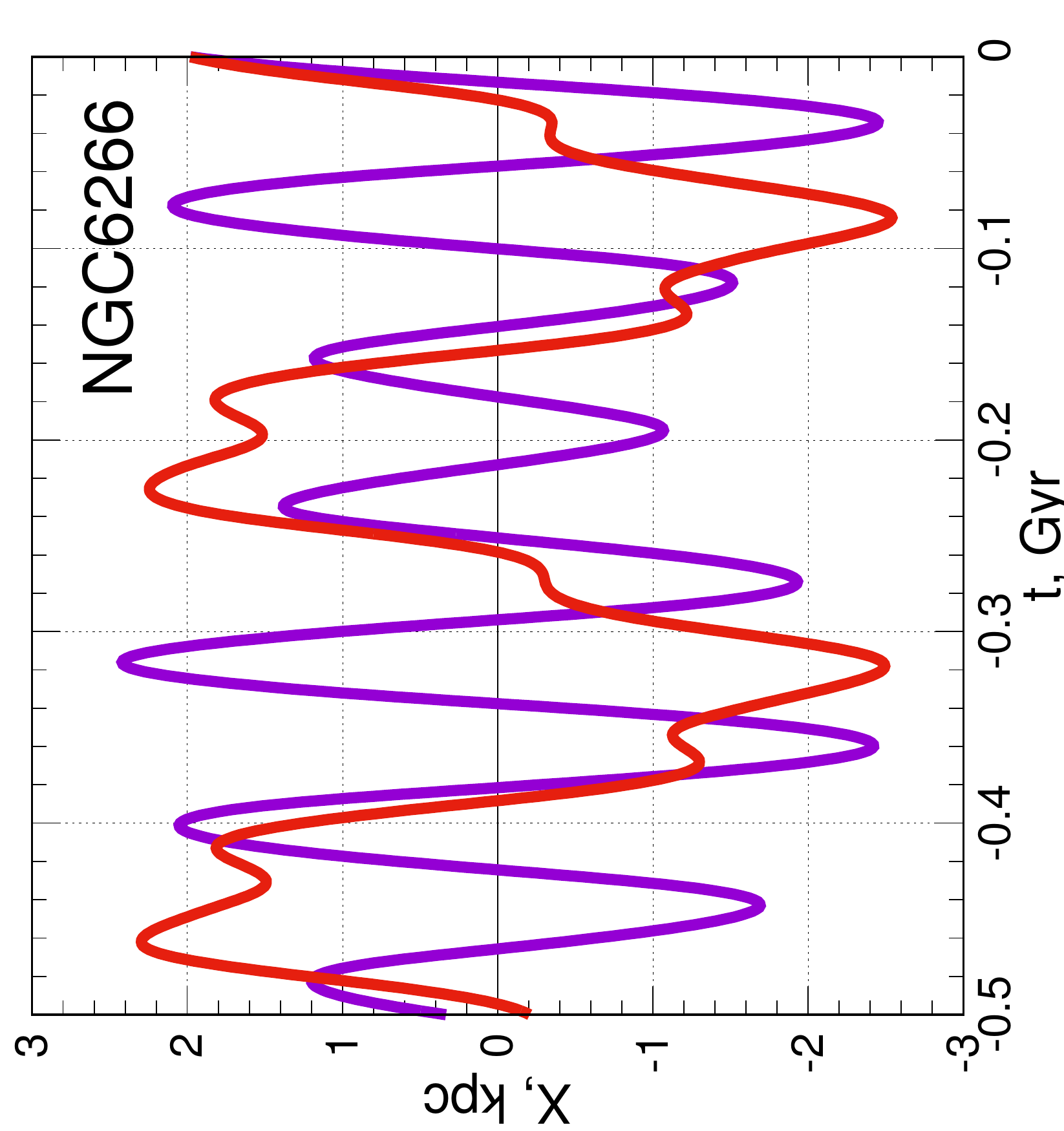}
\includegraphics[width=0.175\textwidth,angle=-90]{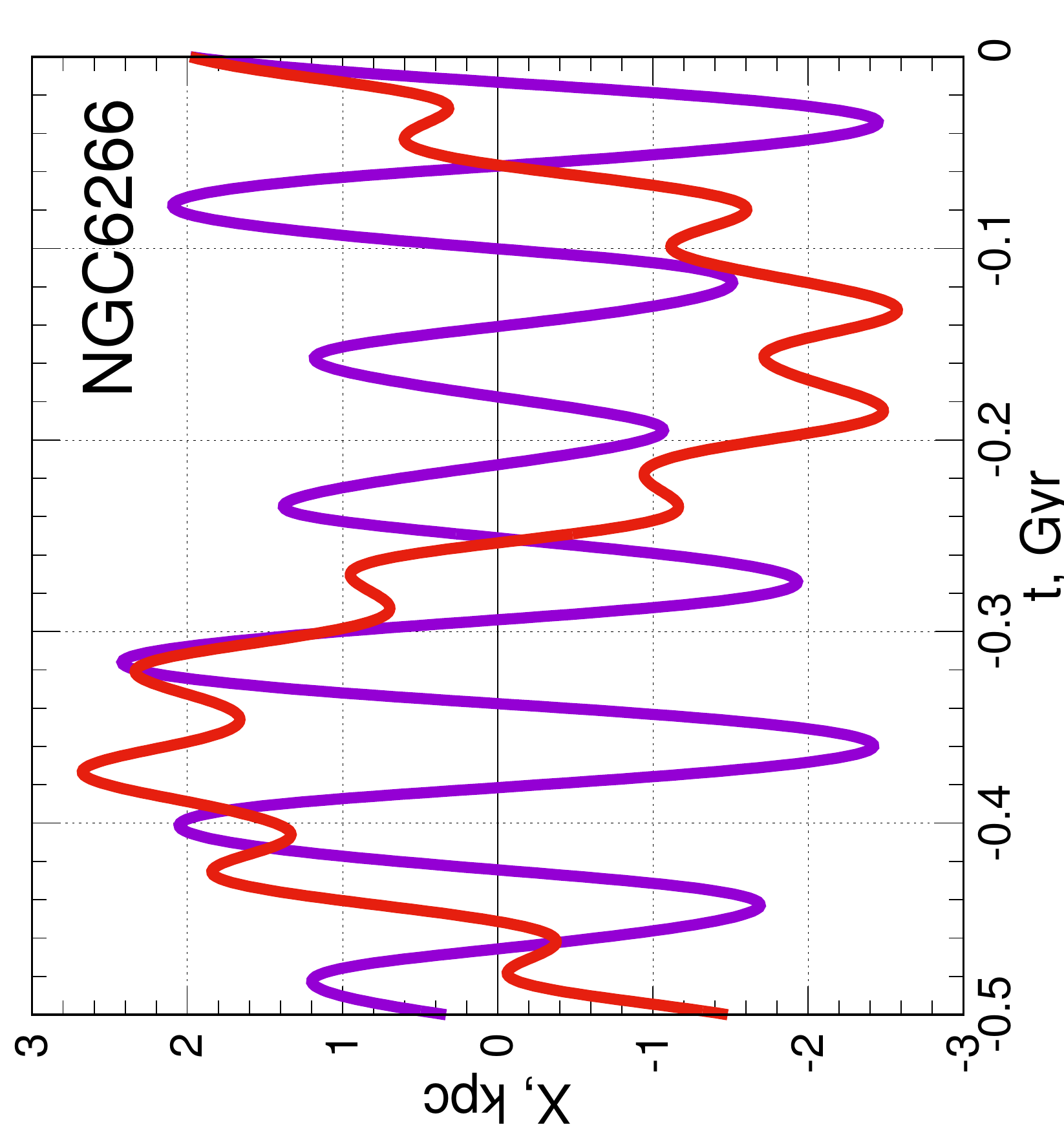}\

\medskip

\includegraphics[width=0.175\textwidth,angle=-90]{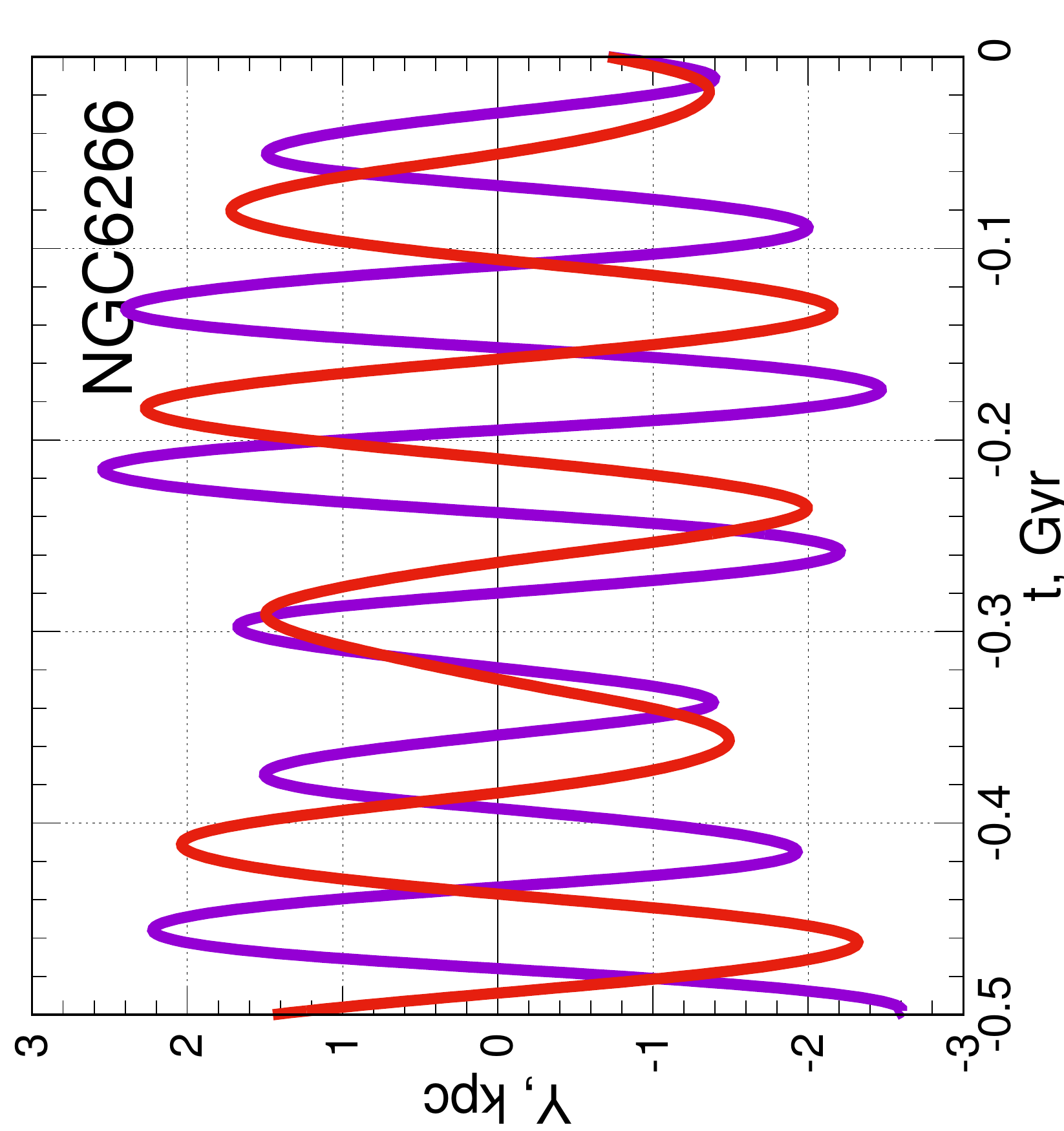}
\includegraphics[width=0.175\textwidth,angle=-90]{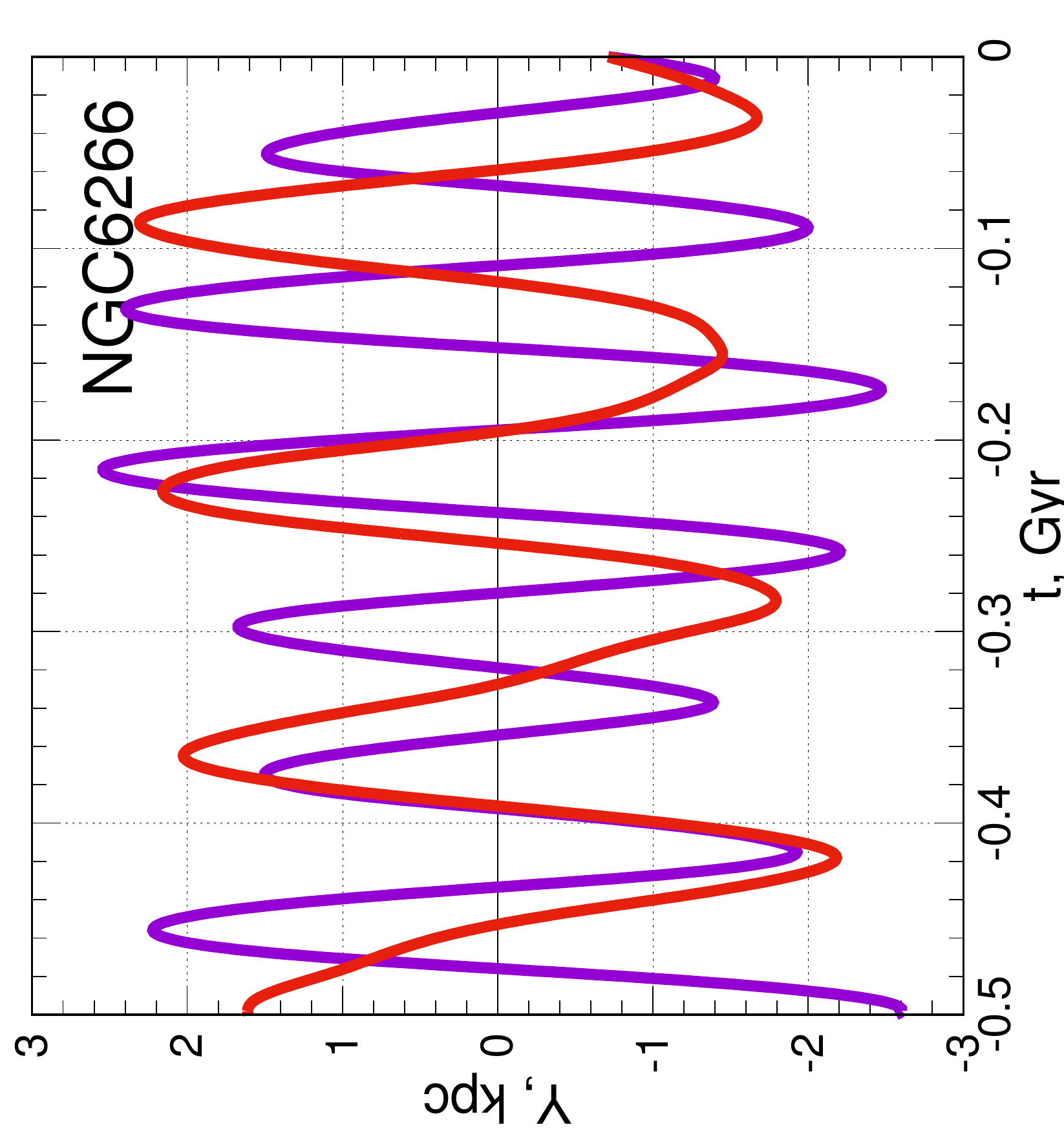}
\includegraphics[width=0.175\textwidth,angle=-90]{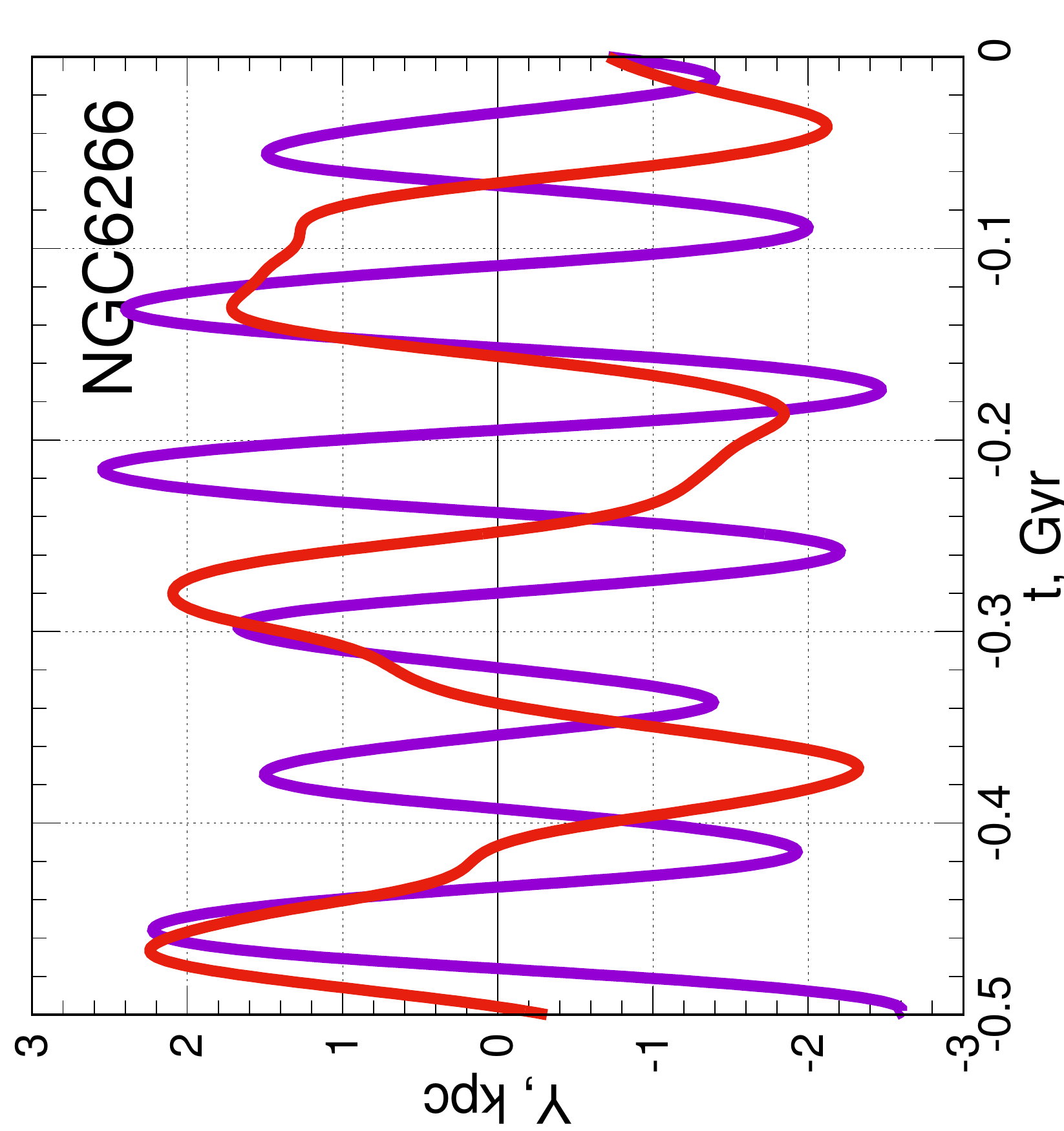}
\includegraphics[width=0.175\textwidth,angle=-90]{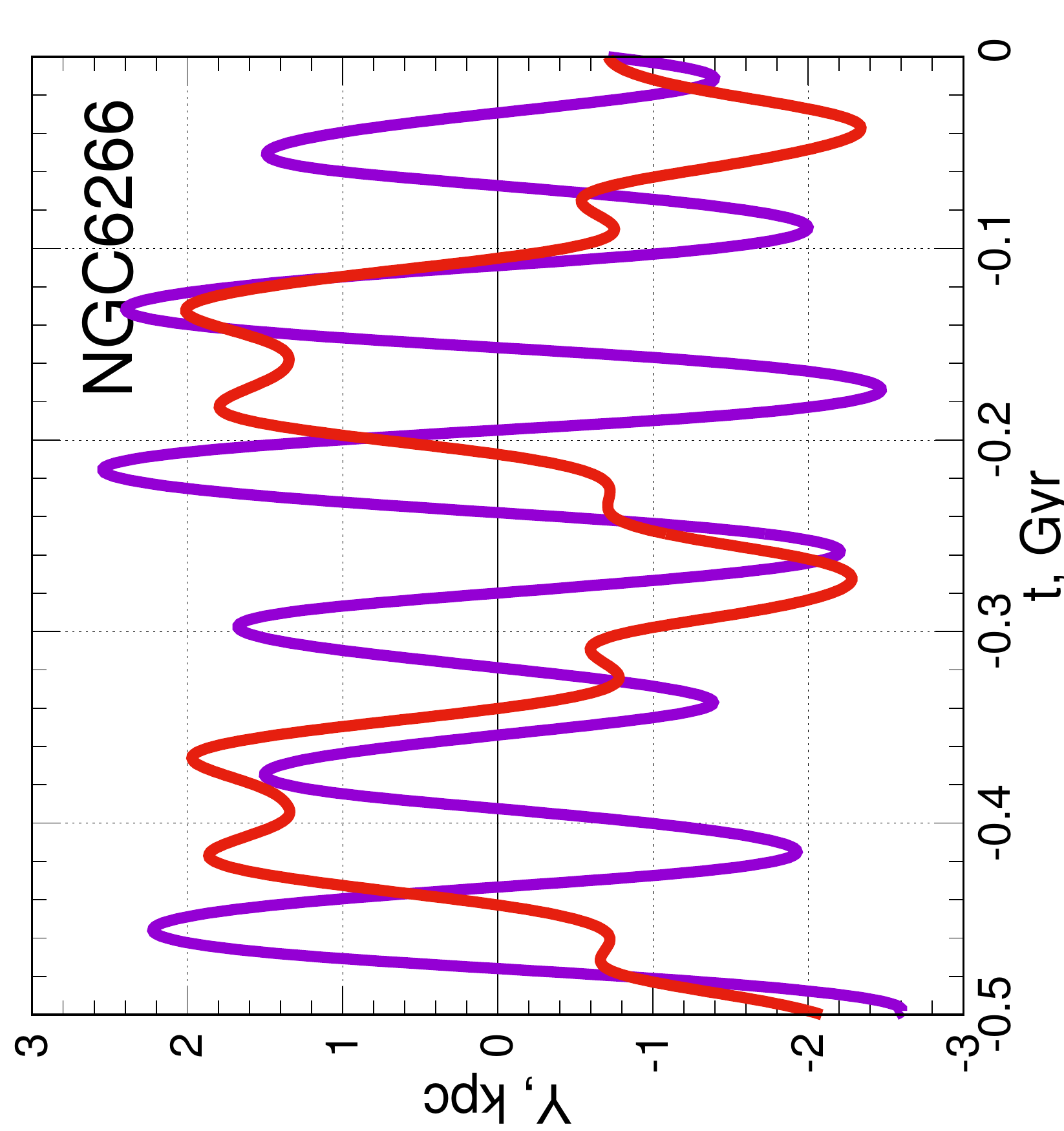}
\includegraphics[width=0.175\textwidth,angle=-90]{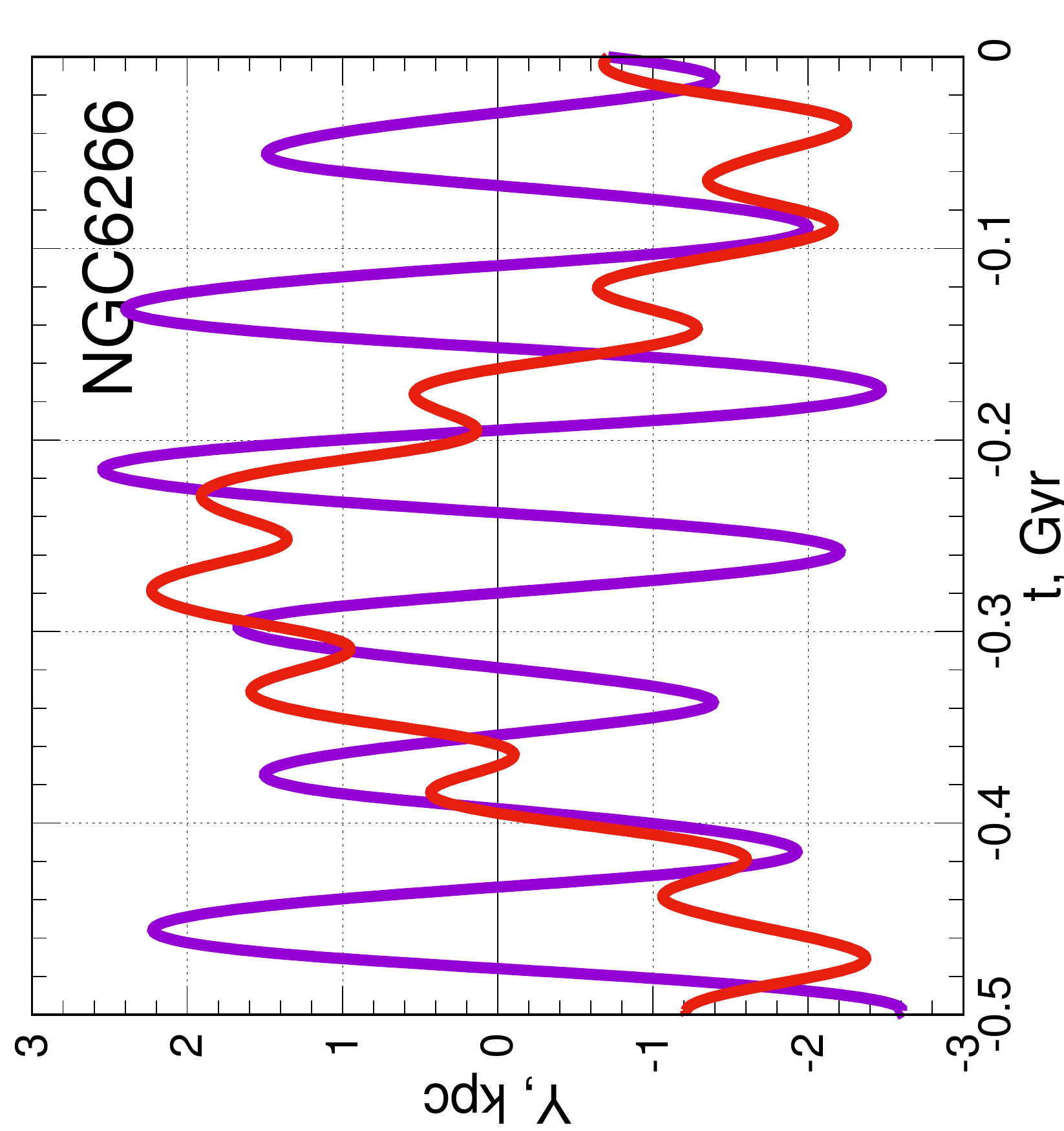}\

\medskip

\includegraphics[width=0.175\textwidth,angle=-90]{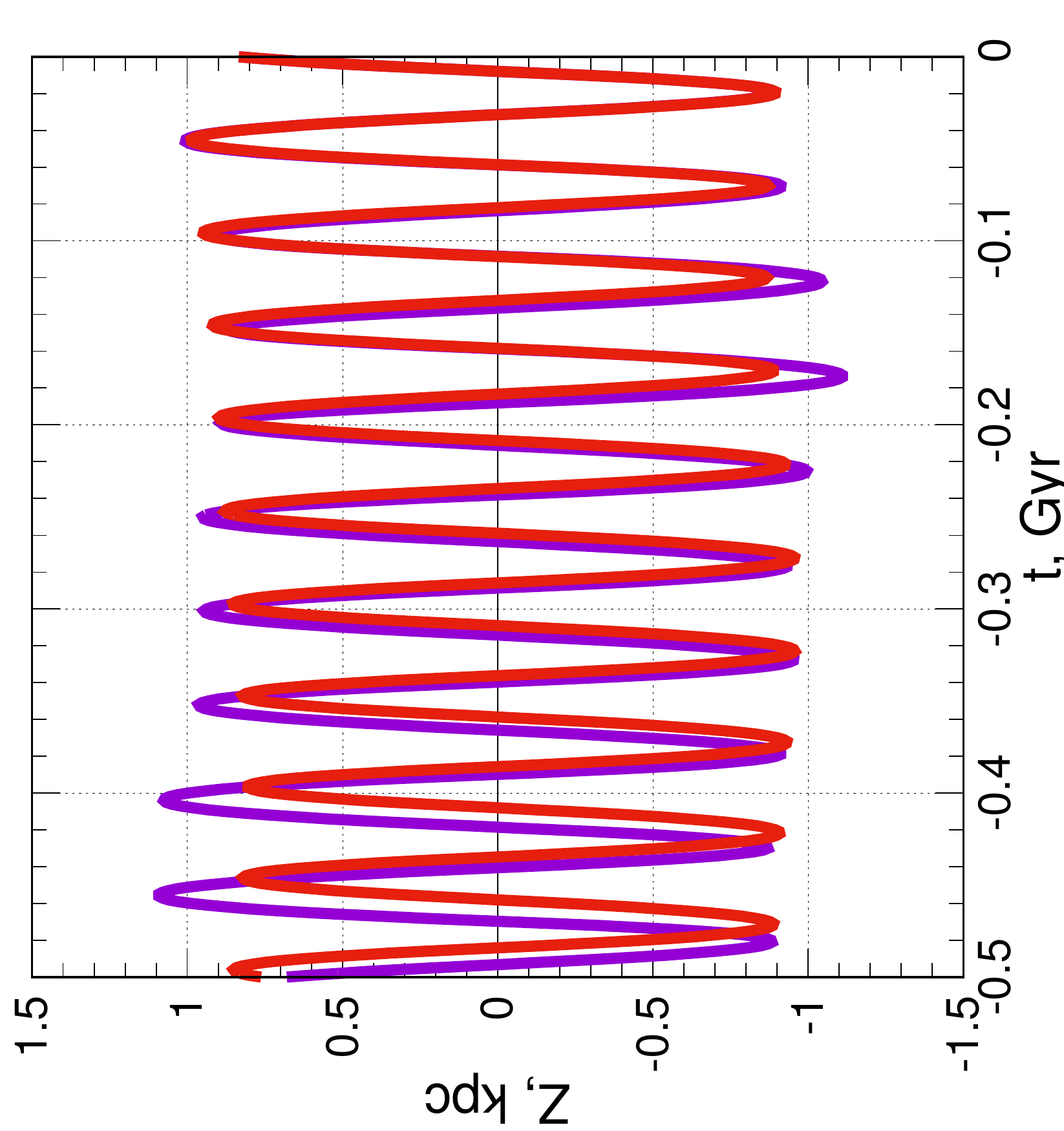}
\includegraphics[width=0.175\textwidth,angle=-90]{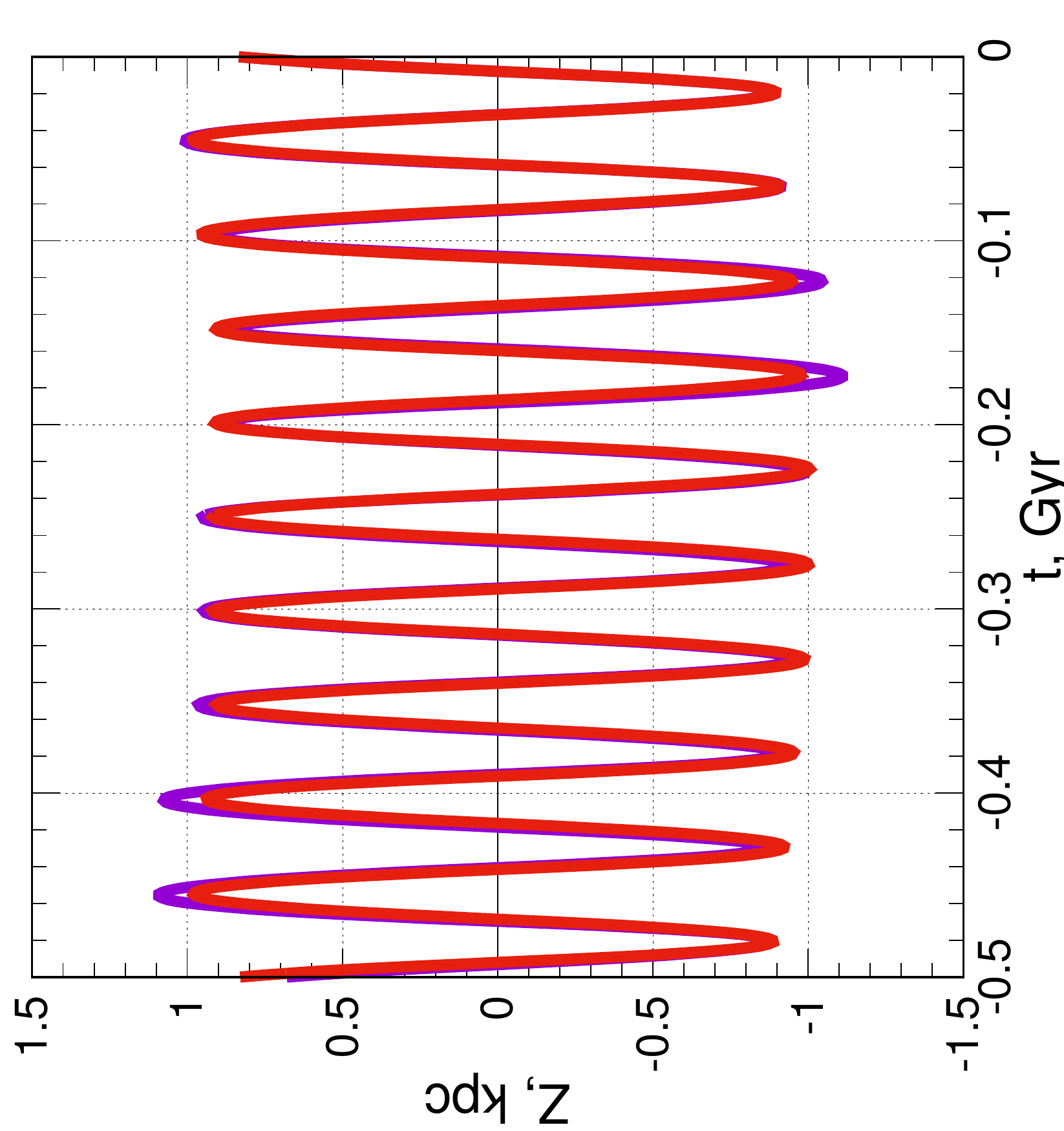}
\includegraphics[width=0.175\textwidth,angle=-90]{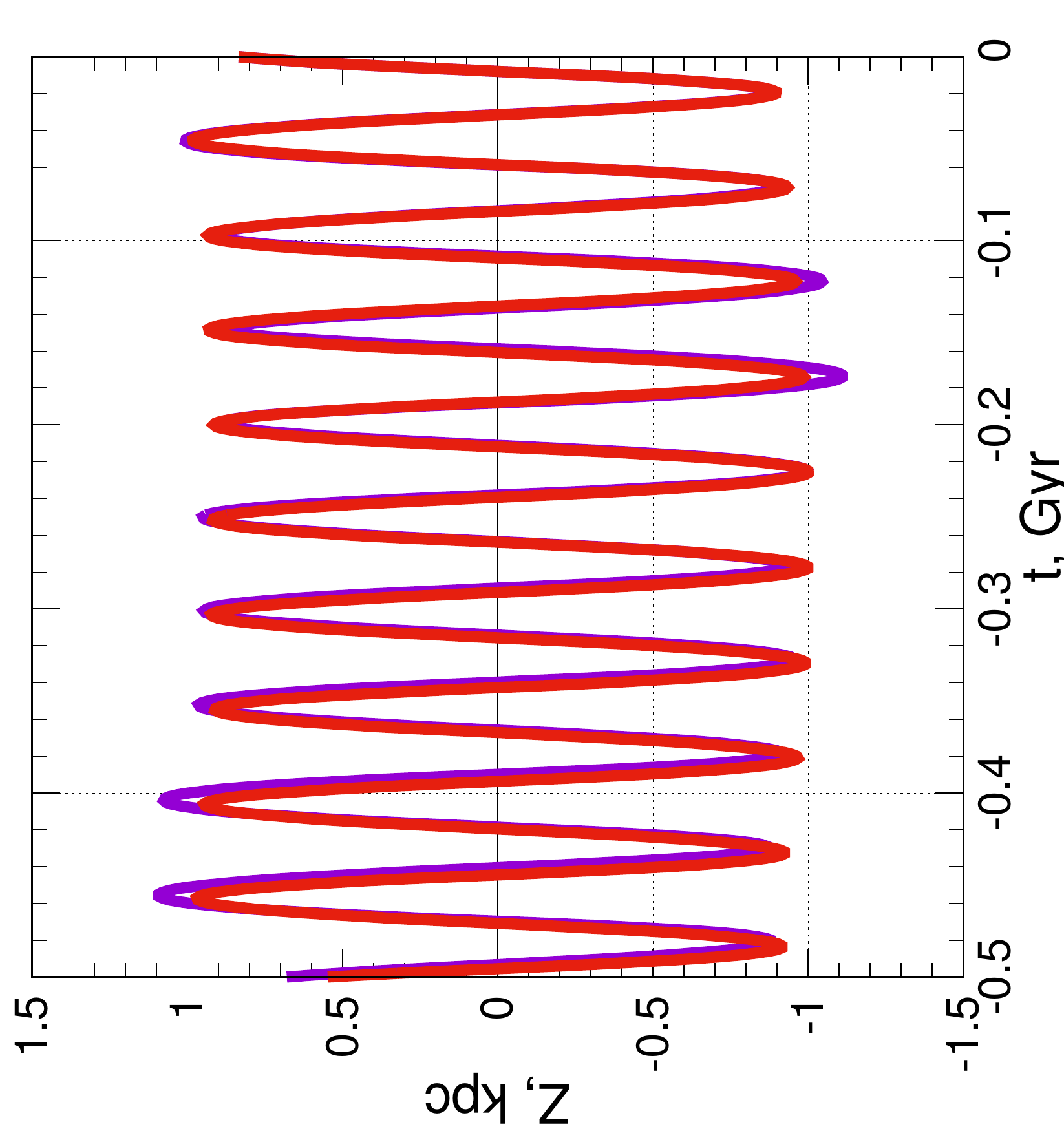}
\includegraphics[width=0.175\textwidth,angle=-90]{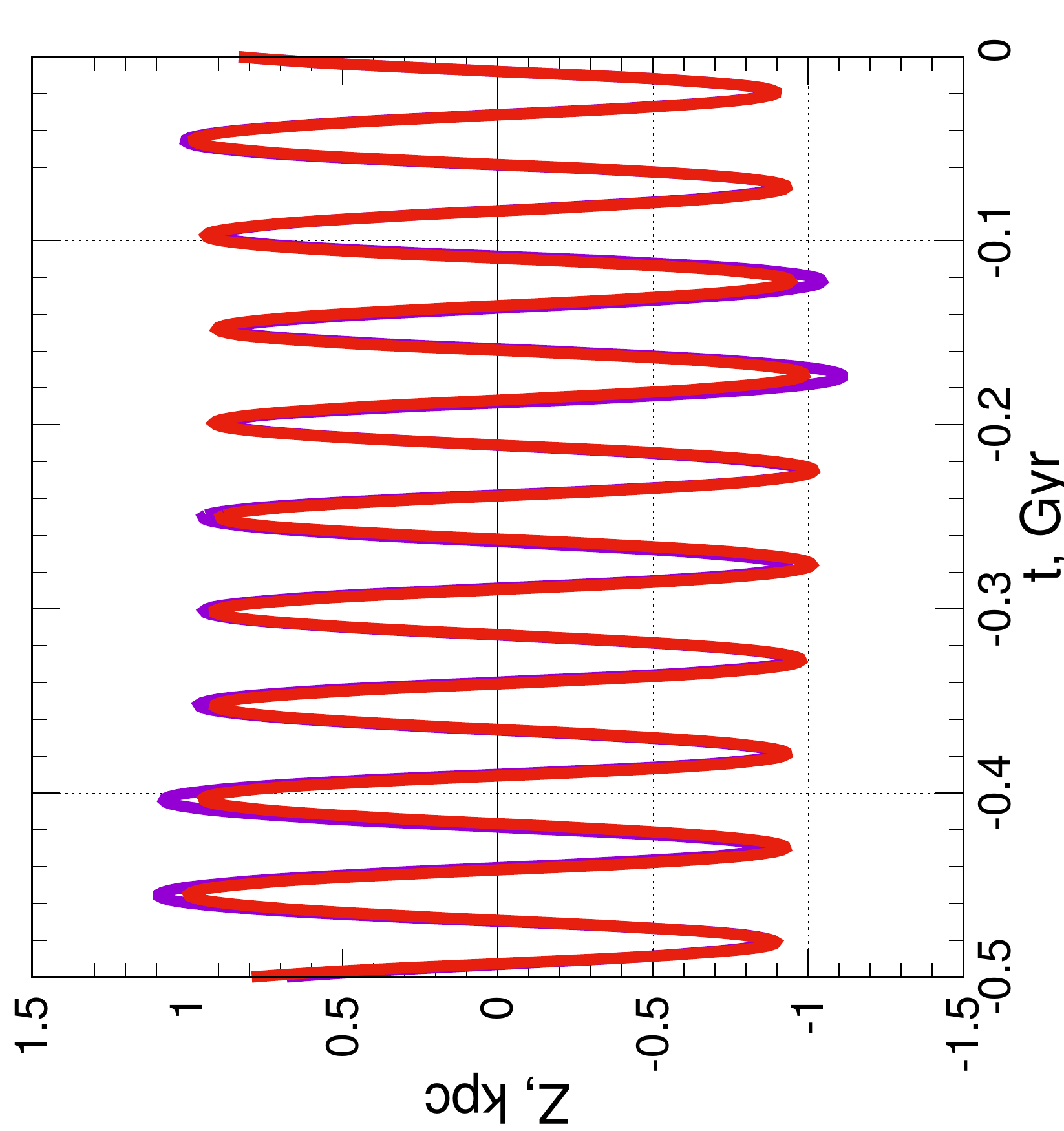}
\includegraphics[width=0.175\textwidth,angle=-90]{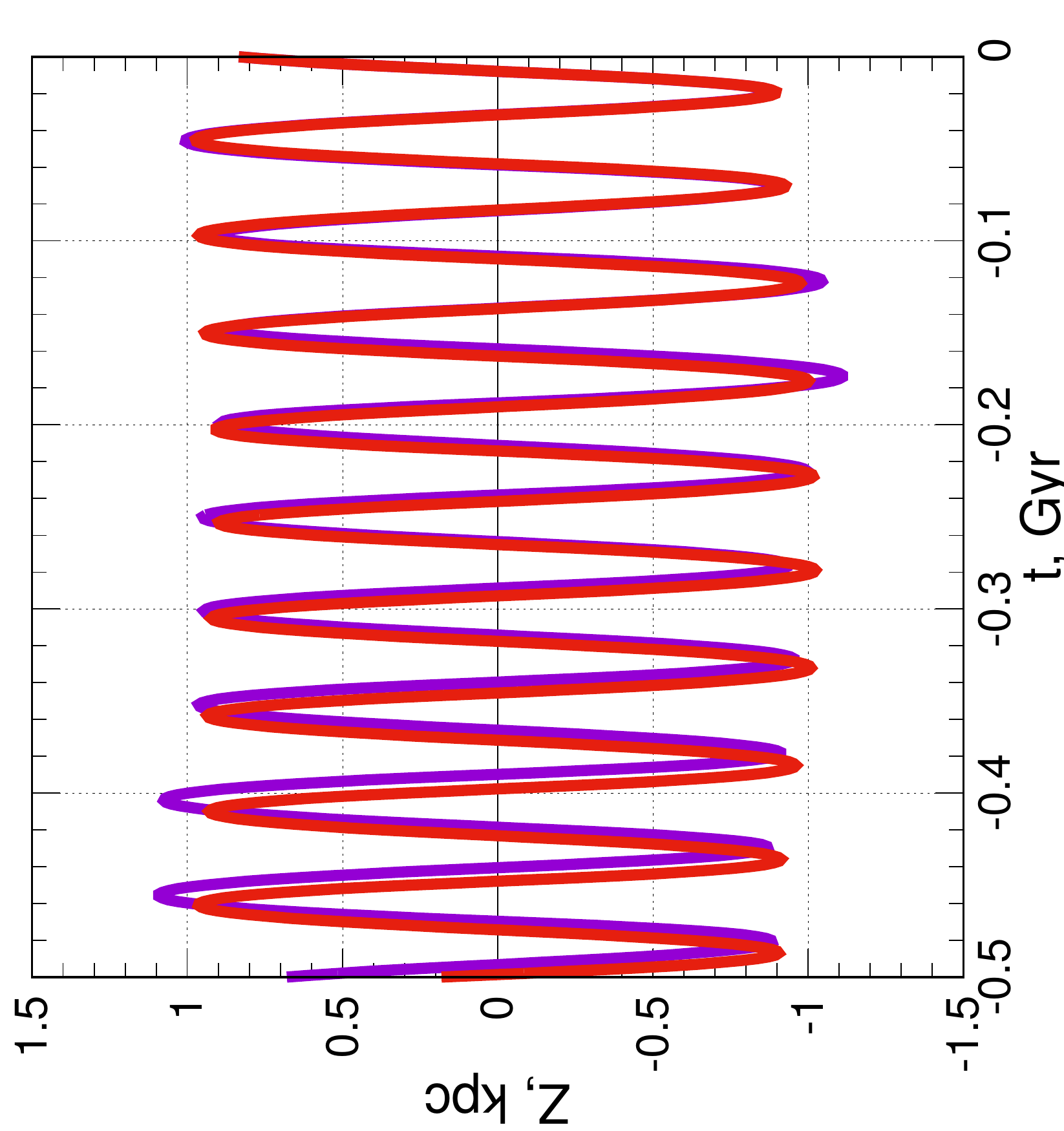}\
\caption{\small Dependences of $R(t), x(t), y(t), z(t)$ (from top to bottom) of the orbits of the globular cluster NGC6266 on the angular velocity of rotation of the bar
$\Omega_b =  20, 30, 40, 50, 60$ km s$^{-1}$ kpc$^{-1}$ (from left to right) are shown in red. For comparison, similar dependencies for $\Omega_b = 0$ km s$^{-1}$ kpc$^{-1}$ are shown in purple.}
\label{fcomp3}
\end{center}}
\end{figure*}

\begin{figure*}
{\begin{center}

\includegraphics[width=0.275\textwidth,angle=-90]{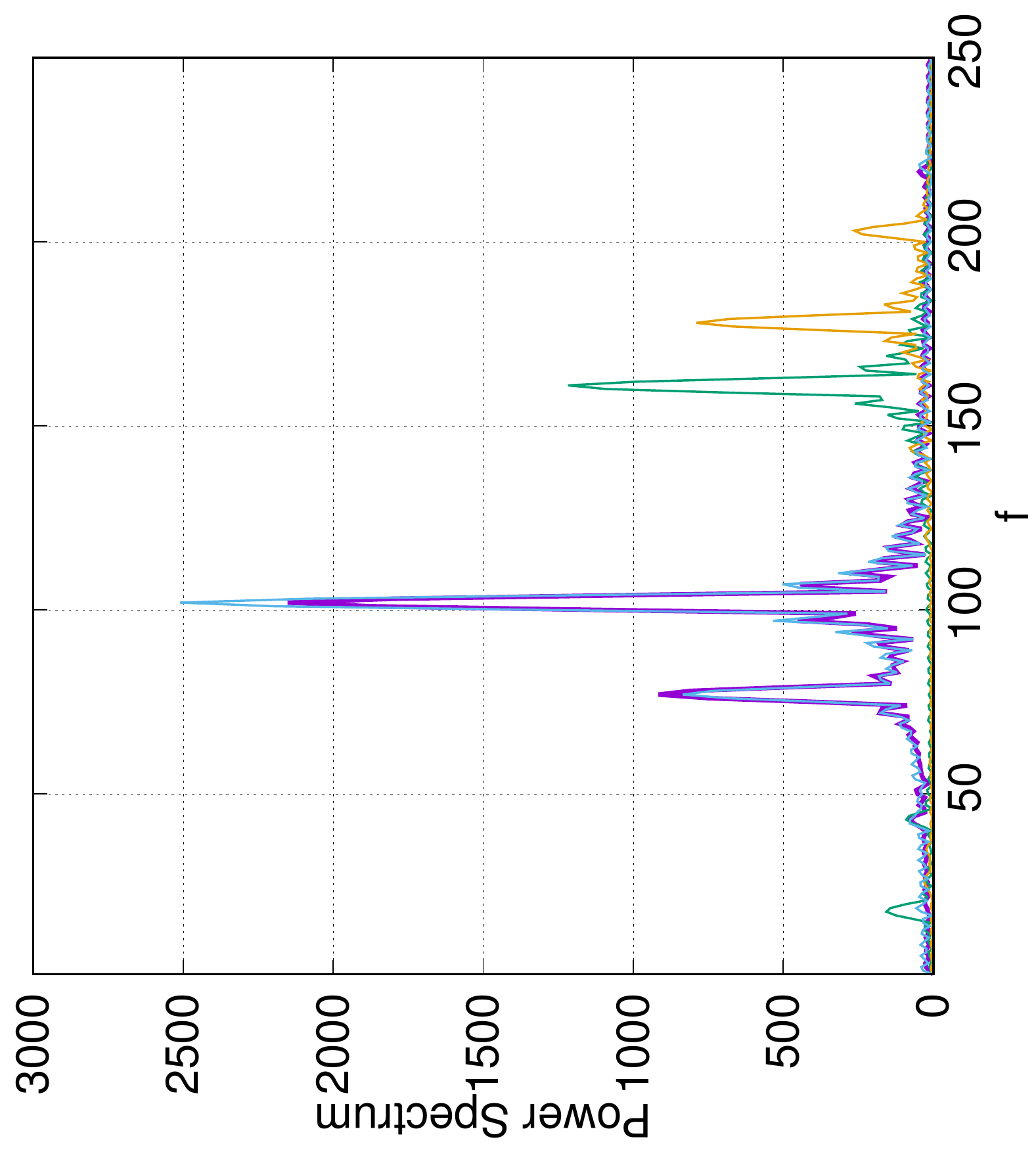}
\includegraphics[width=0.275\textwidth,angle=-90]{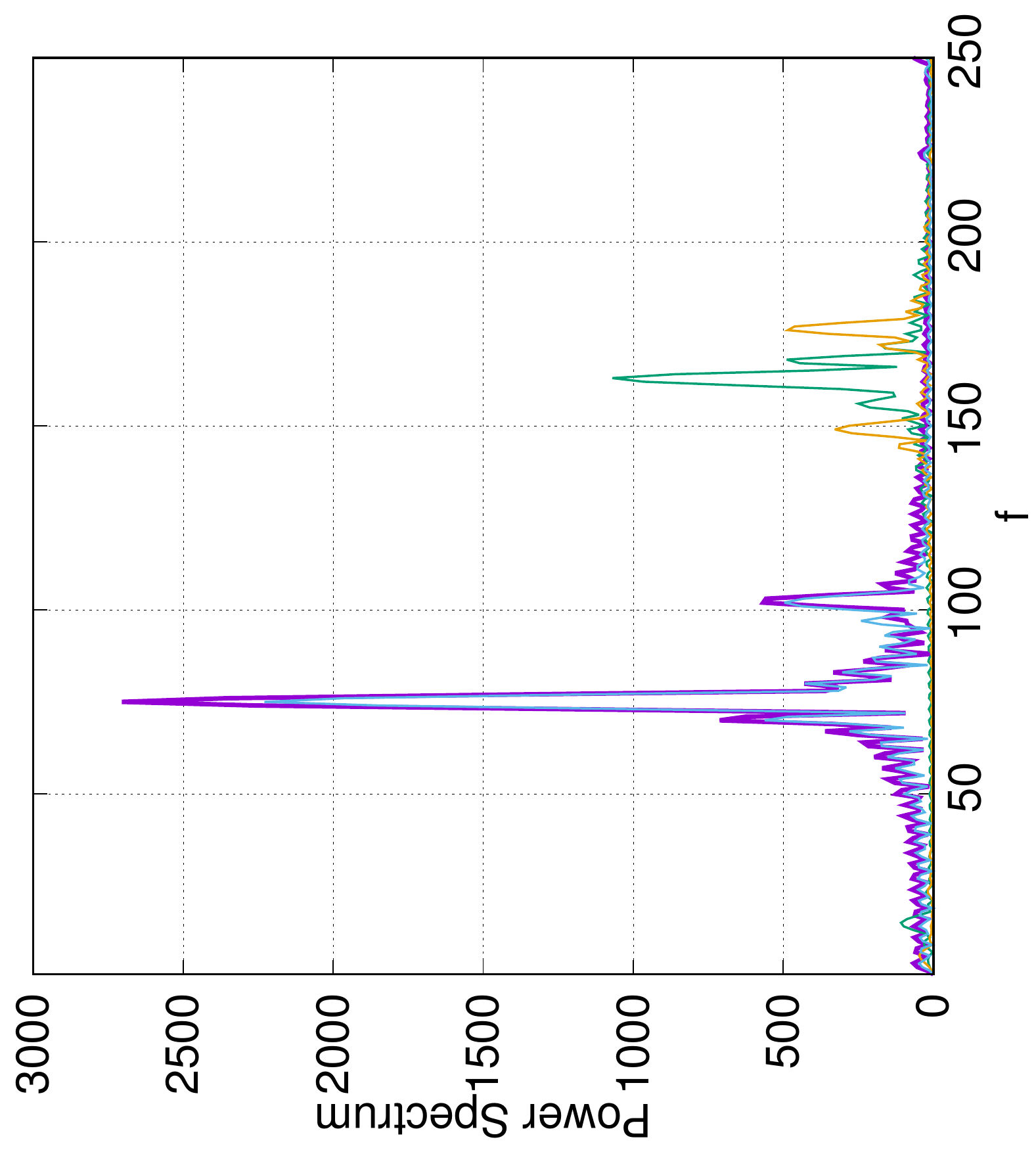}
\includegraphics[width=0.275\textwidth,angle=-90]{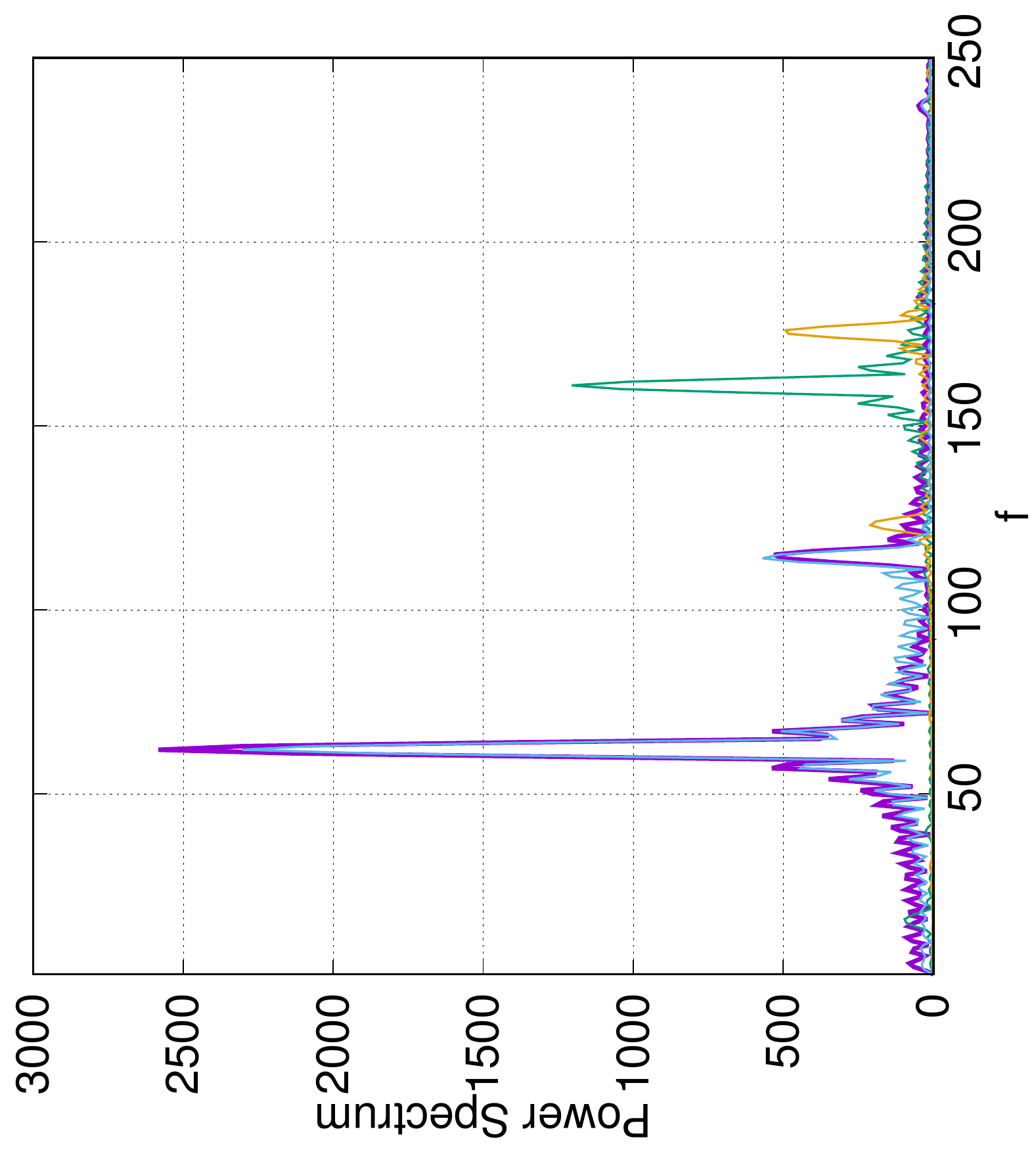}\

\medskip

\includegraphics[width=0.275\textwidth,angle=-90]{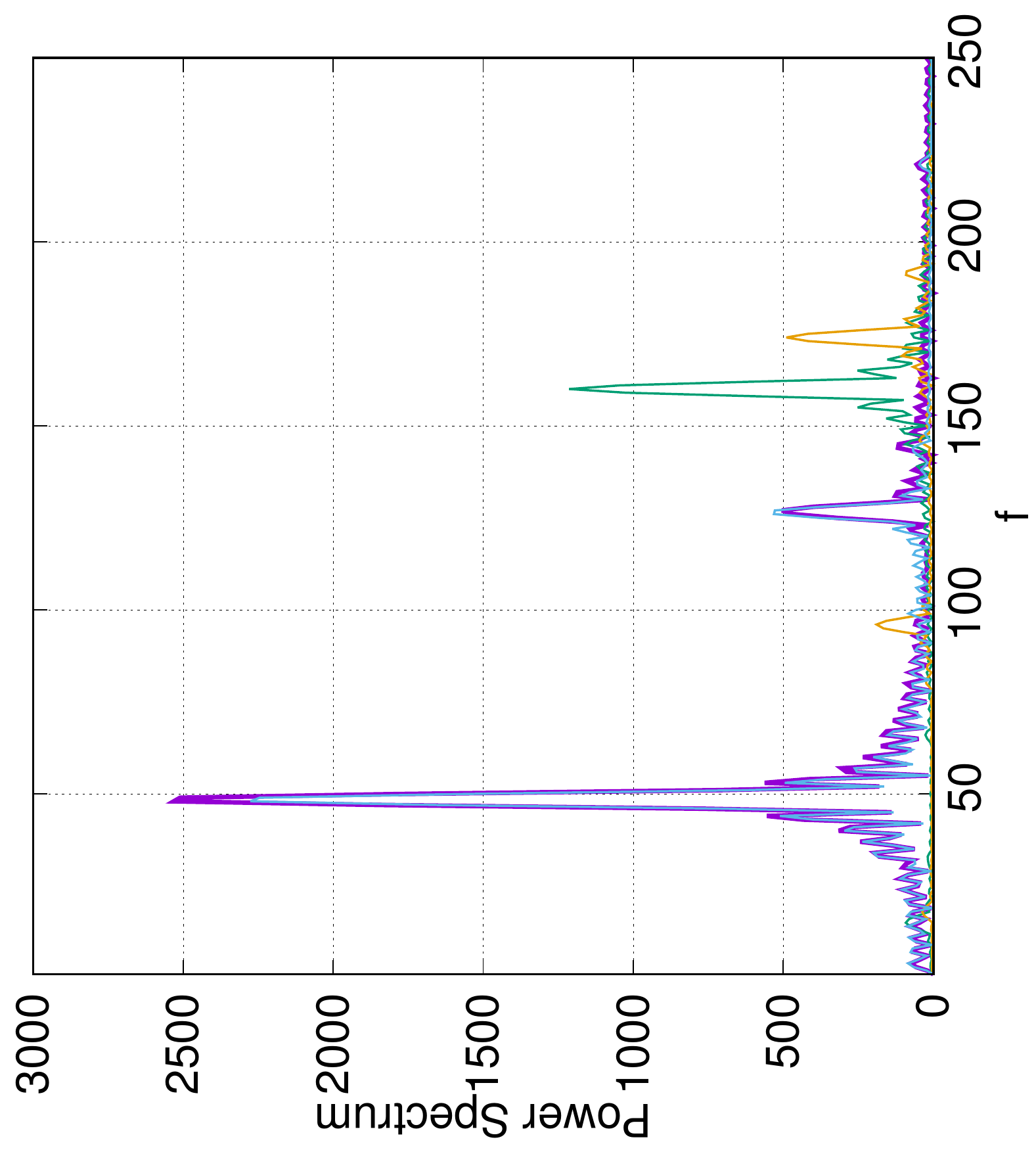}
\includegraphics[width=0.275\textwidth,angle=-90]{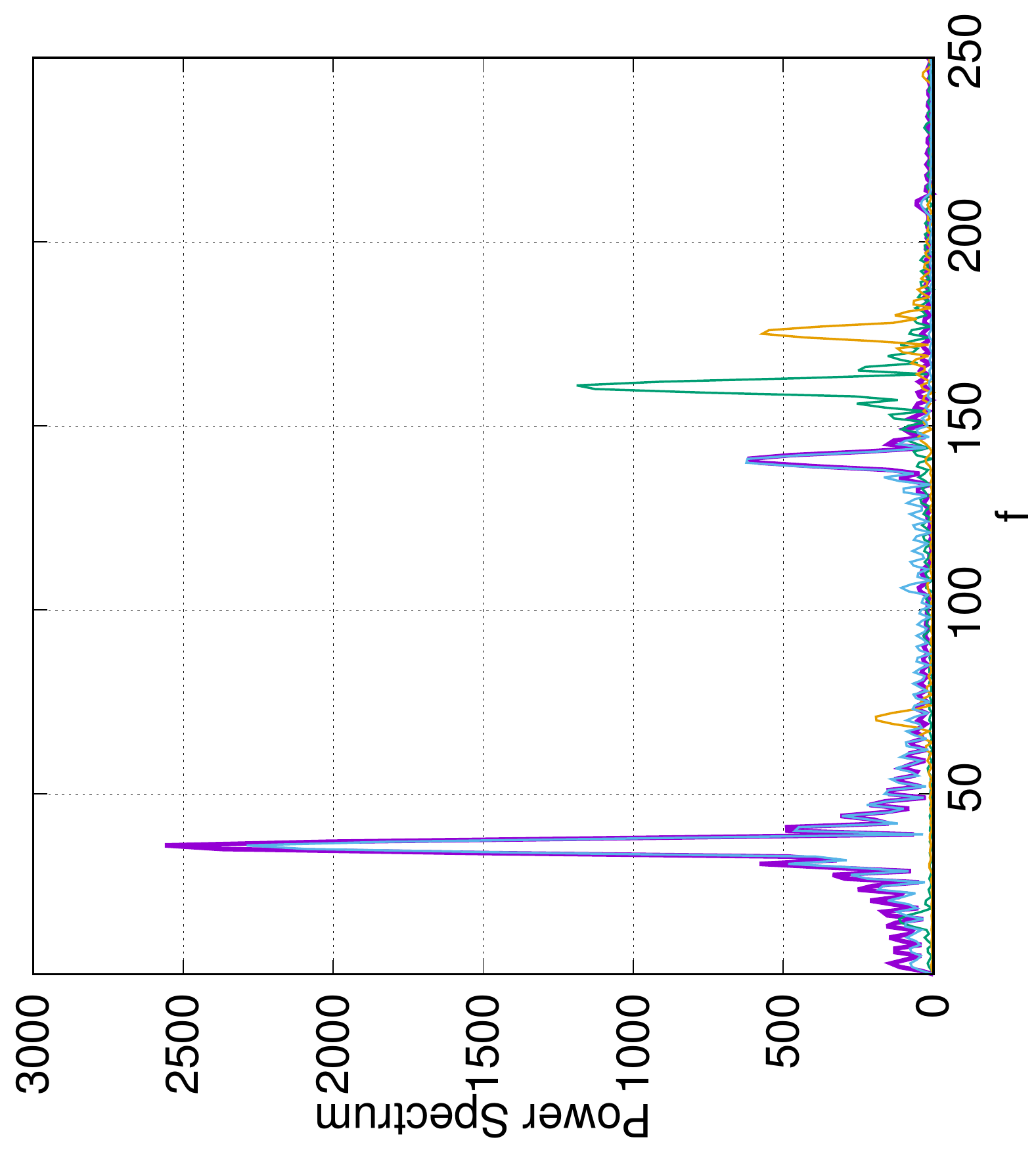}
\includegraphics[width=0.275\textwidth,angle=-90]{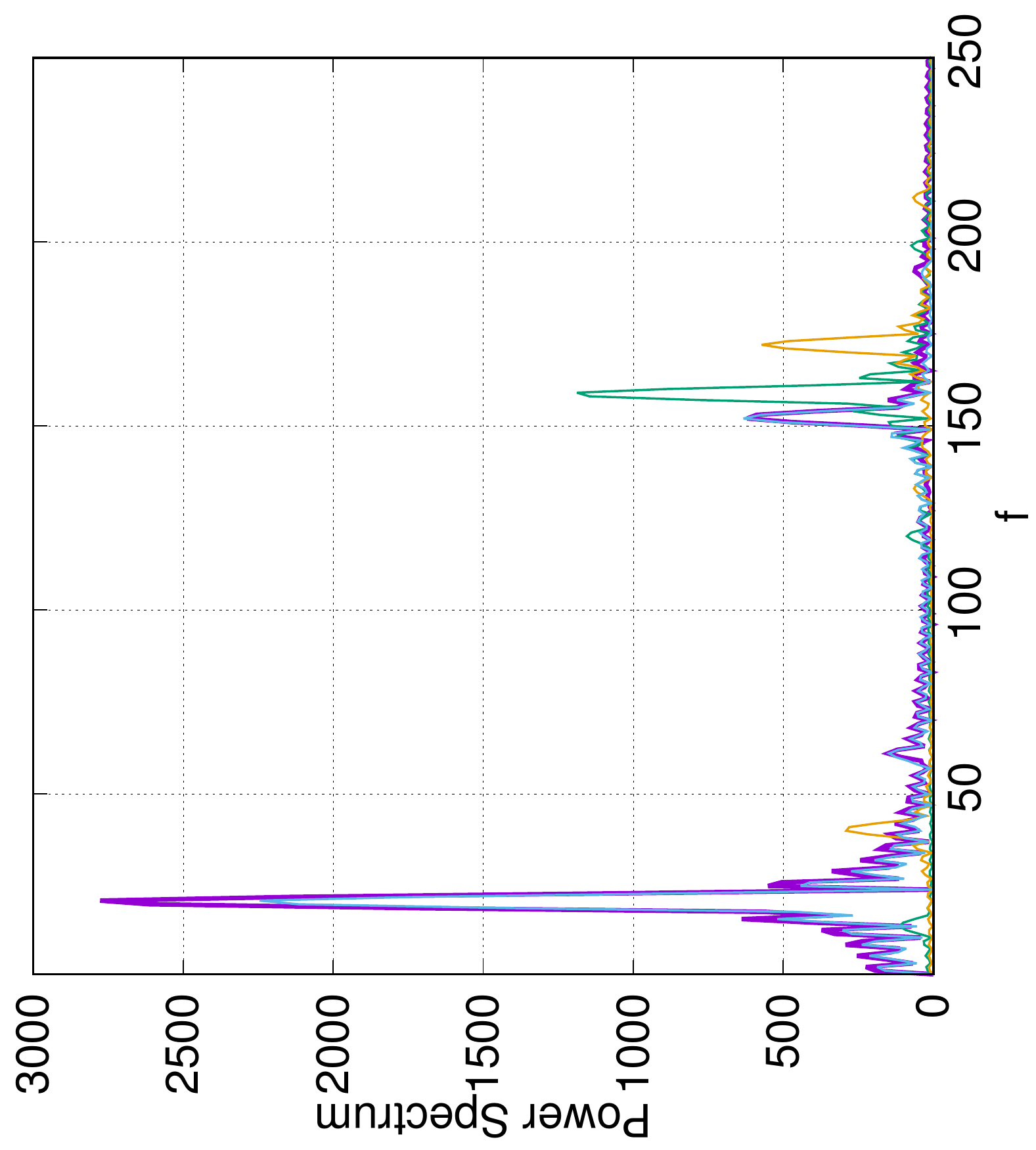}\
\caption{\small  Power spectra of the orbits of the globular cluster NGC6266 ($M_{bar}=430\times M_G$, $q_b=5$ kpc). The power spectra of $X_k, Y_k, Z_k$ and $R_k$ are shown in violet, cyan, green and yellow, respectively. Bar speed is $\Omega_b =  0, 20, 30, 40, 50, 60$ km s$^{-1}$ kpc$^{-1}$ (from left to right and from top to bottom).}
\label{fcomp4}
\end{center}}
\end{figure*}

Modeling the orbits of all GCs in our sample with varying such an important parameter as the bar rotation speed, which is known today with great uncertainty, allowed us to identify the following properties and patterns of dominant orbital frequencies calculated in the rotating bar system:

1)for most GCs, the dominant frequency $f_R$ practically does not change with changes in the bar
rotation speed;

2)the dominant frequency $f_X$ depends on the angular velocity of rotation of the bar $\Omega_b$ according to the law:
\begin{equation}
\label{shift}
f_X(\Omega_b) \approx f_X(0) \pm K\times\Omega_b,
\end{equation}
where $K = 0.1587$ kpc s km$^{-1}$ Gyr$^{-1}$, which leads to a systematic shift in the frequency ratio  $f_R/f_X$ with changes in $\Omega_b$;

3)for most GCs the relation $f_X=f_Y$ is satisfied;

4)the dominant frequency $f_Z$ practically does not change with changes in the rotation speed of the bar.

Let us illustrate these properties using the example of the globular cluster NGC6266. Table~\ref{t:ff} shows the values of dominant frequencies (DFT sample numbers) depending on the value of $\Omega_b$ at $M_{bar}=430\times M_G$, $q_b=5$ kpc. As you can see, the dominant frequencies $f_R$ and $f_Z$ are practically independent of the bar rotation speed and vary within $175\pm 3$ and $161\pm 2$, respectively. Moreover, the spread in values is mainly due to the effect of sampling of the analyzed sequences. The relation $f_X=f_Y$ holds exactly. As for the dominant frequency $f_X$, its change when the angular velocity of the bar changes by 10 units is $13\pm 1$ units in DFT counts, or approximately $13\times 0.122$ in Gyr$^{-1}$ units, i.e. expression (\ref{shift}) is true. In the case of NGC6266, the second term on the right side of expression (\ref{shift}) appears with a plus sign, but, as modeling of the orbits of other GCs shows, this term can also appear with a minus sign, and the resulting value of the coefficient $K$ is valid for all GCs. We also note that the value of the coefficient $K$ does not depend on the mass and length of the bar. A change in $f_X$ with a change in $\Omega_b$, provided that $f_R$ is practically independent of $\Omega_b$, leads to a systematic shift in the $f_R/f_X$ ratio, which we see in the last column of Table~\ref{t:ff}. As follows from the obtained values of $f_R/f_X$, the globular cluster NGC6266 with a given mass and length of the bar can be captured by the bar only if its angular velocity is 10 km s$^{-1}$ kpc$^{-1}$.

 {\begin{table}[t]                                    
 \caption[]
 {\small\baselineskip=1.0ex
Values of dominant frequencies (DFT sample numbers) for the globular cluster NGC6266 depending on
the angular velocity of the bar $\Omega_b$.
  }
 \label{t:ff}
 \begin{center}\begin{tabular}{|c|r|r|r|r|r|}\hline
 $\Omega_b$     & $f_X$ & $f_Y$  & $f_Z$ & $f_R$ & $f_R/f_X$ \\
 км/с/кпк       &       &        &       &       &   \\\hline
 0  &102 &102 & 161 & 178 & 1.75 \\\hline
 10 & 87 & 87 & 165 & 173 & 1.99 \\\hline
 20 & 75 & 75 & 163 & 176 & 2.35 \\\hline
 30 & 62 & 62 & 161 & 176 & 2.84 \\\hline
 40 & 48 & 48 & 160 & 174 & 3.63 \\\hline
 50 & 36 & 36 & 161 & 175 & 4.86 \\\hline
 60 & 21 & 21 & 159 & 172 & 8.19 \\\hline
 \end{tabular}\end{center}\end{table}}

A graphical illustration of the results of modeling the orbits of NGC6266 at various bar velocities is presented in Fig.~\ref{fcomp2}--\ref{fcomp4}.

Figure~\ref{fcomp2} shows the orbits of NGC6266 constructed both in an axisymmetric potential and in a potential with a bar rotating with an angular velocity $\Omega_b = 40$ km s$^{-1}$ kpc$^{-1}$. Functions of orbital coordinates as a function of time $R(t), x(t), y(t), z(t)$ for various angular velocities of rotation of the bar $\Omega_b$ are shown in Fig.~\ref{fcomp3}. The power spectra of discrete sequences of orbital coordinates $x_n, y_n, z_n$ and $R_n$, calculated using formula (\ref{Four}) for different$\Omega_b$, are shown in Fig.~\ref{fcomp4}.

\begin{figure*}
{\begin{center}
\includegraphics[width=0.4\textwidth,angle=-90]{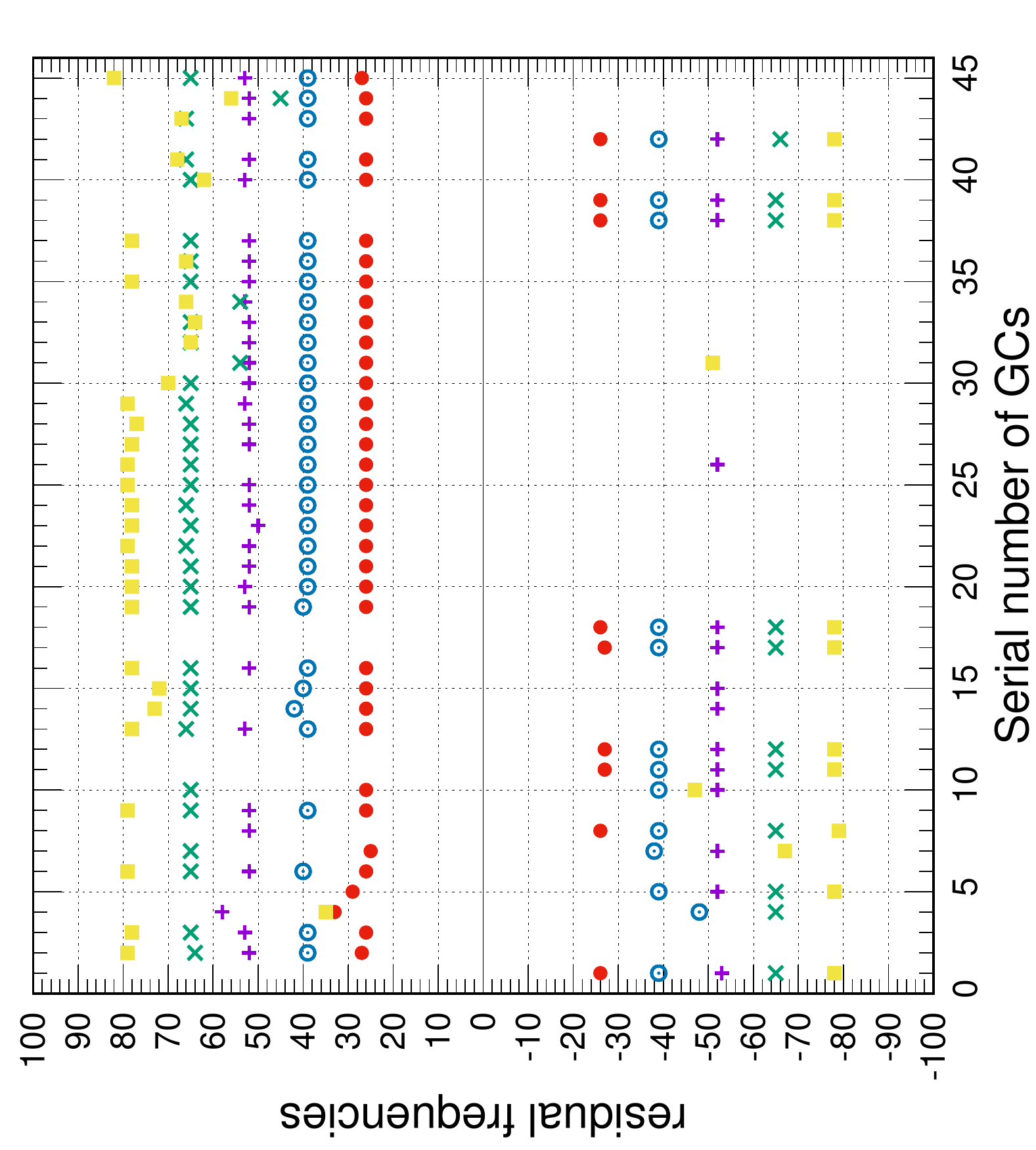}\
\caption{\small Shift of the dominant frequency $f_X(\Omega_b)$ relative to the
frequency $\Delta_x=f_X(\Omega_b) - f_X(0)$ (along the ordinate) for
the entire sample of 45 GCs at the angular velocity of bar rotation
$\Omega_b = 20, 30, 40, 50$ and $60$ km s$^{-1}$ kpc$^{-1}$, indicated by red, blue,
purple, green and yellow, respectively. The abscissa axis shows
the serial numbers of GCs in accordance with Table~\ref{t:f}.}
\label{fcomp5}
\end{center}}
\end{figure*}

A graphic illustration of the effect of shifting the dominant frequency $f_X$ with a change in
$\Omega_b$ for all 45 GCs is shown in Fig.~\ref{fcomp5}. Here, on the graph, the abscissa axis shows the serial numbers of globular clusters (see Table~\ref{t:f}), and the ordinate axis shows the frequency shift values $\Delta_x=f_X(\Omega_b)-f_X(0)$ (in DFT counts) for different values of the angular velocity of bar rotation. It can be seen that for the vast majority of globular clusters relation (\ref{shift}) is satisfied, and the displacement can be both positive and negative.

\begin{figure*}
{\begin{center}
\includegraphics[width=0.275\textwidth,angle=-90]{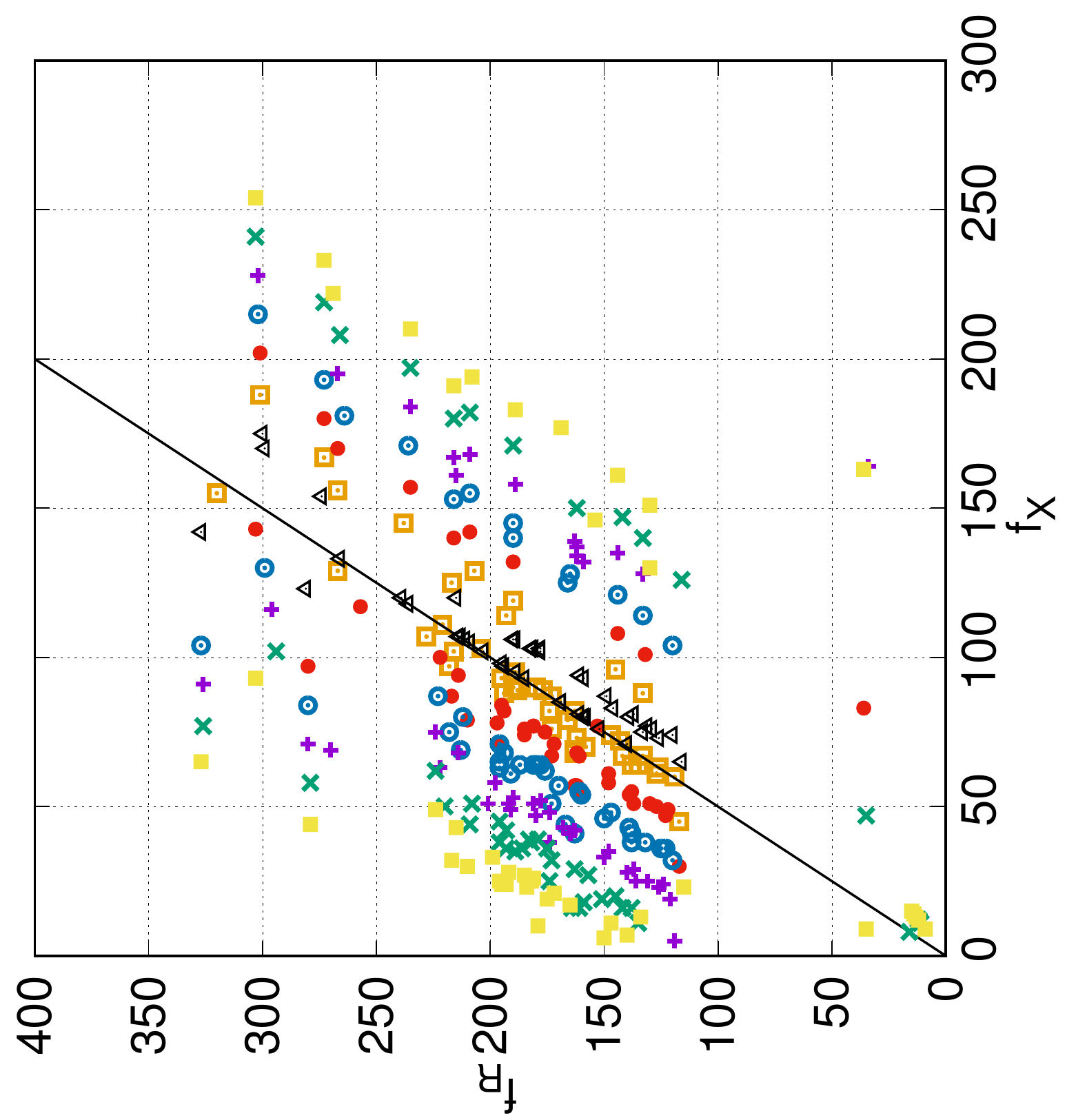}
\includegraphics[width=0.275\textwidth,angle=-90]{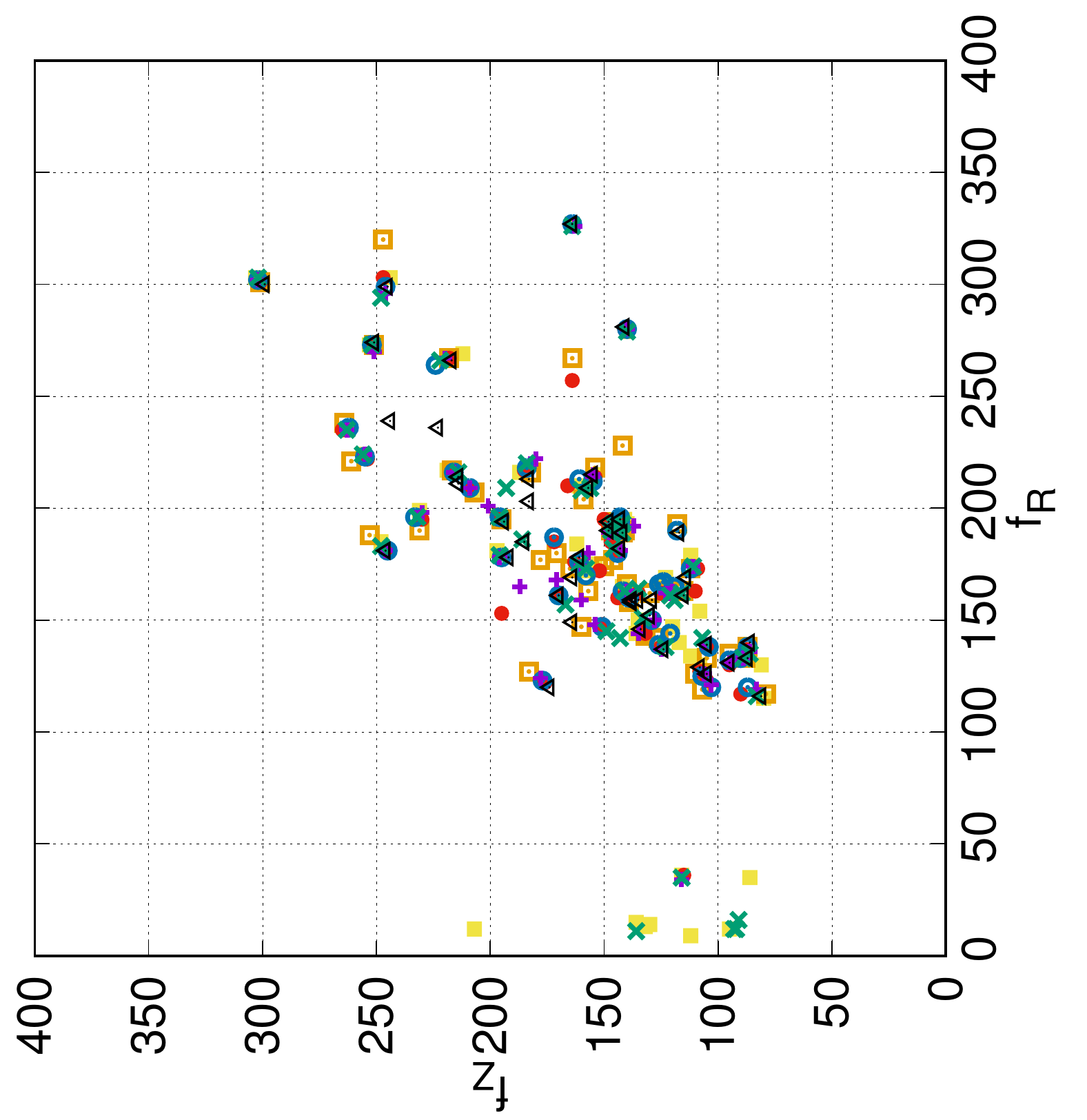}
\includegraphics[width=0.275\textwidth,angle=-90]{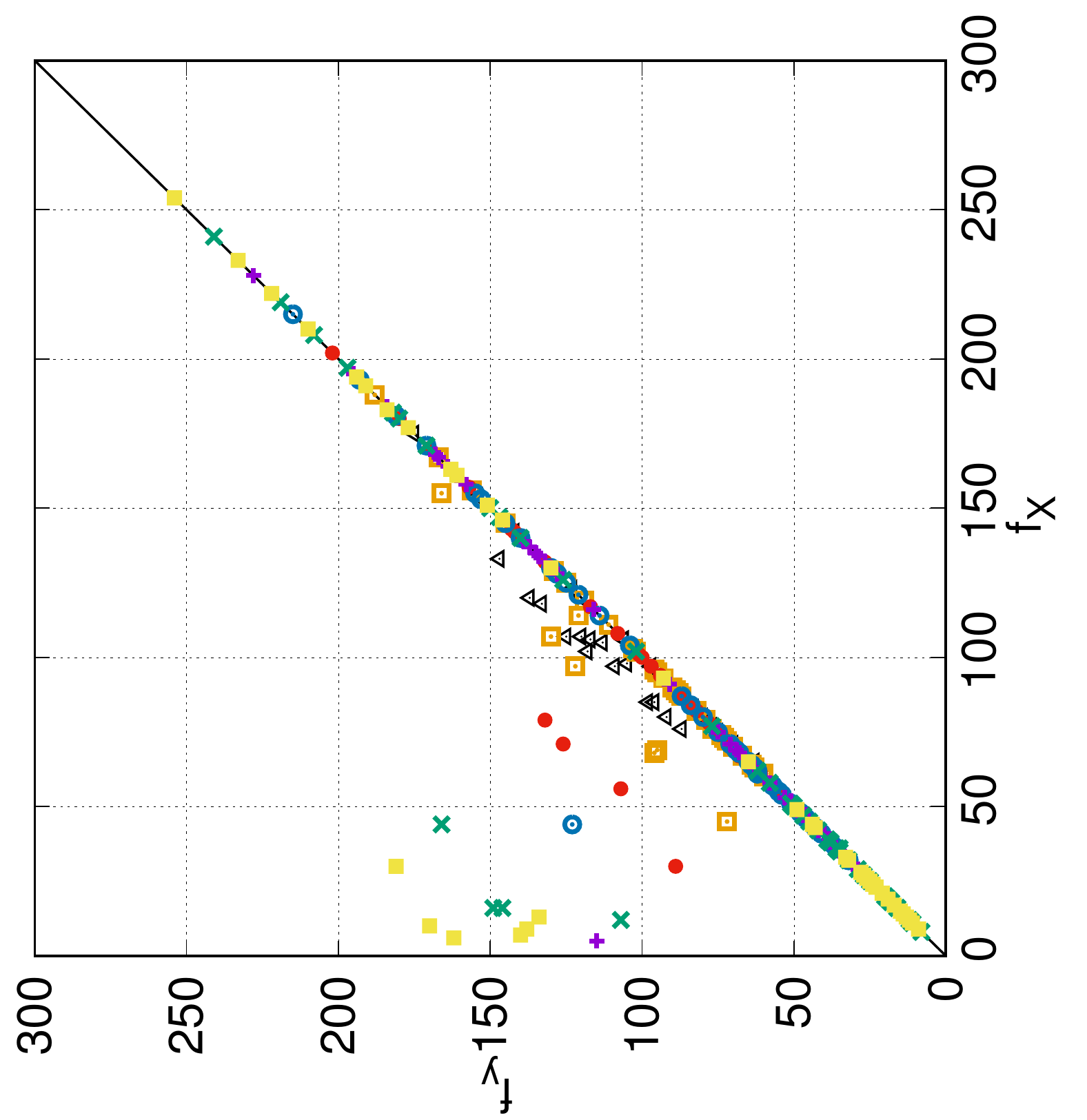}\
\caption{\small Diagrams $"f_X - f_R"$, $"f_R - f_Z"$ and $"f_X - f_Y"$ (from left to right) for values of the angular velocity of bar rotation $\Omega_b = 0, 10, 20, 30, 40, 50, 60$ km s$^{-1}$ kpc$^{-1}$, shown in black, orange, red, blue, purple, green, yellow respectively ($M_{bar}=430\times M_G$, $q_b=5$ kpc). The lines in left panel show the straight line corresponding to the ratio $f_R/f_X = 2$, in right panel -- the ratio $f_X/f_Y=1.0$.}
\label{fcomp6}
\end{center}}
\end{figure*}

Figure~\ref{fcomp6} shows the diagrams $"f_X - f_R"$, $"f_R - f_Z"$ and $"f_X - f_Y"$ for all 45 globular clusters at bar rotation velocities $\Omega_b = 0, 10, 20, 30, 40, 50, 60$ km s$^{-1}$ kpc$^{-1}$, confirming, with a few exceptions, the properties of dominant frequencies listed above. Let us note that these exceptions are of independent interest and require a separate study in the future.

\bigskip

 {\begin{table}[t]                                    
 \caption[]
 {\small\baselineskip=1.0ex
Probabilities ($P$) of GC capture by the bar when varying the bar parameters according to a uniform distribution law, taking into account the uncertainties of the GC 6D phase space and the peculiar velocity of the Sun using the Monte Carlo method}
 \label{t:fff}
 \begin{center}\begin{tabular}{|l|l||l|l|}\hline
  ID       & $P$     & ID       & $P$     \\\hline
 Terzan 3  & 0.14  & NGC 6522 & 0.38   \\\hline
 NGC 6256  & 0.17  & NGC 6539 & 0.11   \\\hline
 NGC 6266  & 0.20  & NGC 6540 & 0.68   \\\hline
 NGC 6304  & 0.08  & NGC 6569 & 0.20   \\\hline
 NGC 6342  & 0.17  & ESO 456-SC78 & 0.10 \\\hline
 Terzan 4  & 0.11  & NGC 6717 & 0.19   \\\hline
 Terzan 5  & 0.24  & NGC 6723 & 0.04   \\\hline
 \end{tabular}\end{center}\end{table}}

\section{FREQUENCY ANALYSIS OF ORBITS CONSTRUCTED AT DIFFERENT BAR PARAMETERS}

In the previous Section, the dependence of dominant frequencies on the angular velocity of bar
rotation was studied. As modeling has shown, the values of the dominant frequencies also depend
on the mass and size of the bar. We simulated the orbital motion of all 45 GCs on a fairly dense
grid of bar parameters, namely: the bar mass $M_{bar}$ changed in the interval $[110 - 430]\times M_G$ with a step of $10\times M_G$, bar length $q_b$ -- in the interval $[2-5]$ kpc with a step of 0.25 kpc, bar angular velocity $\Omega_b$ -- in range of values $[10-60]$ km s$^{-1}$ kpc$^{-1}$ with a step of 2.5 km s$^{-1}$ kpc$^{-1}$ (see also Table~\ref{t:model-III}).

\begin{figure*}
{\begin{center}
\includegraphics[width=0.275\textwidth,angle=-90]{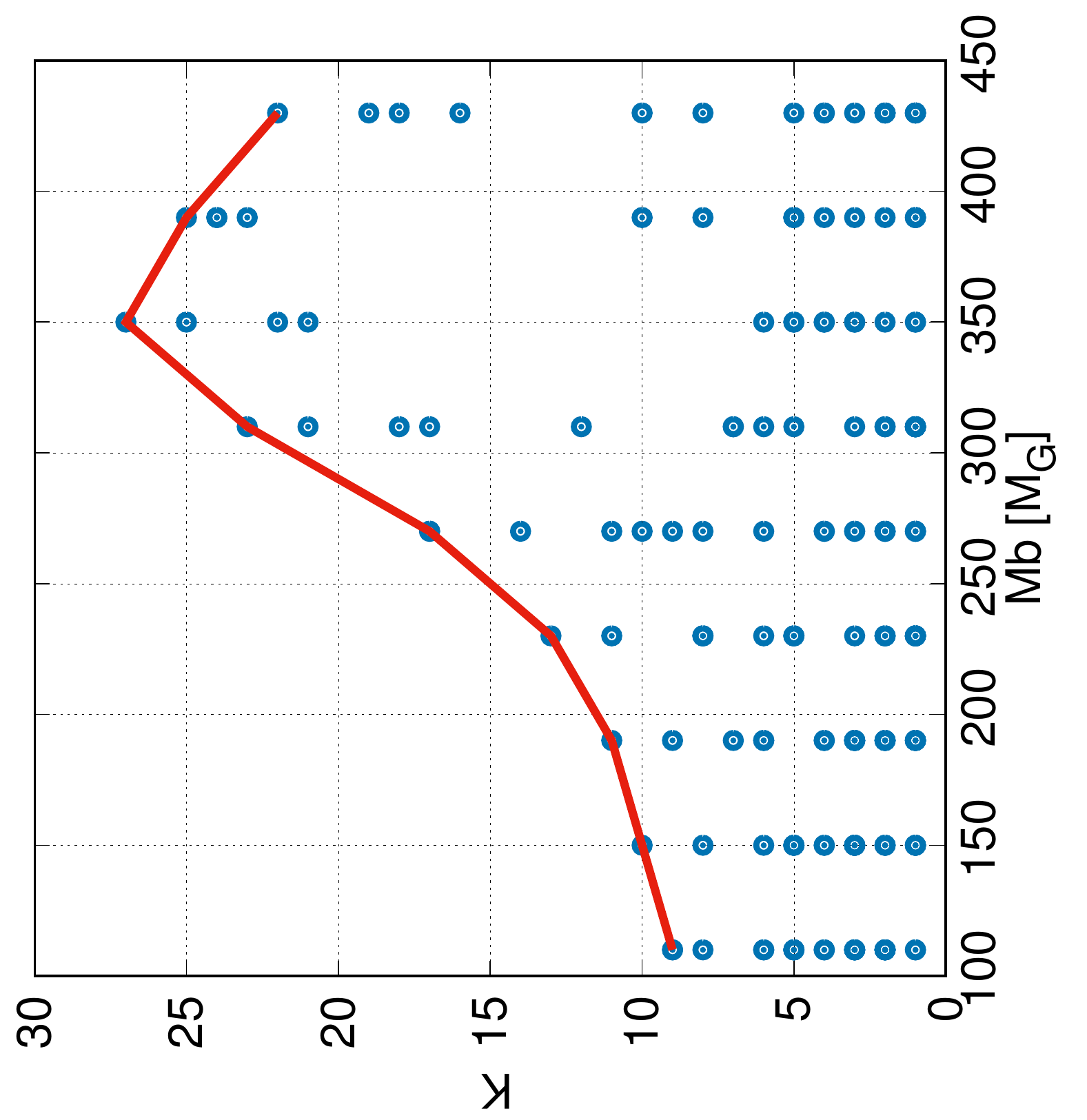}
\includegraphics[width=0.275\textwidth,angle=-90]{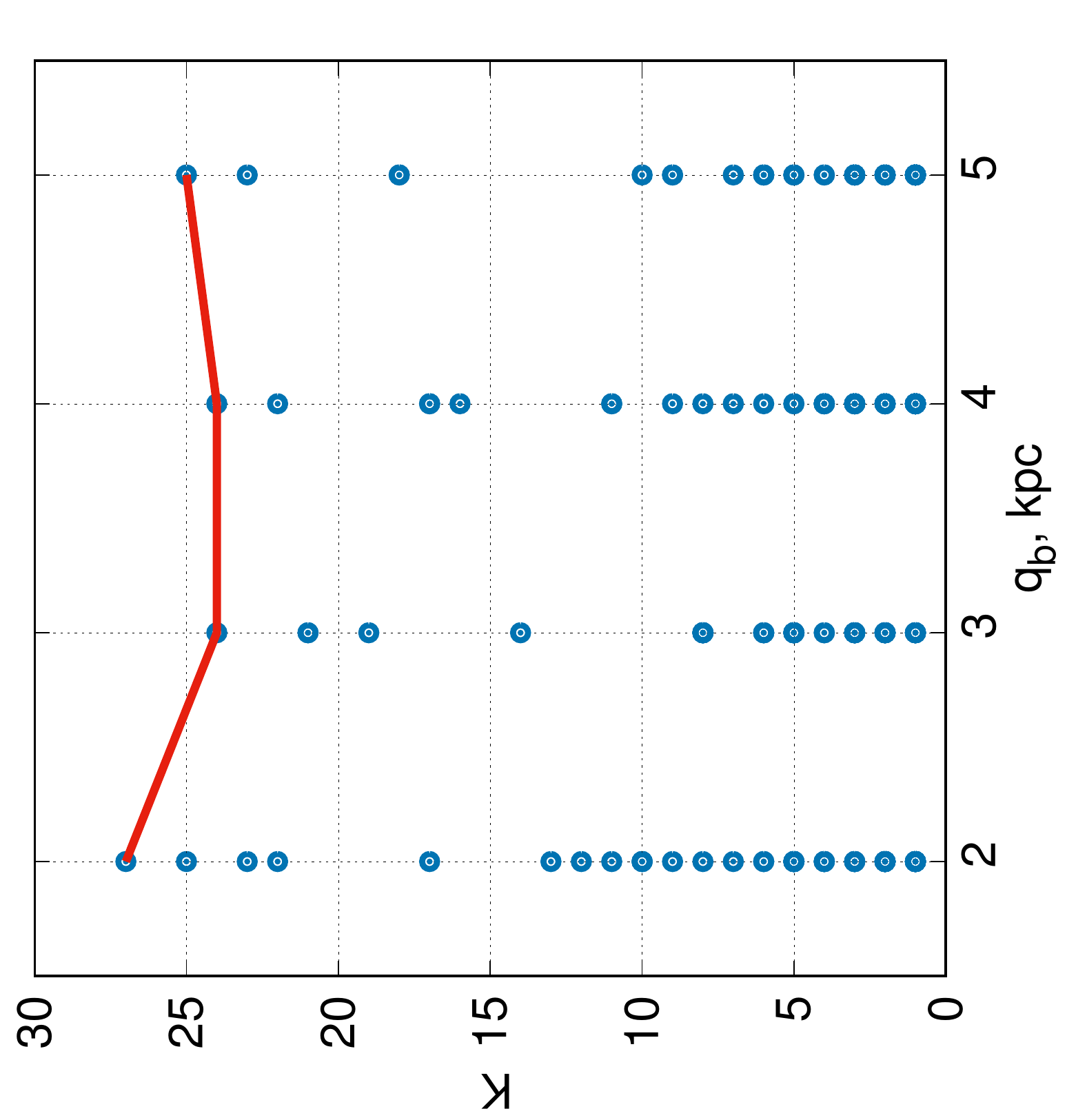}
\includegraphics[width=0.275\textwidth,angle=-90]{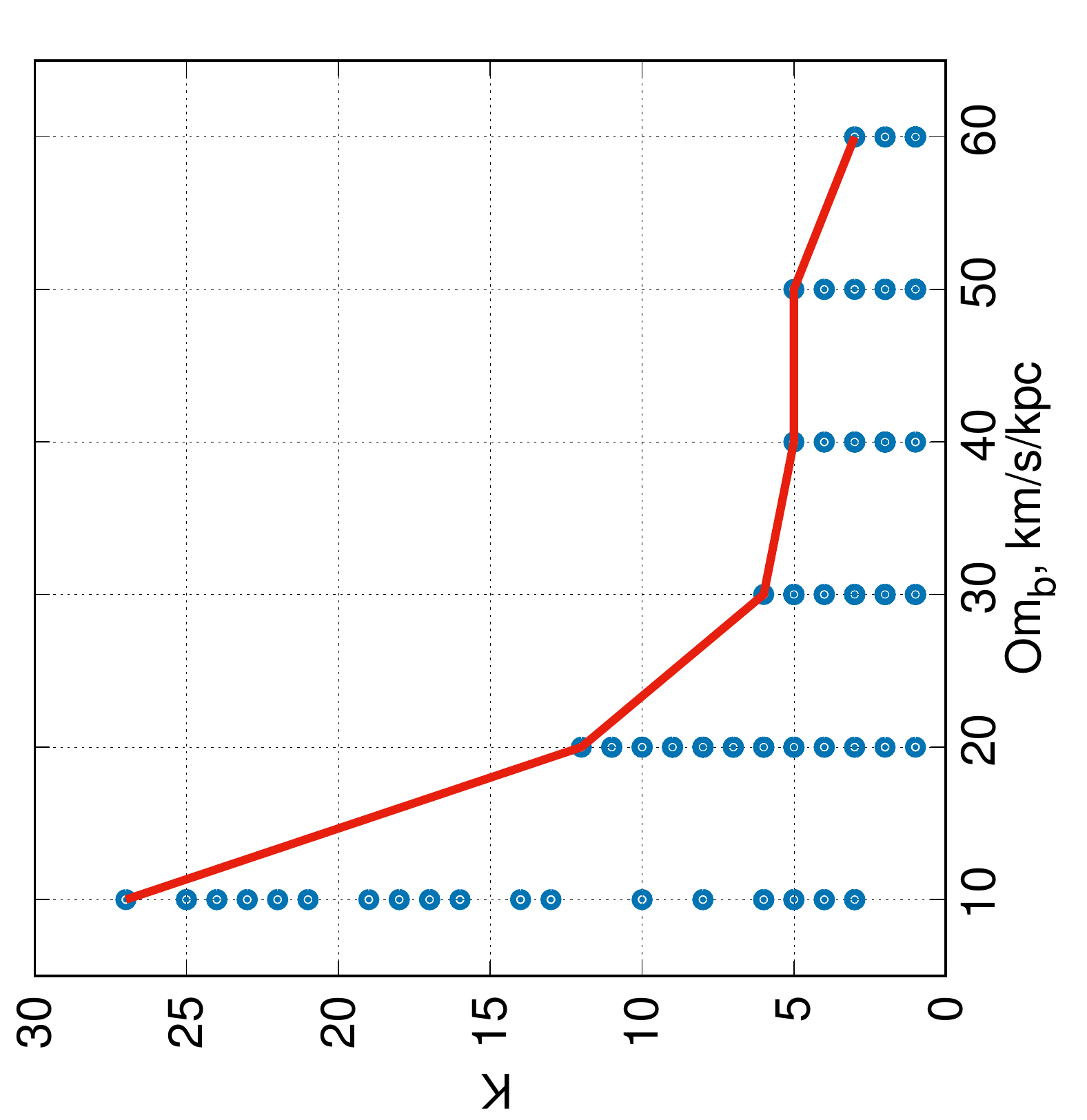}\
\caption{\small Dependence of the number of globular clusters with the frequency ratio $f_R/f_X \approx 2$ on the value of the bar parameters
$M_{bar}, q_b, \Omega_b$ (panels from left to right).}
\label{fcomp7}
\end{center}}
\end{figure*}

\begin{figure*}
{\begin{center}
\includegraphics[width=0.2\textwidth,angle=-90]{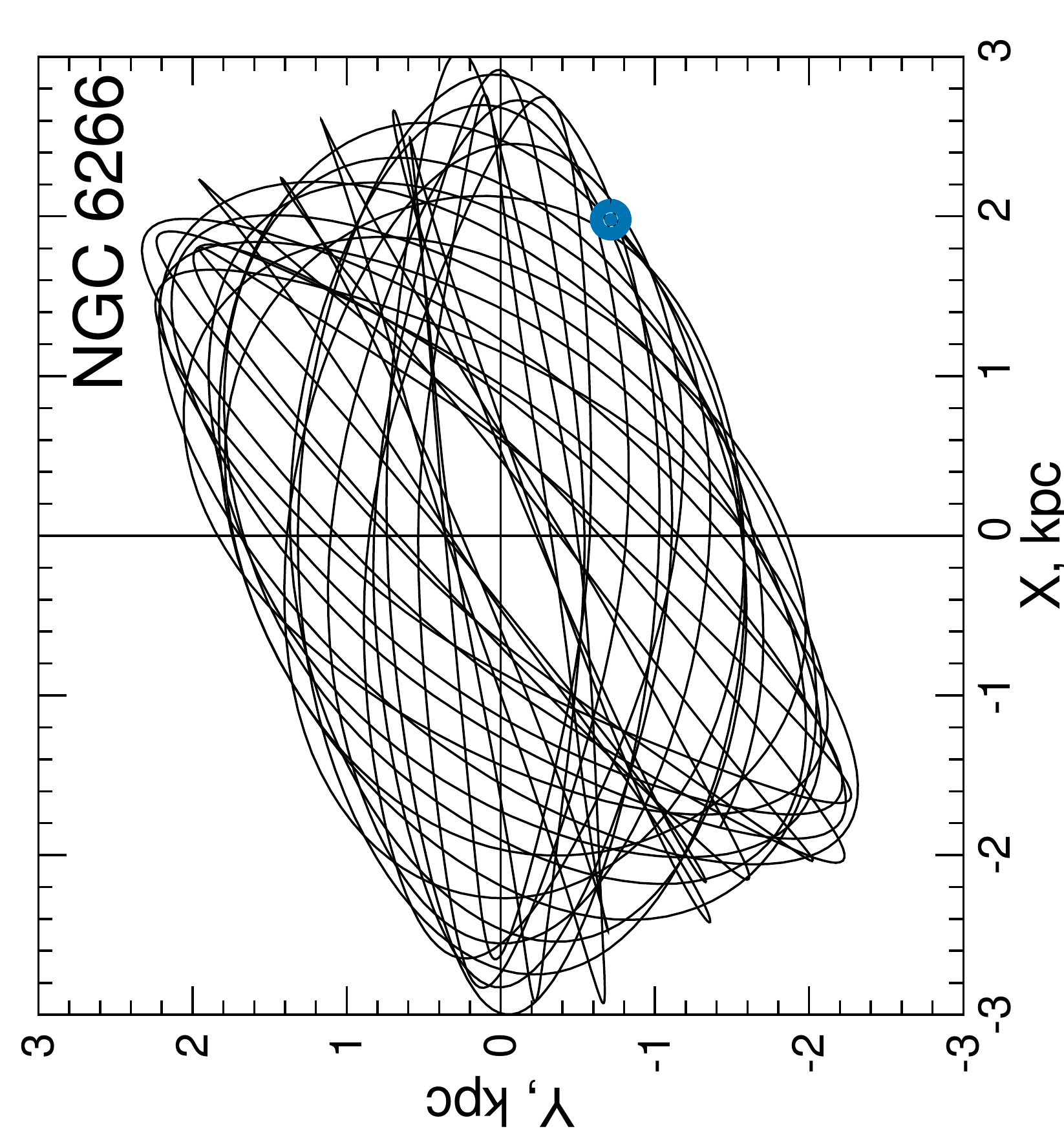}
\includegraphics[width=0.2\textwidth,angle=-90]{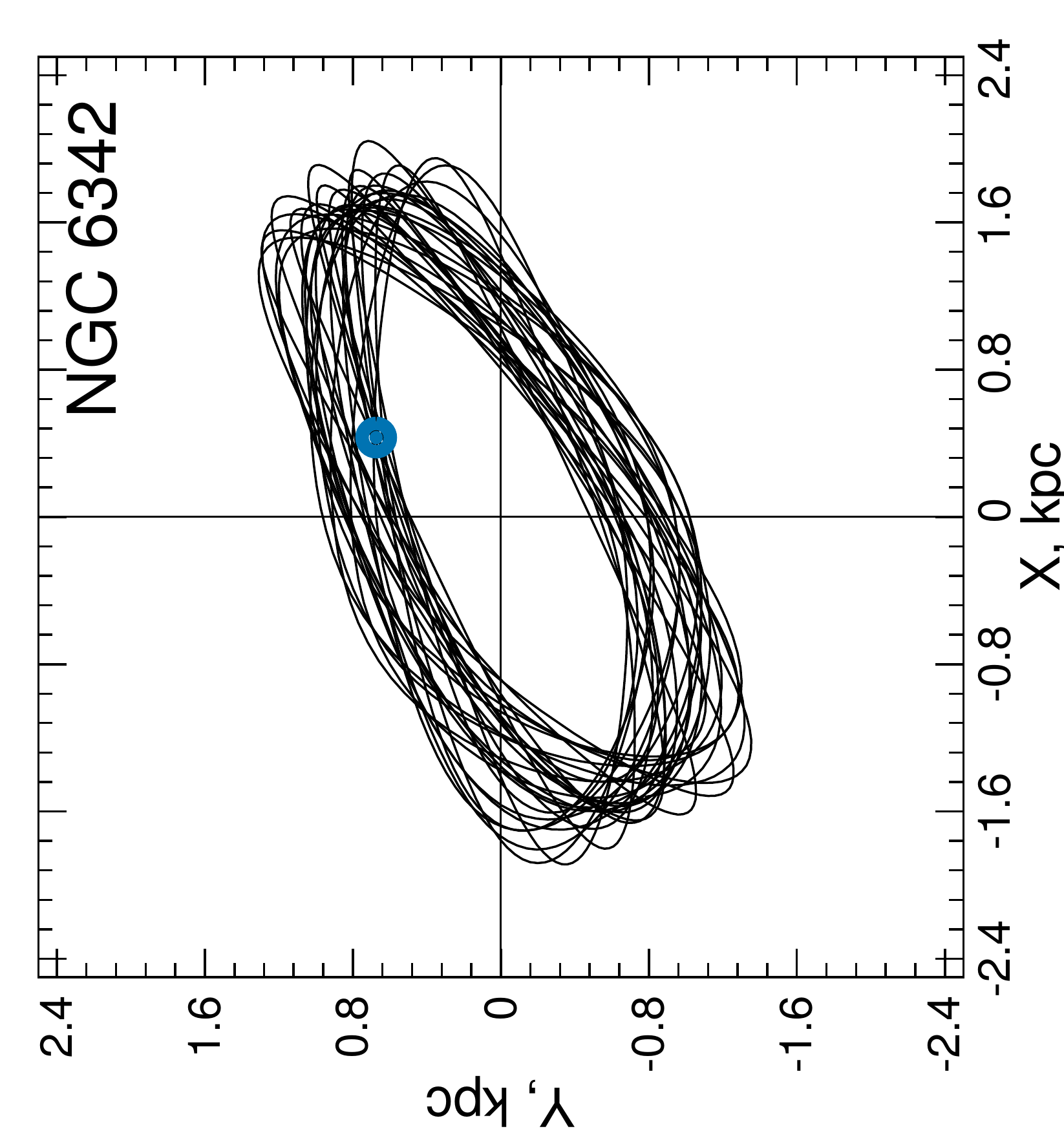}
\includegraphics[width=0.2\textwidth,angle=-90]{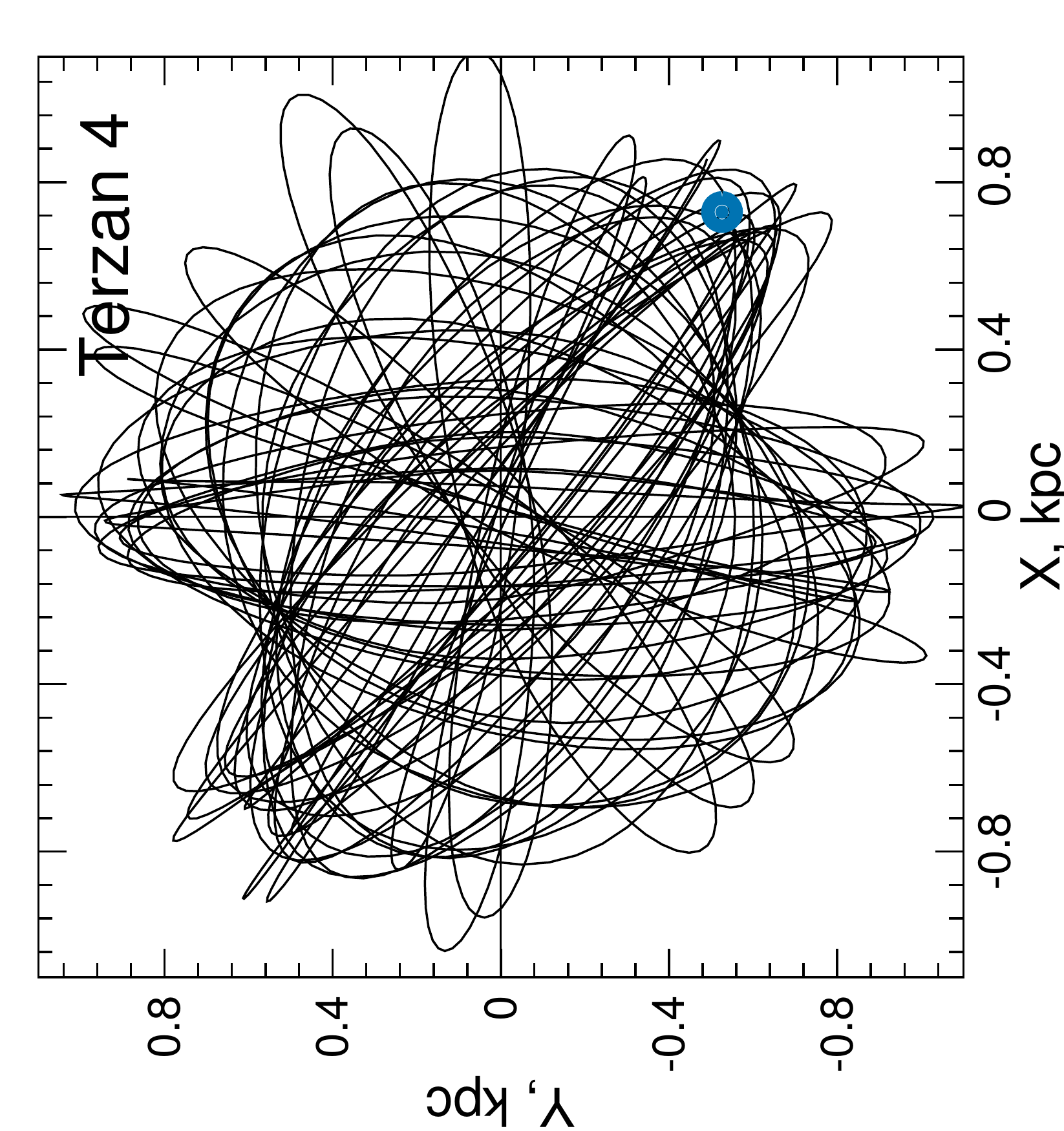}
\includegraphics[width=0.2\textwidth,angle=-90]{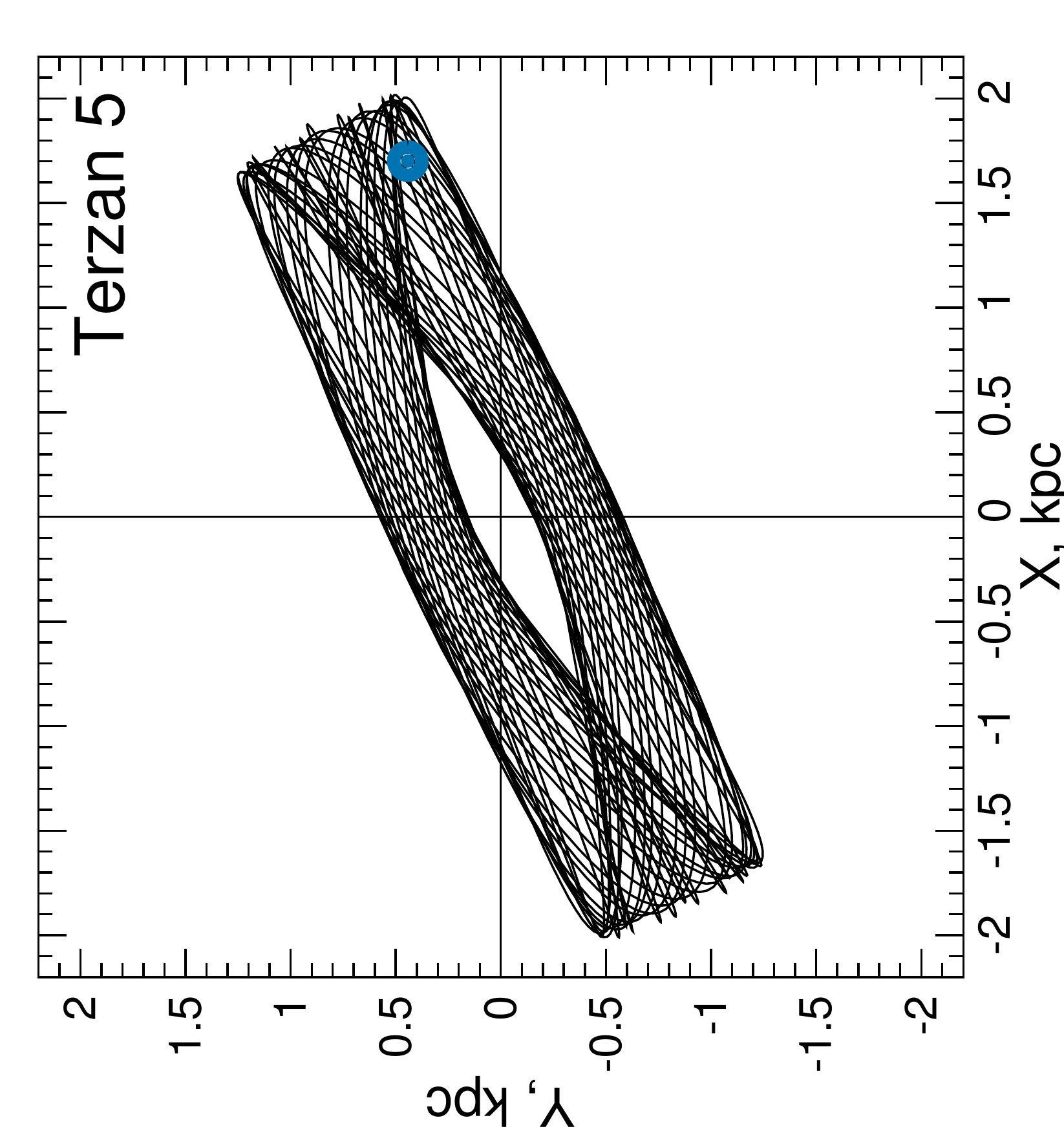}\

\bigskip

\includegraphics[width=0.2\textwidth,angle=-90]{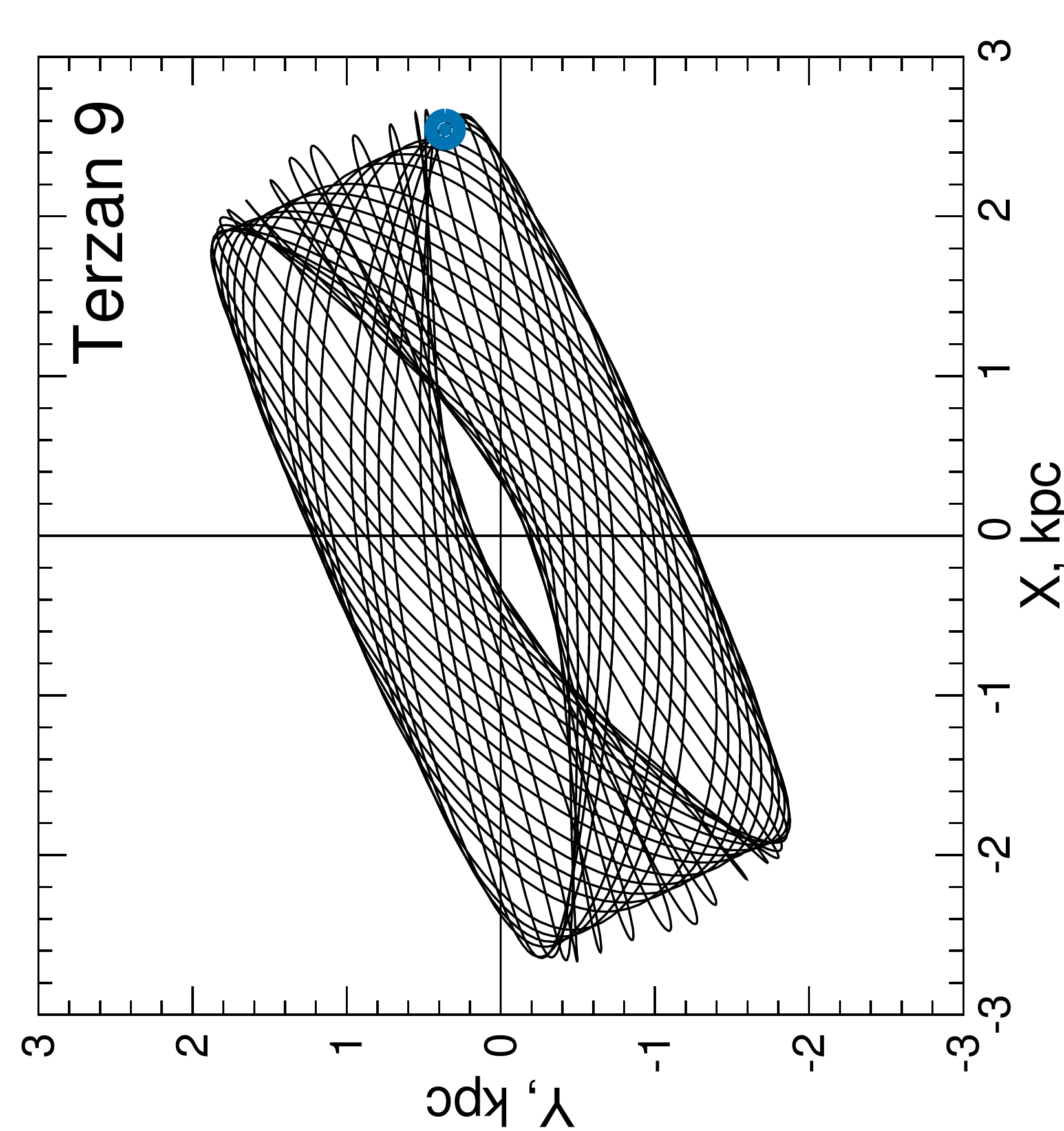}
\includegraphics[width=0.2\textwidth,angle=-90]{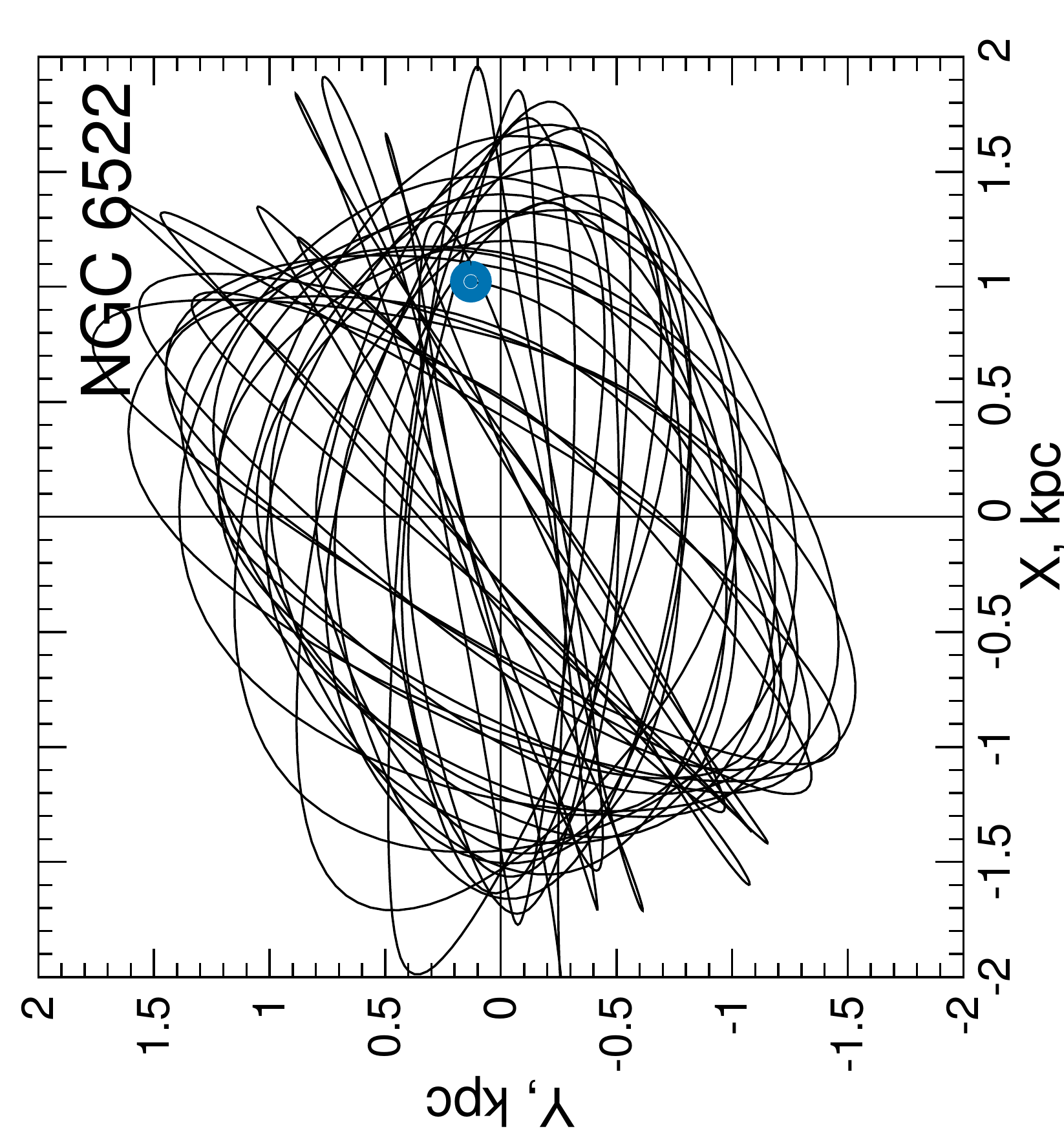}
\includegraphics[width=0.2\textwidth,angle=-90]{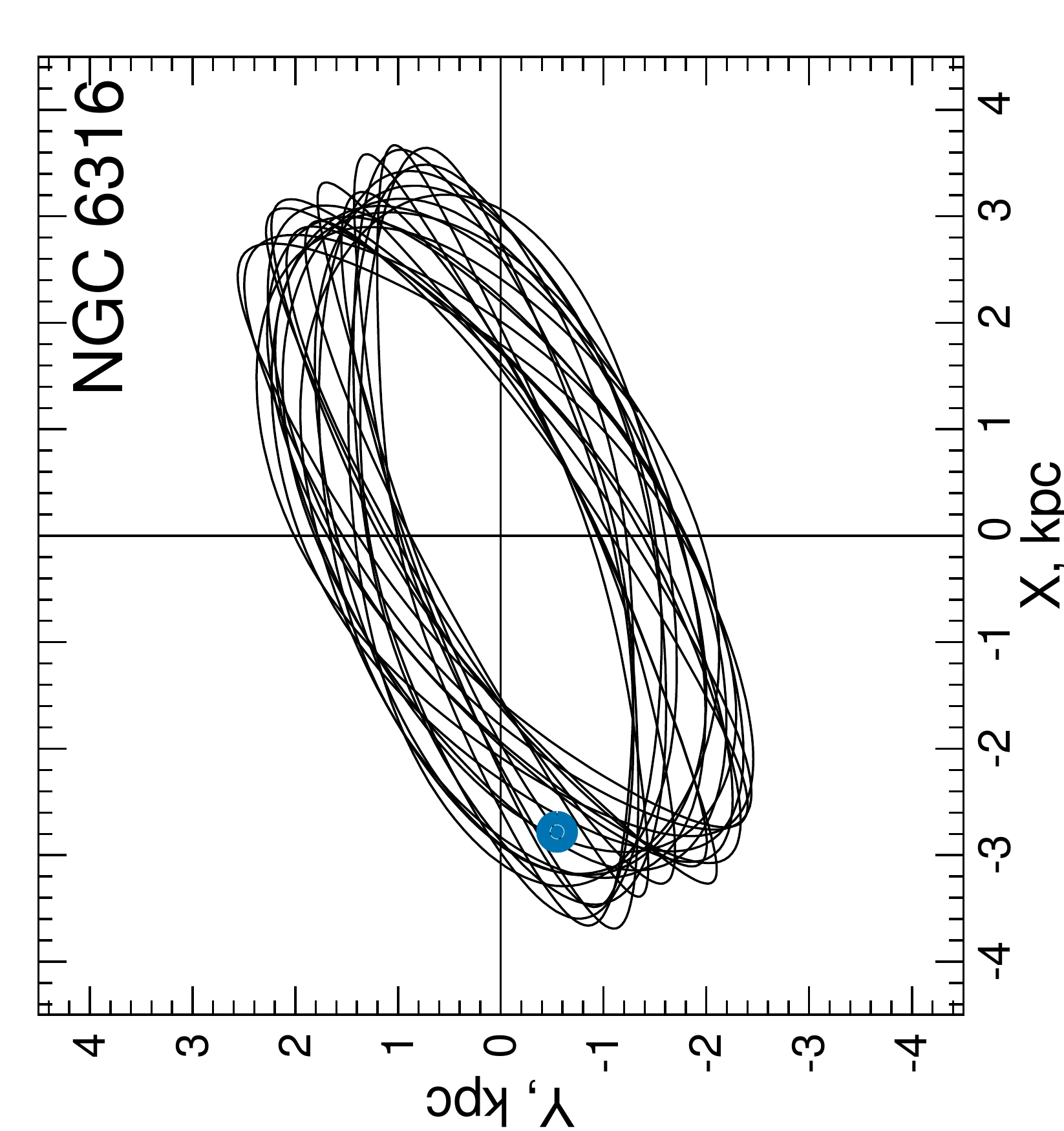}
\includegraphics[width=0.2\textwidth,angle=-90]{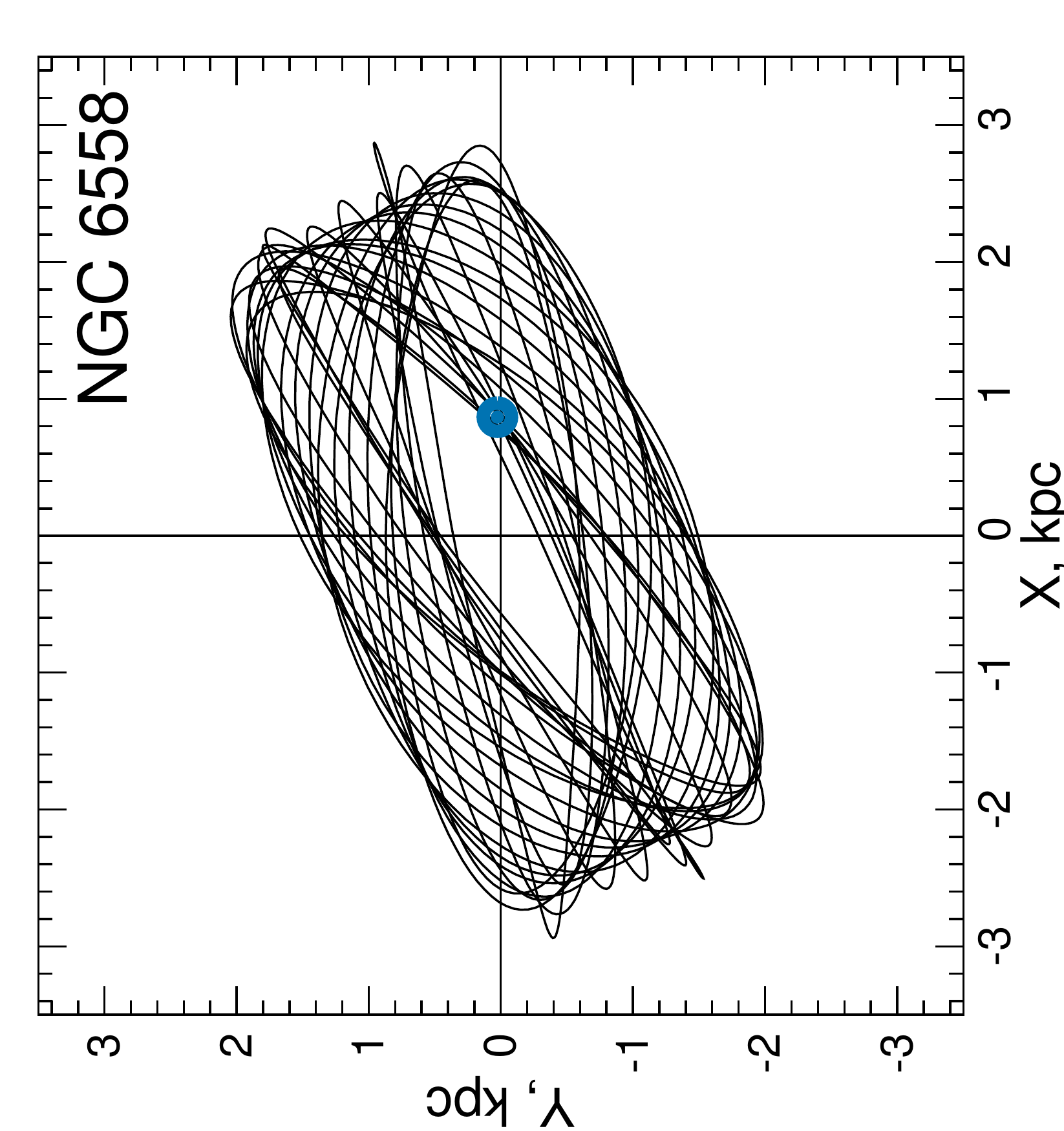}\

\bigskip

\includegraphics[width=0.2\textwidth,angle=-90]{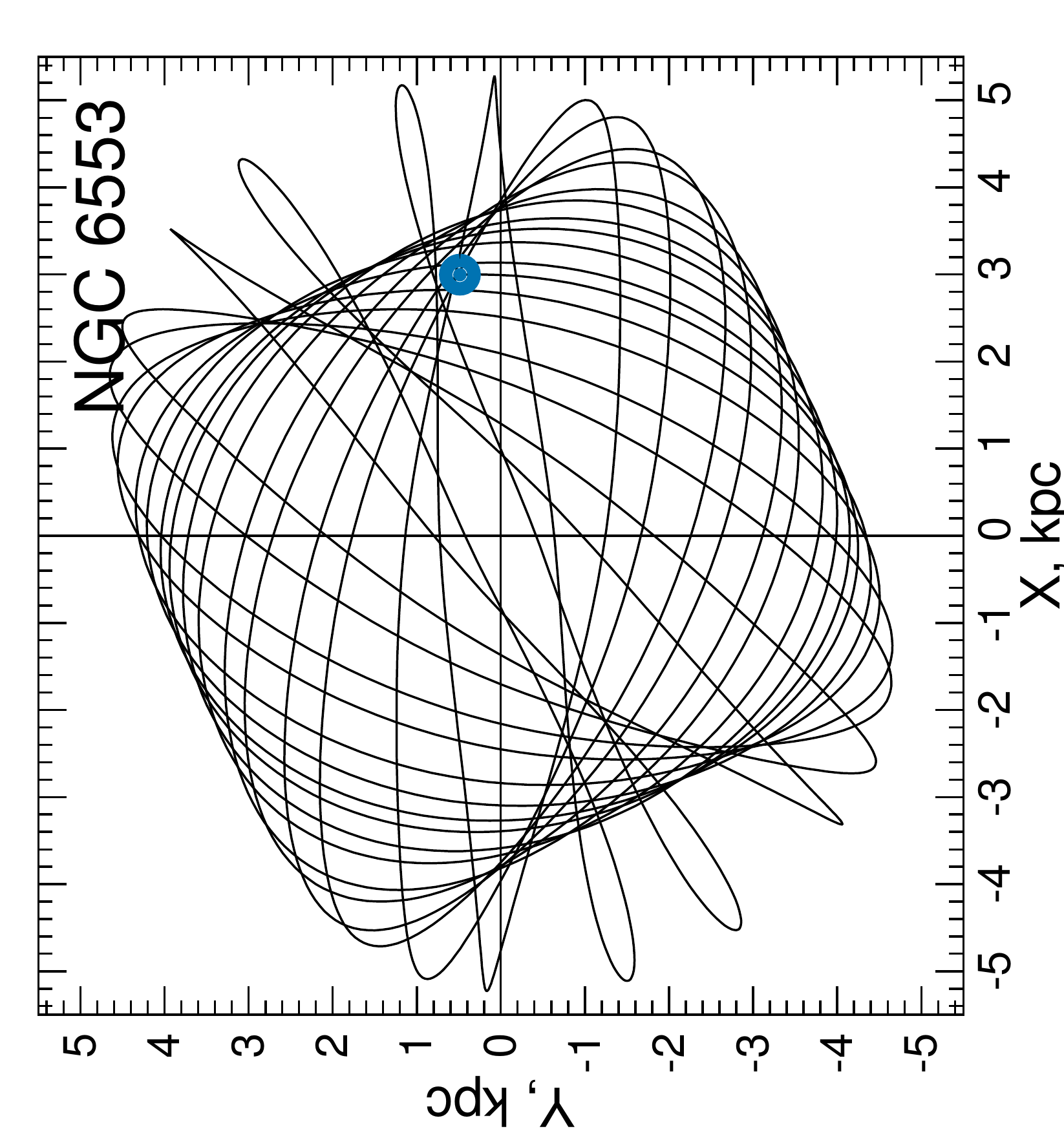}
\includegraphics[width=0.2\textwidth,angle=-90]{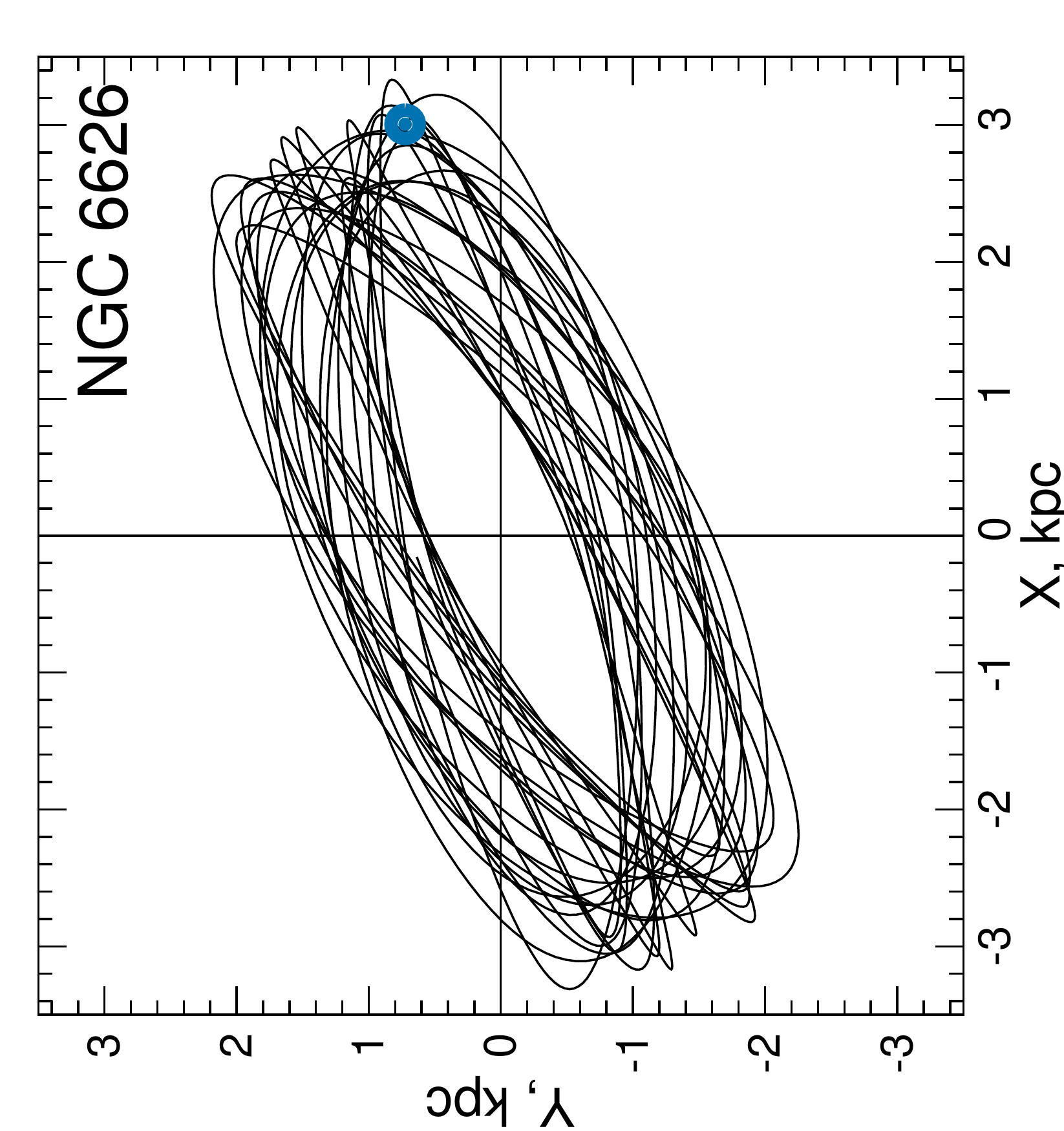}
\includegraphics[width=0.2\textwidth,angle=-90]{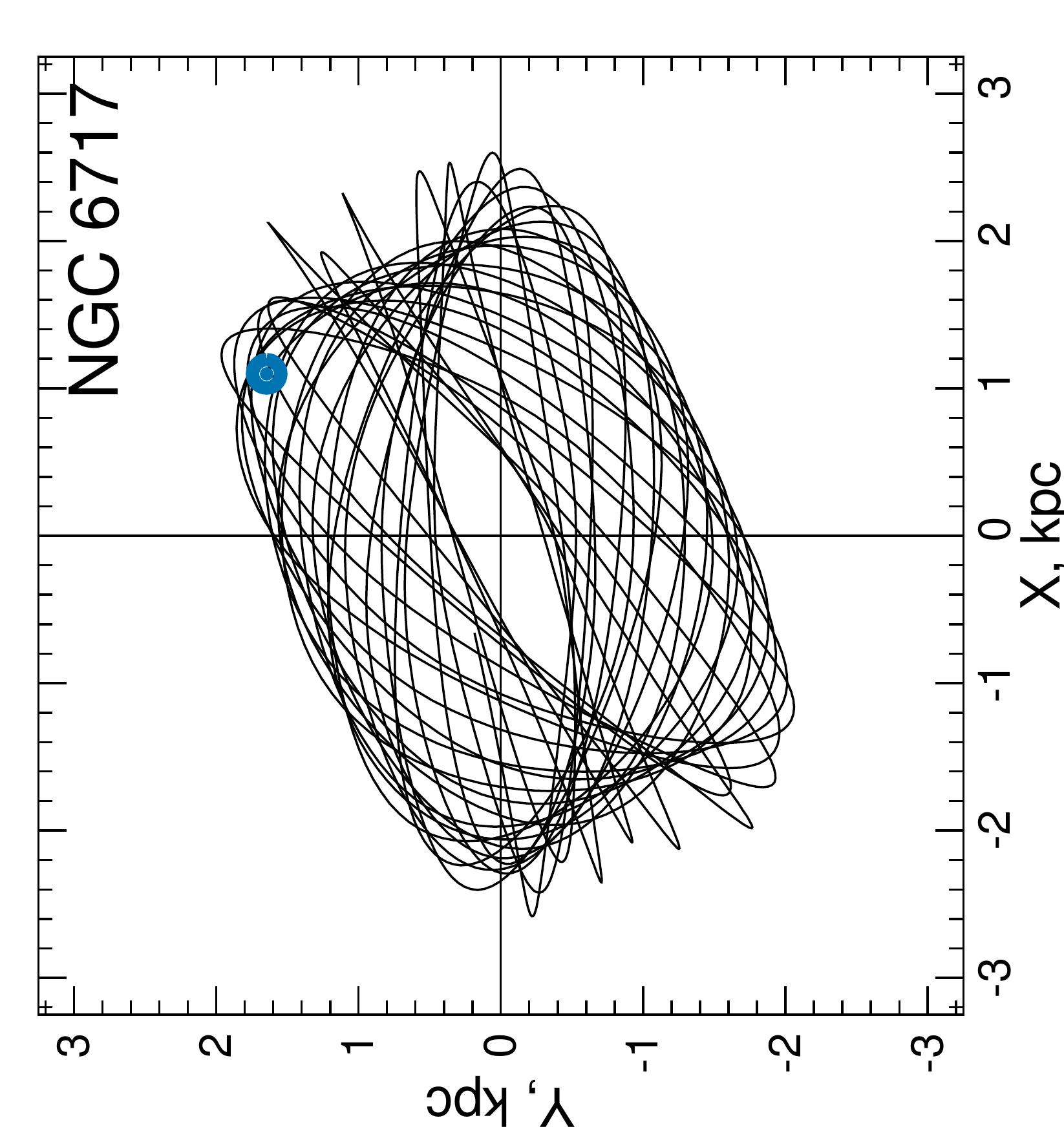}
\includegraphics[width=0.2\textwidth,angle=-90]{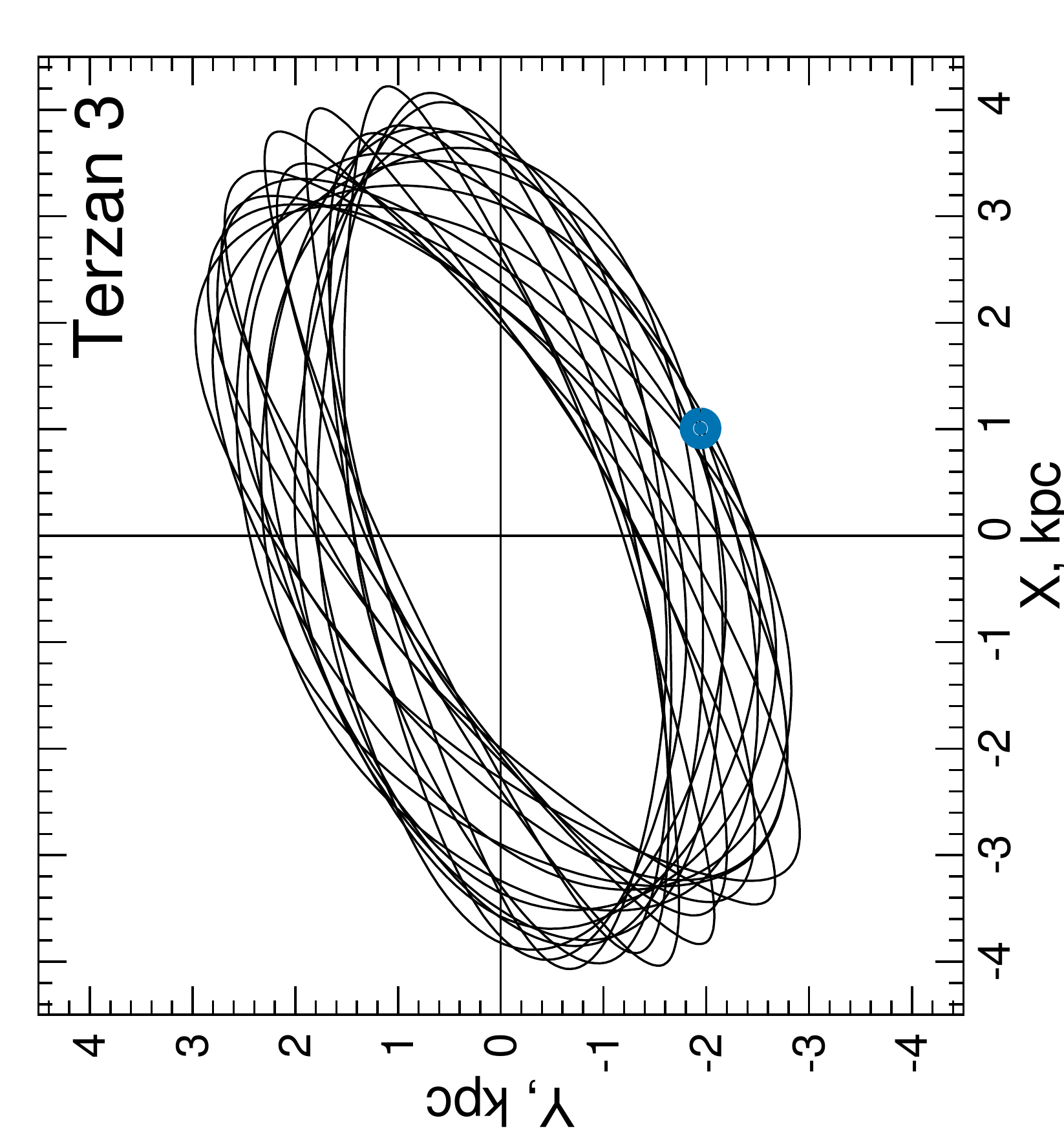}\

\bigskip

\includegraphics[width=0.2\textwidth,angle=-90]{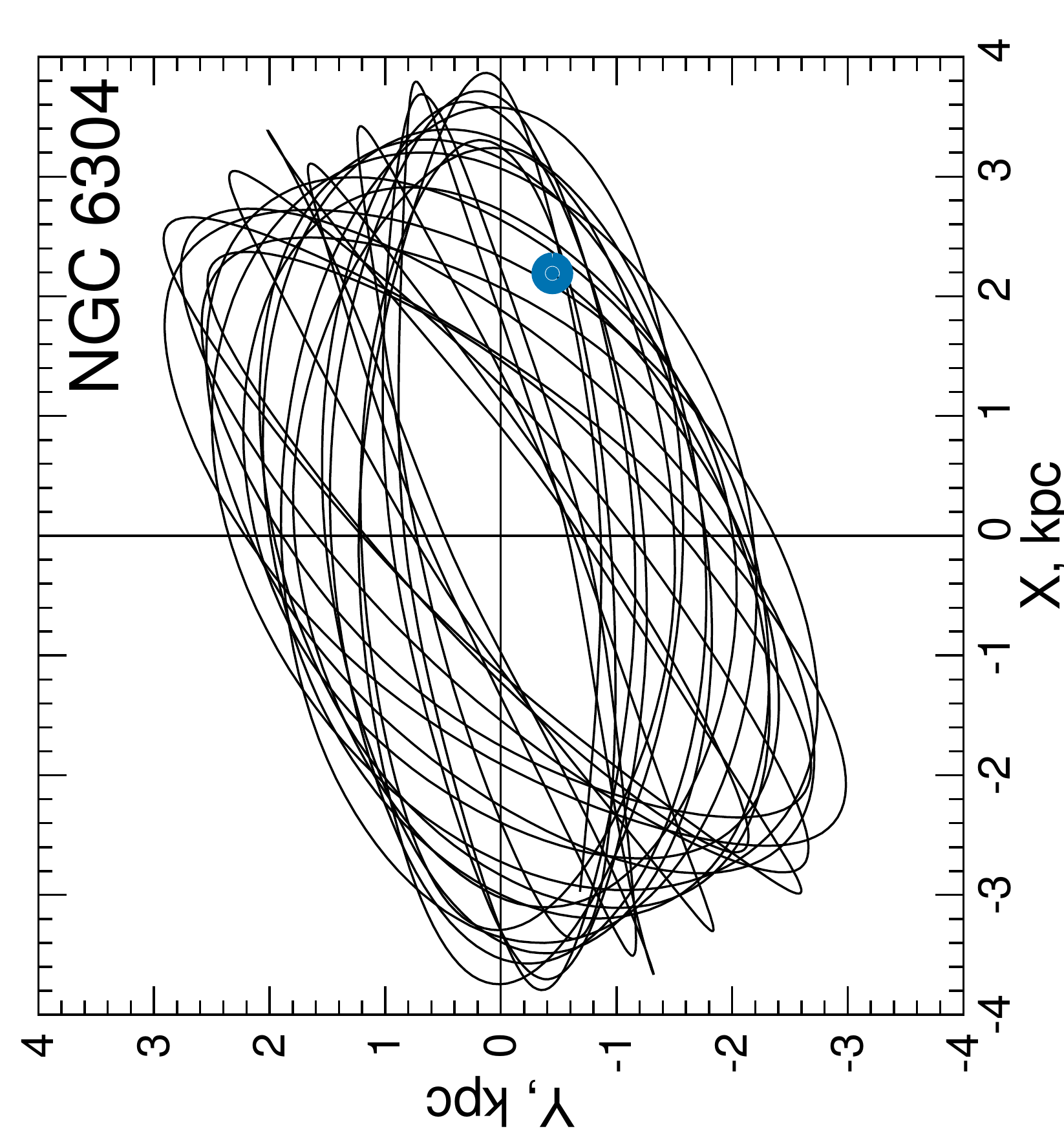}
\includegraphics[width=0.2\textwidth,angle=-90]{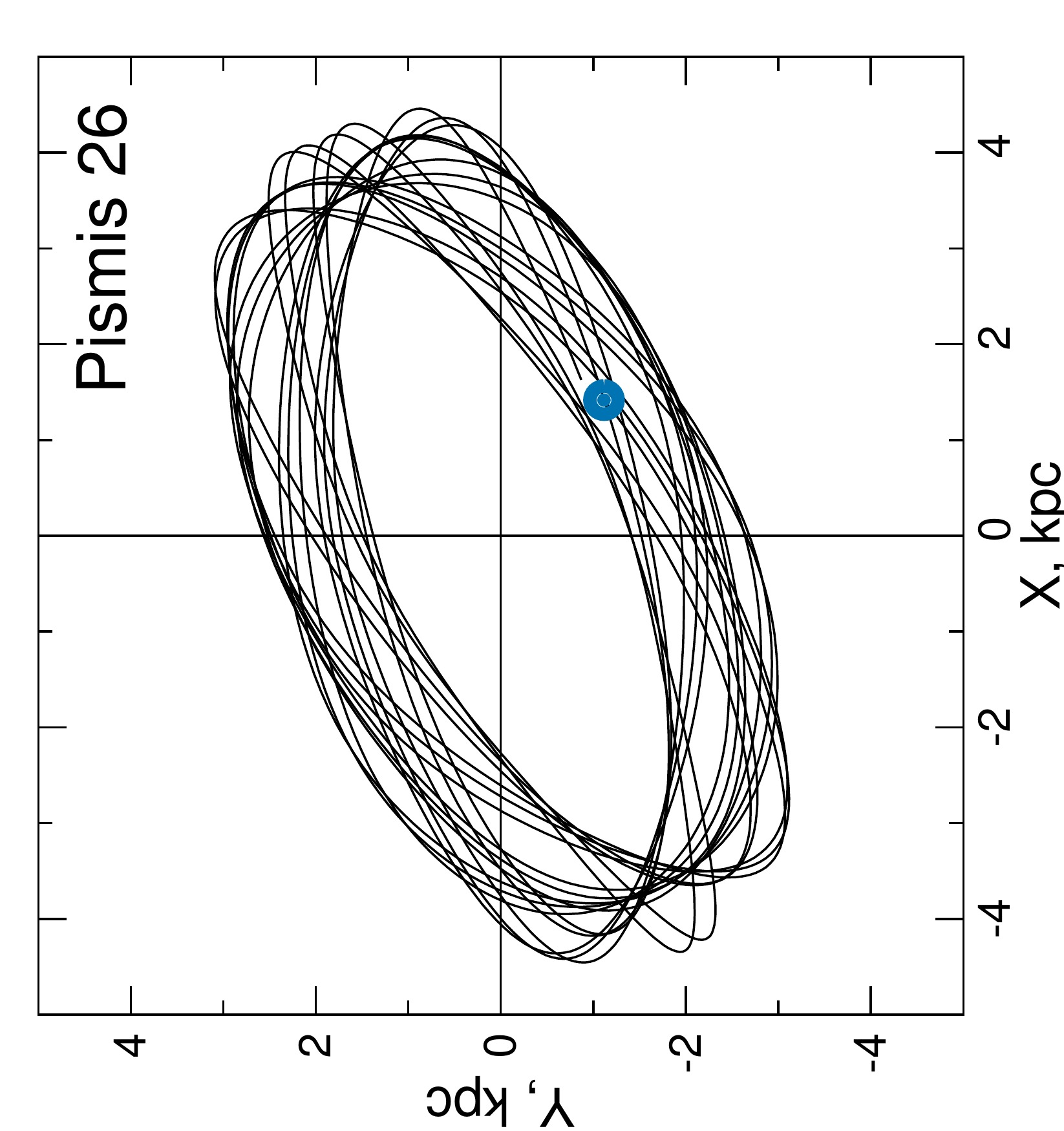}
\includegraphics[width=0.2\textwidth,angle=-90]{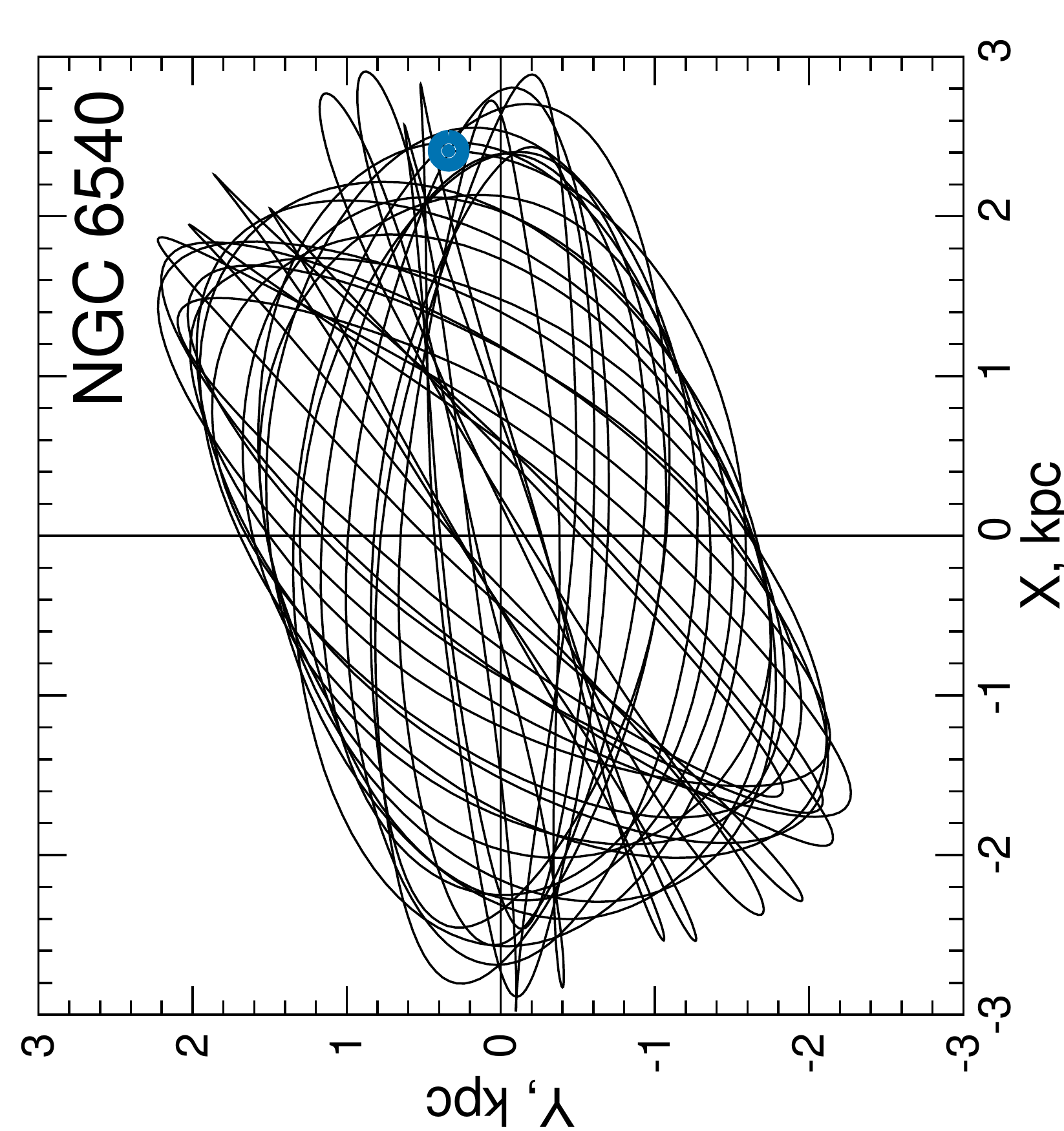}
\includegraphics[width=0.2\textwidth,angle=-90]{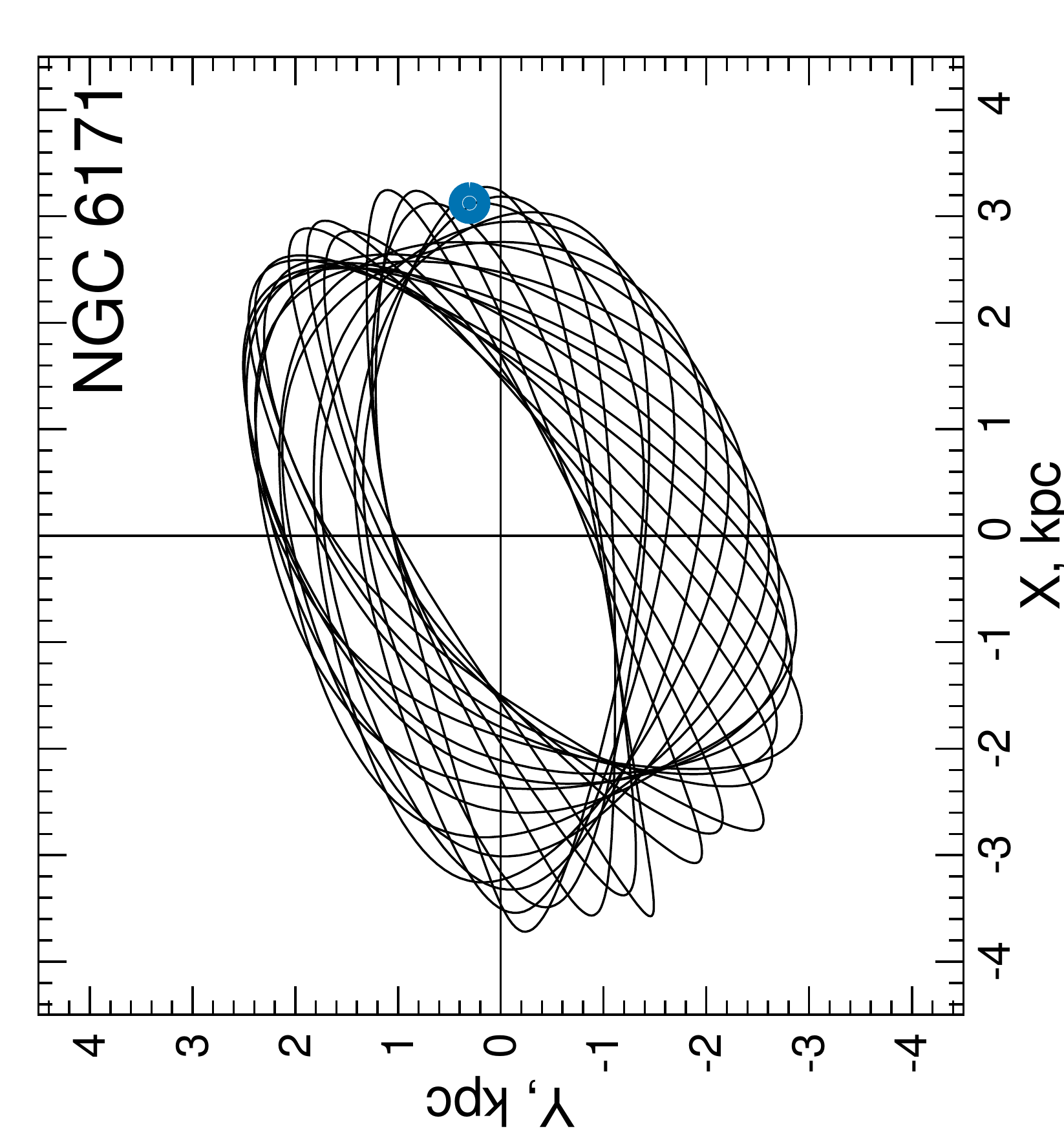}\
\caption{\small Orbits of globular clusters NGC6266, NGC6342, NGC6522, NGC6558, NGC6626, NGC6717, NGC6304, NGC6540, NGC6171, NGC6316, NGC6553, Terzan 3, Terzan 4, Terzan 5, Terzan 9, Pismis 26, built in a rotating bar system with a speed of $\Omega_b=10$ km s$^{-1}$ kpc$^{-1}$, mass $M_{bar}=430 \times M_G$ and length $q_b=5$ kpc, in projection onto the galactic plane $X-Y$, satisfying the ratio of dominant frequencies $f_R/f_X \approx 2$. The beginning of the orbits is indicated by a blue circle.}
\label{fcomp8}
\end{center}}
\end{figure*}

Next, for each orbit, the ratio of dominant frequencies $f_R/f_X$ was calculated in order to determine whether the orbit was captured by the bar with the appropriate parameters or not according to the criterion $f_R/f_X=2\pm 0.1$. Below in Fig.~\ref{fcomp7} the obtained dependences of the number of globular clusters with $f_R/f_X\approx 2$ on the values of the bar parameters -- mass, length and angular velocity of rotation -- are shown.

As can be seen from the figure, the number of globular clusters that satisfy the criterion
$f_R/f_X=2\pm 0.1$ depends most strongly on the mass and rotation speed of the bar and reaches a maximum at $M_{bar}=350\times M_G$ and $\Omega_b=10$ km s$^{-1}$ kpc$^{-1}$. The dependence on the bar length is less significant, however, it should be noted that the largest number of GCs with a frequency ratio $f_R/f_X\approx 2$ is observed at the shortest bar length of 2 kpc.

Thus, as a result of the simulation, we determined for each set of values ($M_{bar}, q_b, \Omega_b$) from the parameter grid indicated above, globular clusters that satisfy the criterion $f_R/f_X=2\pm 0.1$. Let’s give a few examples. Thus, for the most realistic values of the bar mass $M_{bar}=430\times M_G$ and bar length $q_b=5$ kpc at a bar rotation speed $\Omega_b=10$
km s$^{-1}$ kpc$^{-1}$ the criterion $f_R/f_X=2\pm 0.1$ is satisfied by the globular clusters NGC6266, NGC6342, NGC6522, NGC6558, NGC6626, NGC6717, NGC6304, NGC6540, NGC6171, NGC6316, NGC6553, Terzan 3, Terzan 4, Terzan 5, Terzan 9, Pismis 26; at $\Omega_b=15$ km s$^{-1}$ kpc$^{-1}$ -- the GCs NGC6266, NGC6717, NGC6304, NGC6540, Terzan 3, Terzan 4, Terzan 5; at $\Omega_b=20$ km s$^{-1}$ kpc$^{-1}$ -- the GC NGC6540; at $\Omega_b=30$ and $40$ km s$^{-1}$ kpc$^{-1}$ -- there is no GC, at $\Omega_b=50$ km s$^{-1}$ kpc$^{-1}$ -- the GC NGC6539. With the same bar mass, but length $q_b=2.25$ kpc and rotation speed $\Omega_b=45$ km s$^{-1}$ kpc$^{-1}$ the criterion $f_R/f_X=2\pm 0.1$ is satisfied by the GCs NGC6266, NGC6522, NGC6540, NGC6553, Terzan 3, ESO456-SC78.

\begin{figure*}
{\begin{center}
\includegraphics[width=0.275\textwidth,angle=-90]{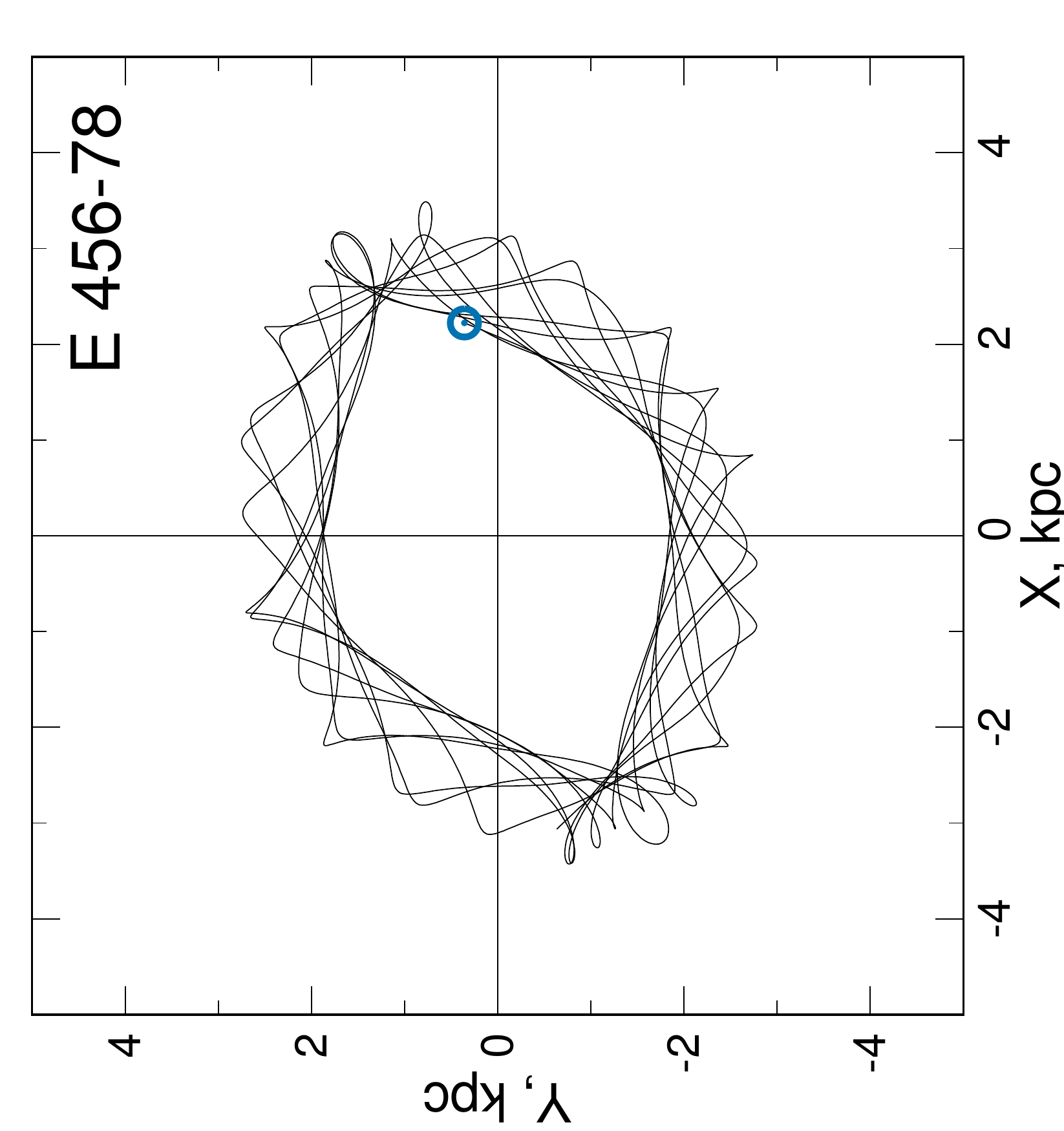}
\includegraphics[width=0.275\textwidth,angle=-90]{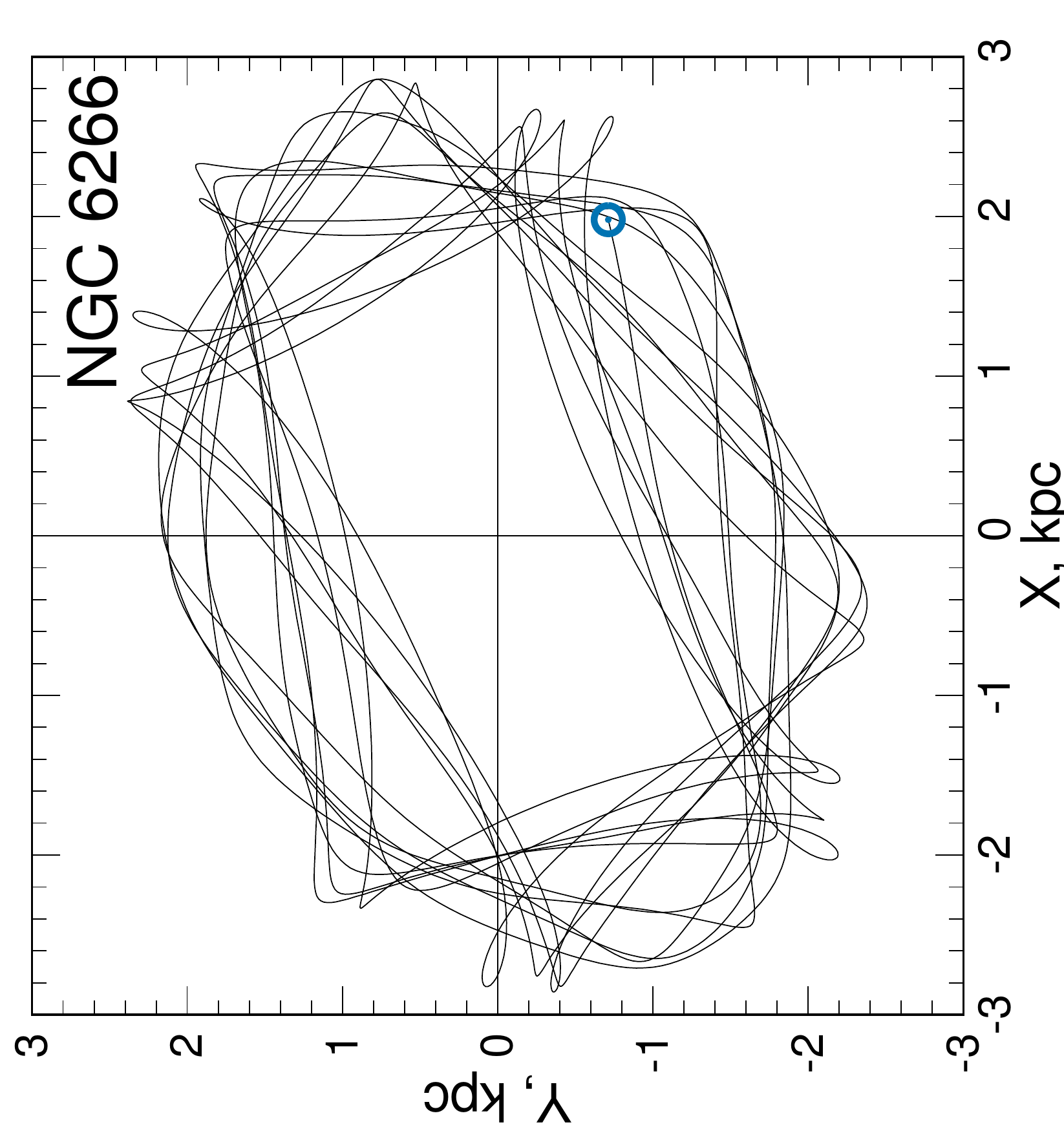}
\includegraphics[width=0.275\textwidth,angle=-90]{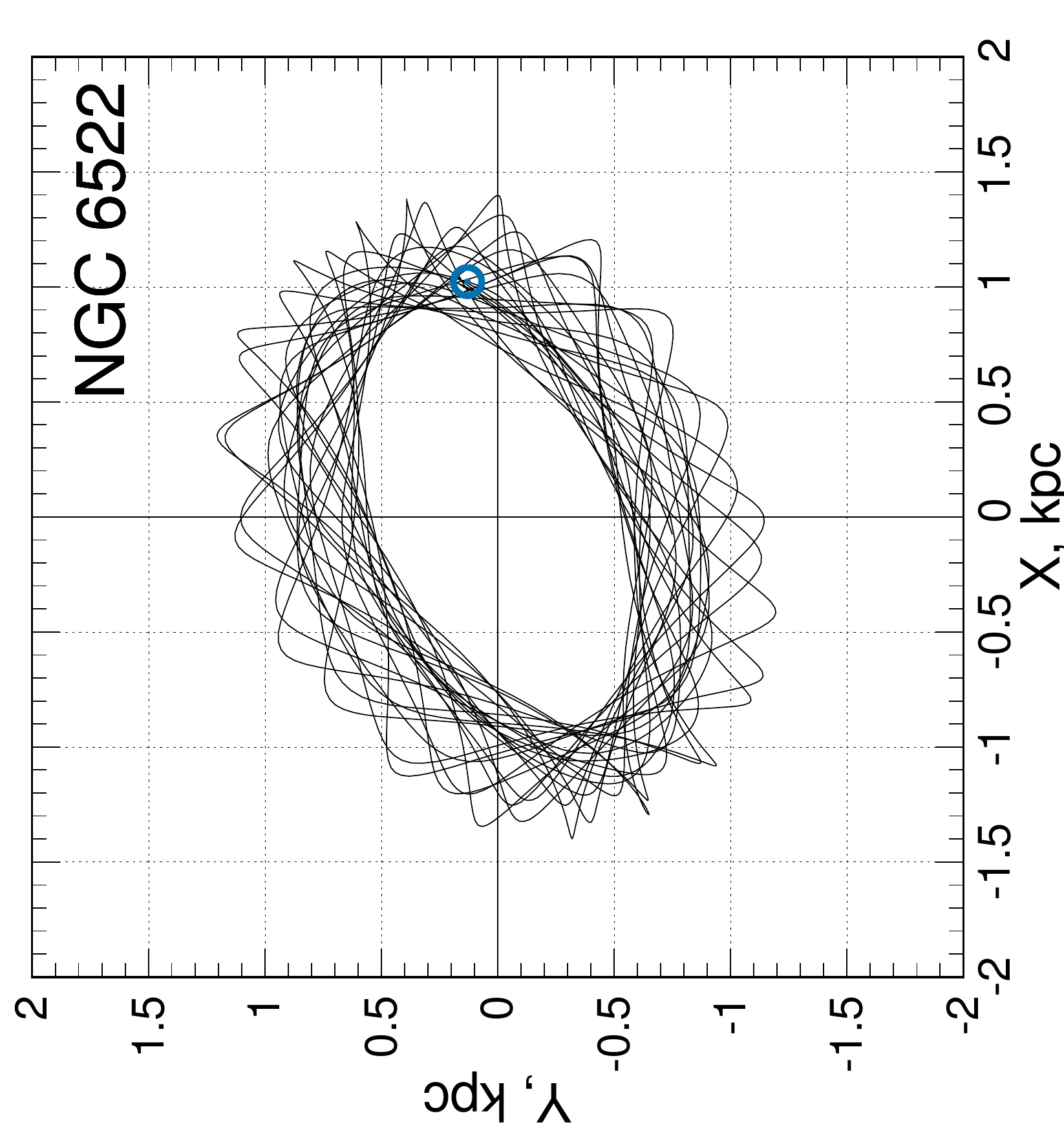}\

\bigskip

\includegraphics[width=0.275\textwidth,angle=-90]{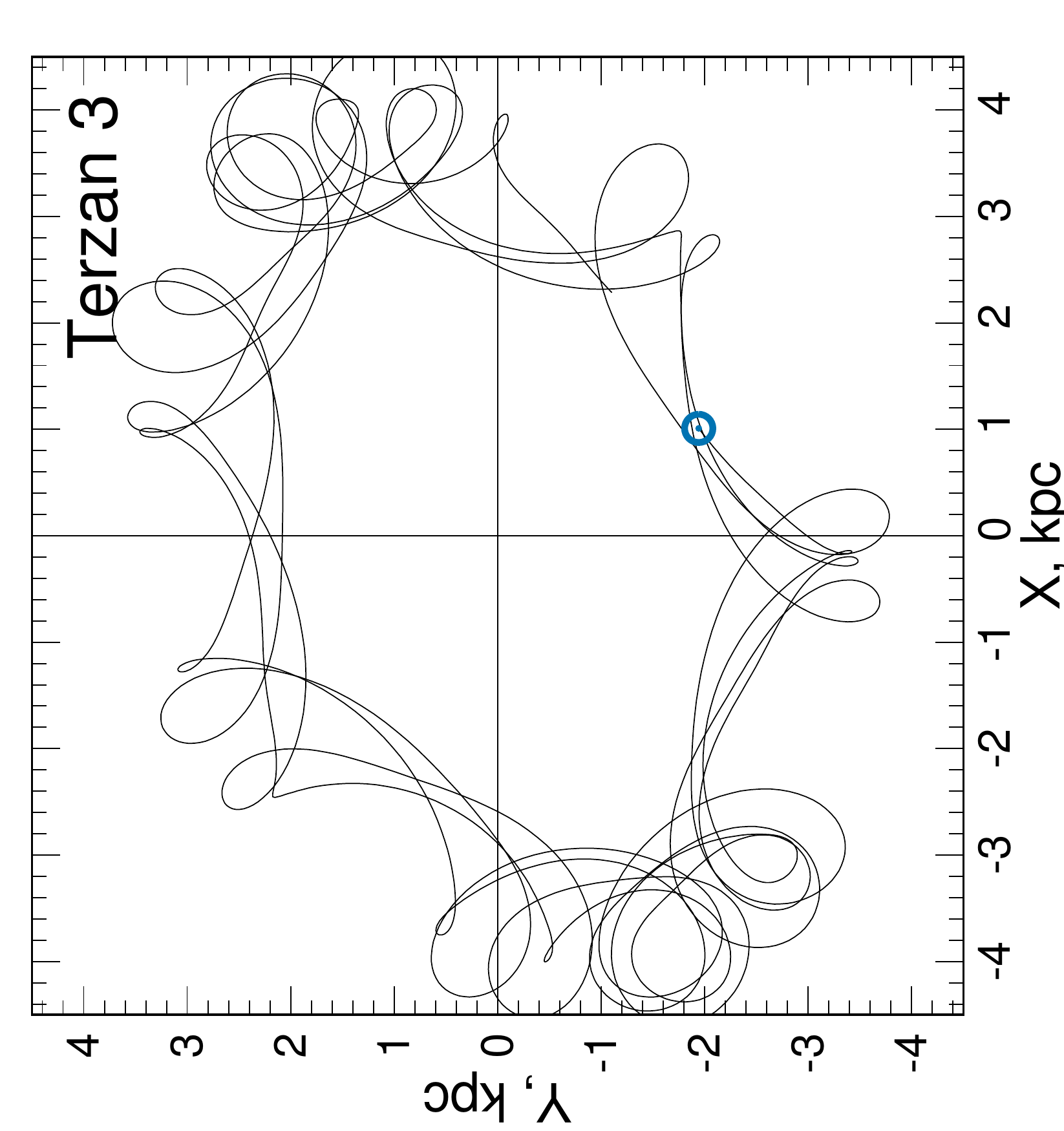}
\includegraphics[width=0.275\textwidth,angle=-90]{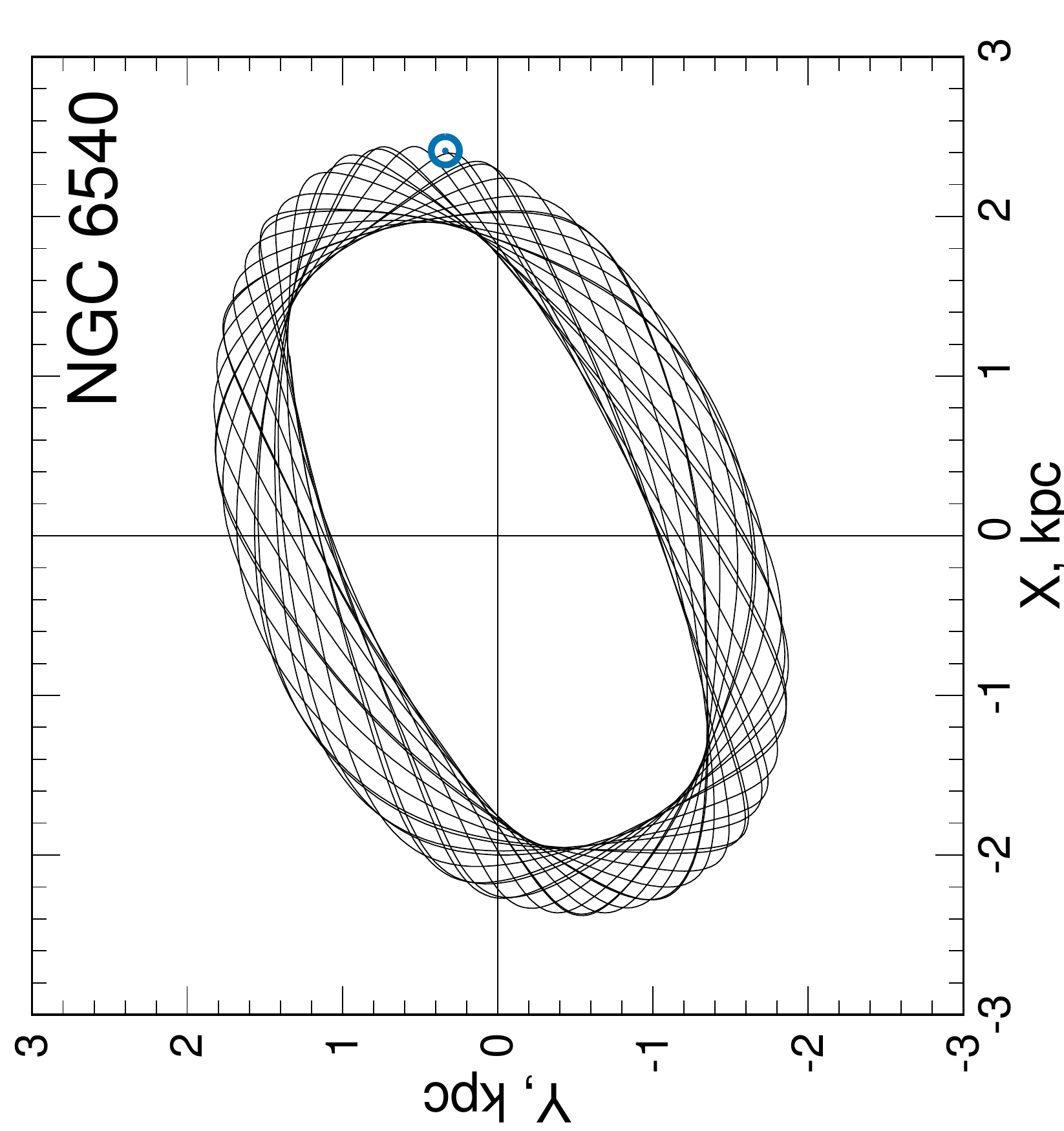}
\includegraphics[width=0.275\textwidth,angle=-90]{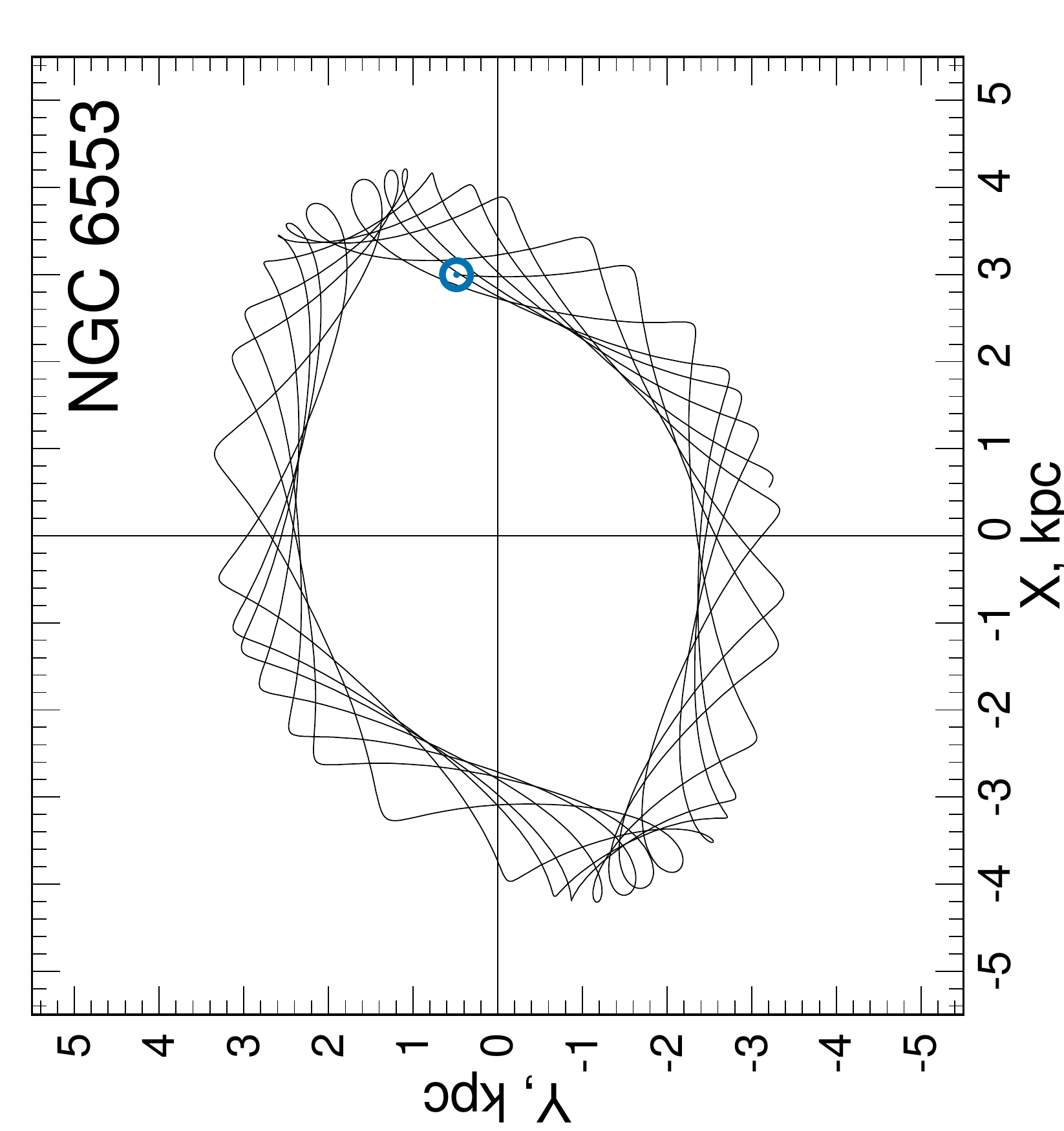}\
\caption{\small Orbits of the globular clusters NGC6266, NGC6522, NGC6540, NGC6553, Terzan 3, ESO456-SC78, built in a rotating bar system with a speed of $\Omega_b=45$ km s$^{-1}$ kpc$^{-1}$, mass $M_{bar}=430\times M_G$ and length $q_b=2.25$ kpc, in projection onto the galactic plane $X-Y$, satisfying the ratio of dominant frequencies $f_R/f_X \approx 2$. The beginning of the orbits is indicated by a blue circle.}
\label{fcomp9}
\end{center}}
\end{figure*}

Orbits in projection onto the galactic plane $X-Y$ of GCs with $f_R/f_X\approx 2$ for the first and last of the considered examples, calculated in the rotating bar system, are shown in Fig.~\ref{fcomp8} and \ref{fcomp9}, respectively. It is clear from the figures that the orbits have an ellipsoidal shape, their orientation relative to the $X$ axis coincides with the given orientation of the bar ($25^o$). So these orbits can be considered captured by the bar with the appropriate parameters.

\bigskip

Further, as a result of the simulation, the probabilities of GC capture by the bar were determined
when the bar parameters were varied randomly according to a uniform distribution law in the following more or less realistic ranges of values: mass $M_{bar}=[330 - 430]\times M_G$, length $q_b=[2-5]$ kpc, angular velocity of rotation $\Omega_b =[30-50]$ km s$^{-1}$ kpc$^{-1}$, degree of bar elongation $a/c=[2-5]$ in projection $X-Y$ ($10^3$ implementations). When integrating the orbits, the uncertainties in the initial positions and velocities of the GCs (6D phase space), as well as the peculiar velocity of the Sun using the Monte Carlo method ($10^2$ implementations) were taken into account. The probability $P$ of the bar capturing a particular GC was calculated as the ratio of episodes $N_c$, when $f_R/f_X=2\pm 0.1$, to the total number of random realizations $N_t$, in this case equal to $10^3 \times 10^2=10^5$, i.e. $P=N_c/N_t$.

Table~\ref{t:fff} provides a list of 14 globular clusters with the most significant capture probabilities $P$, of which the GCs NGC6266, NGC6569, Terzan 5, NGC6522 and NGC6540 show probabilities $P\geq 0.2$, namely 0.2, 0.2, 0.24, 0.38 and 0.68, respectively. In this case, the maximum probability is for the disk object NGC6540. The remaining globular clusters from Table~\ref{t:f}, which are not included in Table~\ref{t:fff}, have capture probabilities close to zero.

In order to assess the degree of stability (regularity) of the orbital motion, approximations of the characteristic Lyapunov exponent (CLE) were calculated for all 45 GCs in a non-axisymmetric potential with the basic bar parameters indicated above. The maximum values of the CLE (MCLE) were calculated by the "shadow" trajectory method using the following formula (see Mel’nikov (2018) and references therein):
\begin{equation}
\label{Lyap}
L(n)=\frac{1}{n \delta t} \sum_{i=1}^{n} \ln{\frac{d_i}{d_{i-1}}},
\end{equation}
where $d_i$ is the distance between the reference and shadow phase points at the $i$-th integration step, $\delta t$ is the time step value. The true value of the MCLE is equal to the limit $L(n)$ at $n \rightarrow \infty$. In practice, $L(n)$ obtained for a large value of $n$ is taken as the MCLE value. In this case, non-zero values of the MCLE indicate a chaotic (unstable), and zero values indicate a regular (stable) nature of the movement.

The dependence $L(n)$ obtained for all 45 globular clusters with the following perturbation of the phase starting point: $x_1=x_0+x_0\times 0.00001,~y_1=y_0+y_0\times 0.00001,~z_1=z_0+z_0\times 0.00001$, shown in Fig.~\ref{Lyap}. It is clear from the graph that the approximations of the Lyapunov exponents tend to zero with increasing $n$, albeit slowly, which confirms the regularity (absence of chaos) of the orbital motion of all GCs in a given potential.

\begin{figure*}
{\begin{center}
\includegraphics[width=0.4\textwidth,angle=-90]{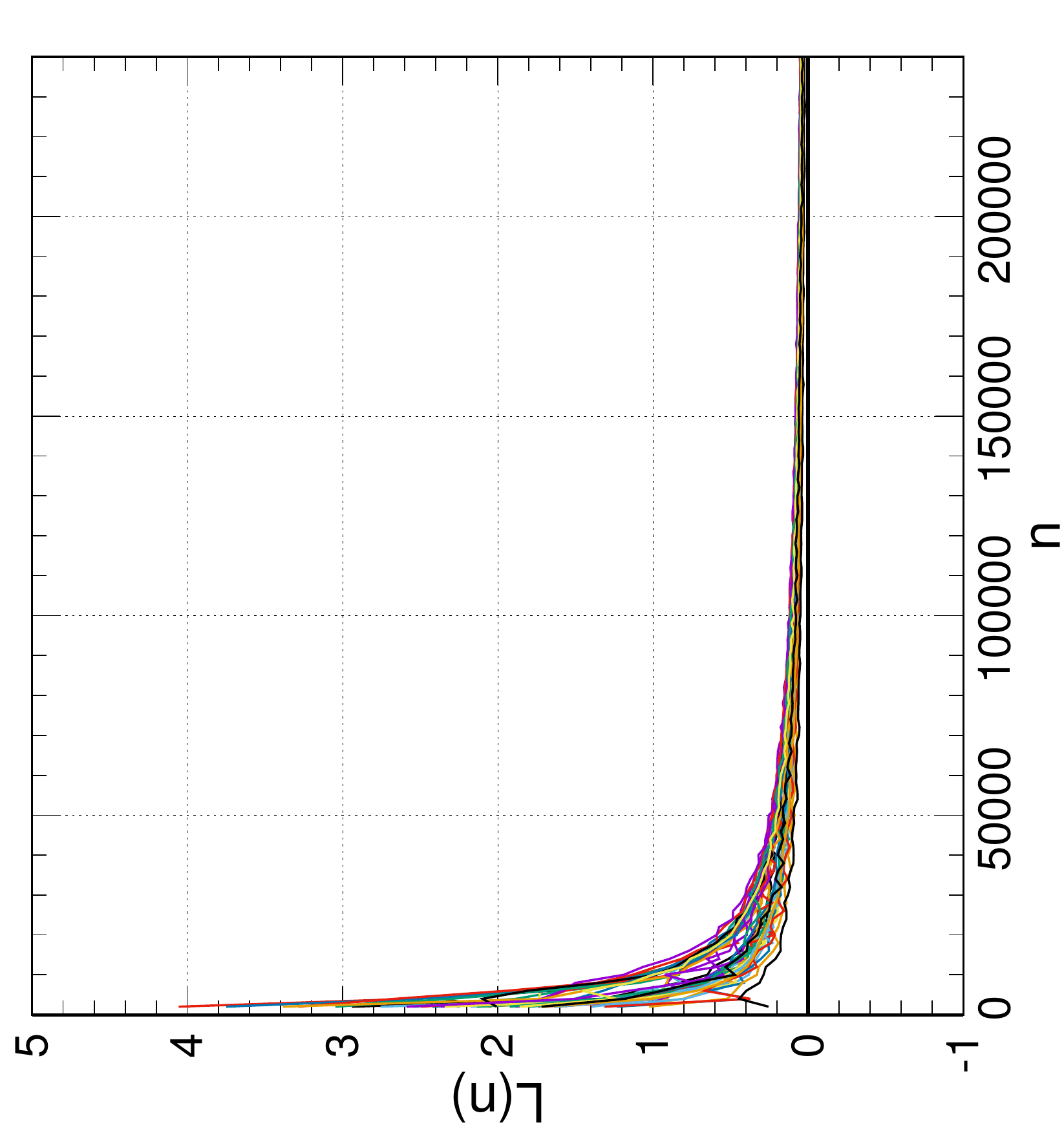}
\caption{\small The dependence $L(n)$ for all 45 globular clusters,
obtained in a non-axisymmetric gravitational potential
with the basic parameters of a triaxial bar.}
\label{Lyap}
\end{center}}
\end{figure*}

\begin{figure*}
{\begin{center}

       \includegraphics[width=0.275\textwidth,angle=-90]{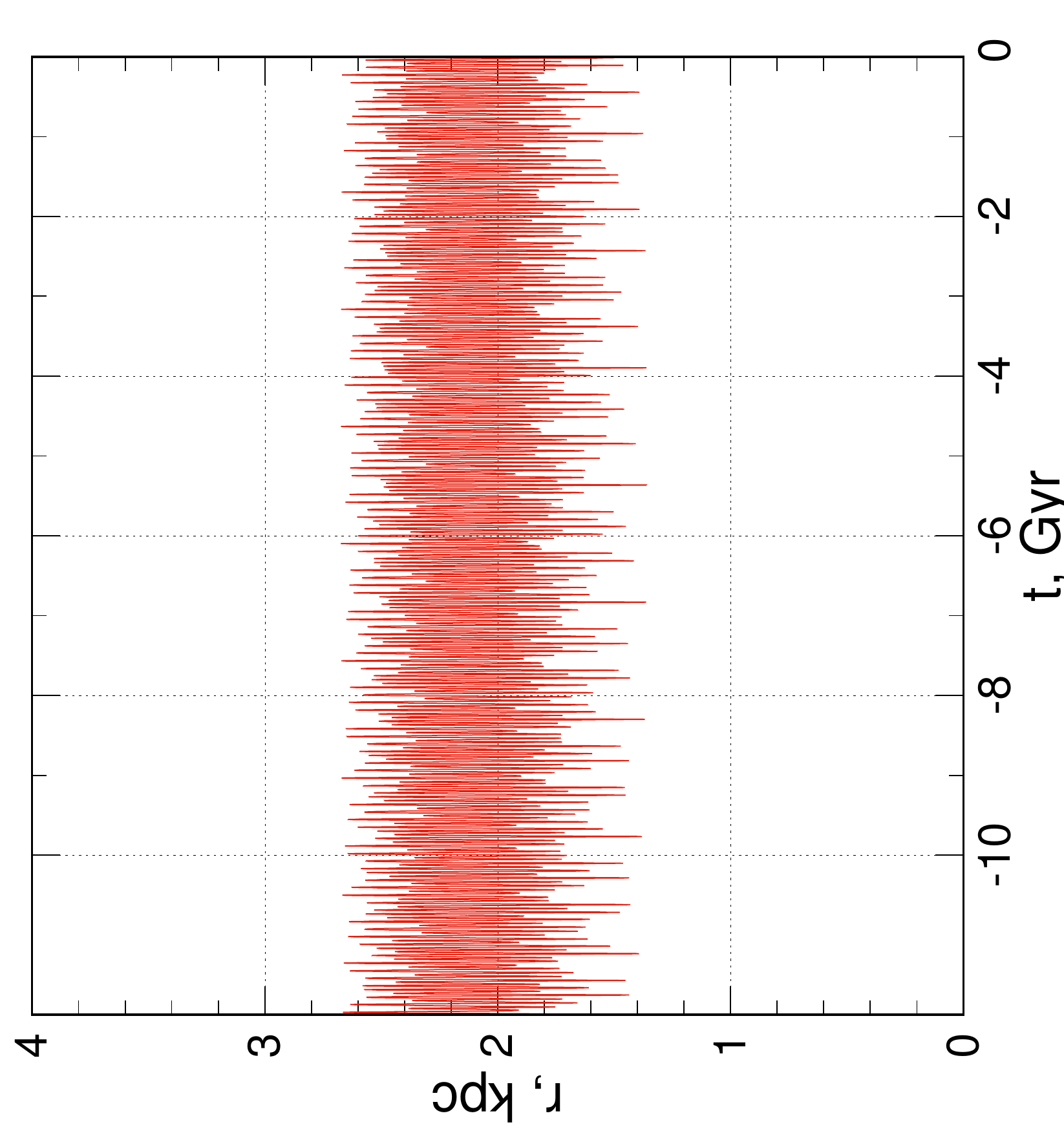}
       \includegraphics[width=0.275\textwidth,angle=-90]{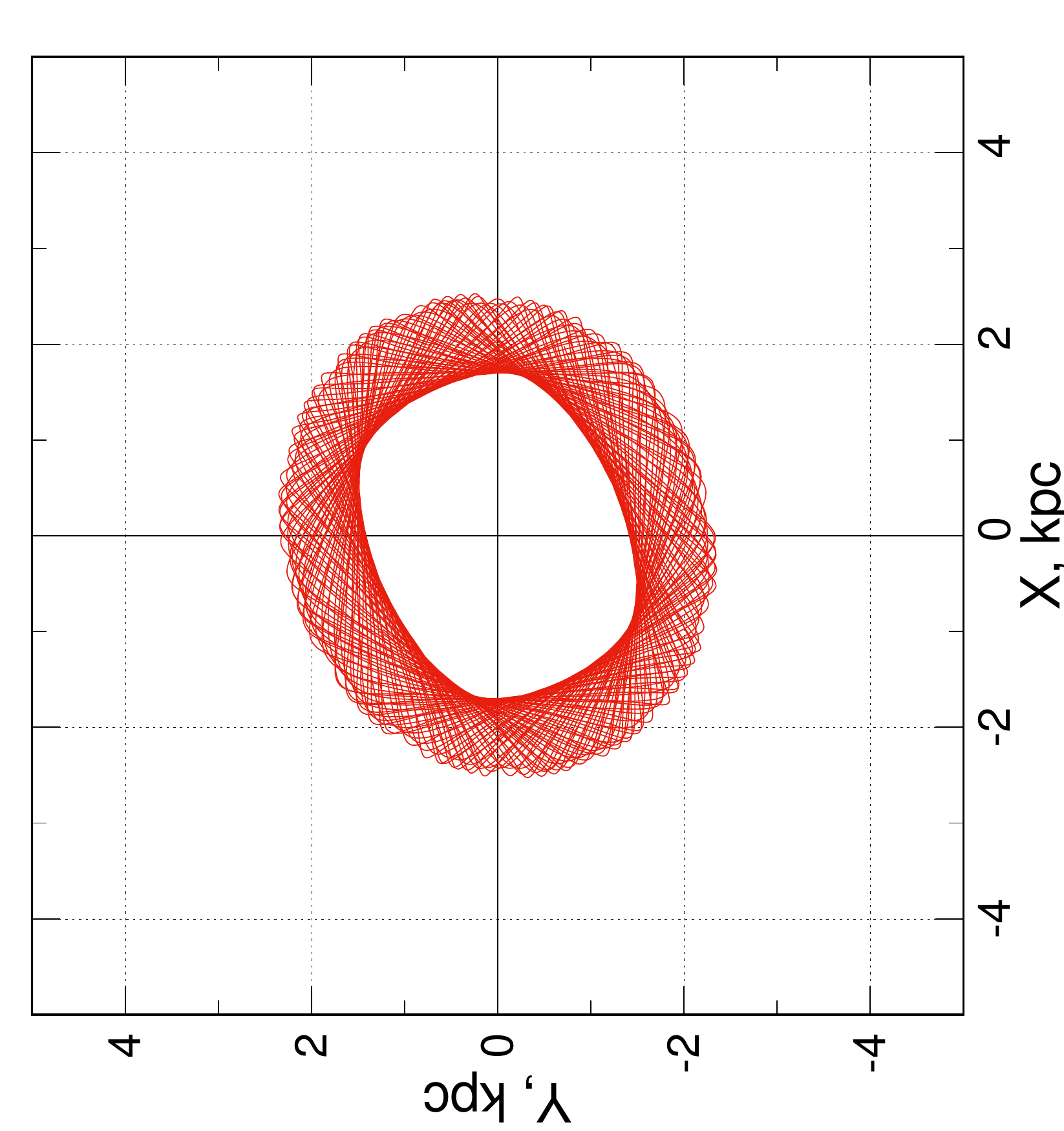}
       \includegraphics[width=0.275\textwidth,angle=-90]{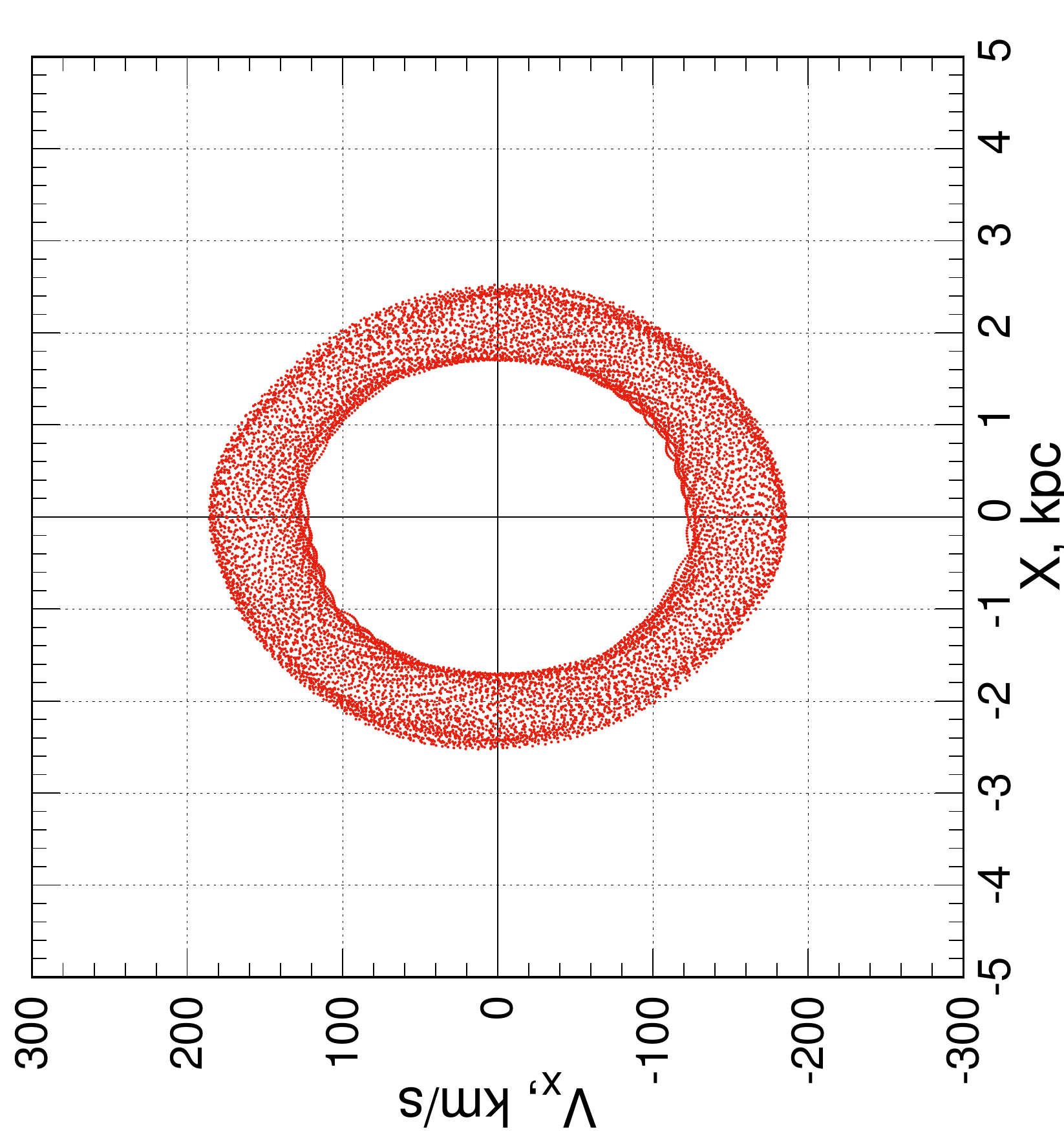}

       \includegraphics[width=0.275\textwidth,angle=-90]{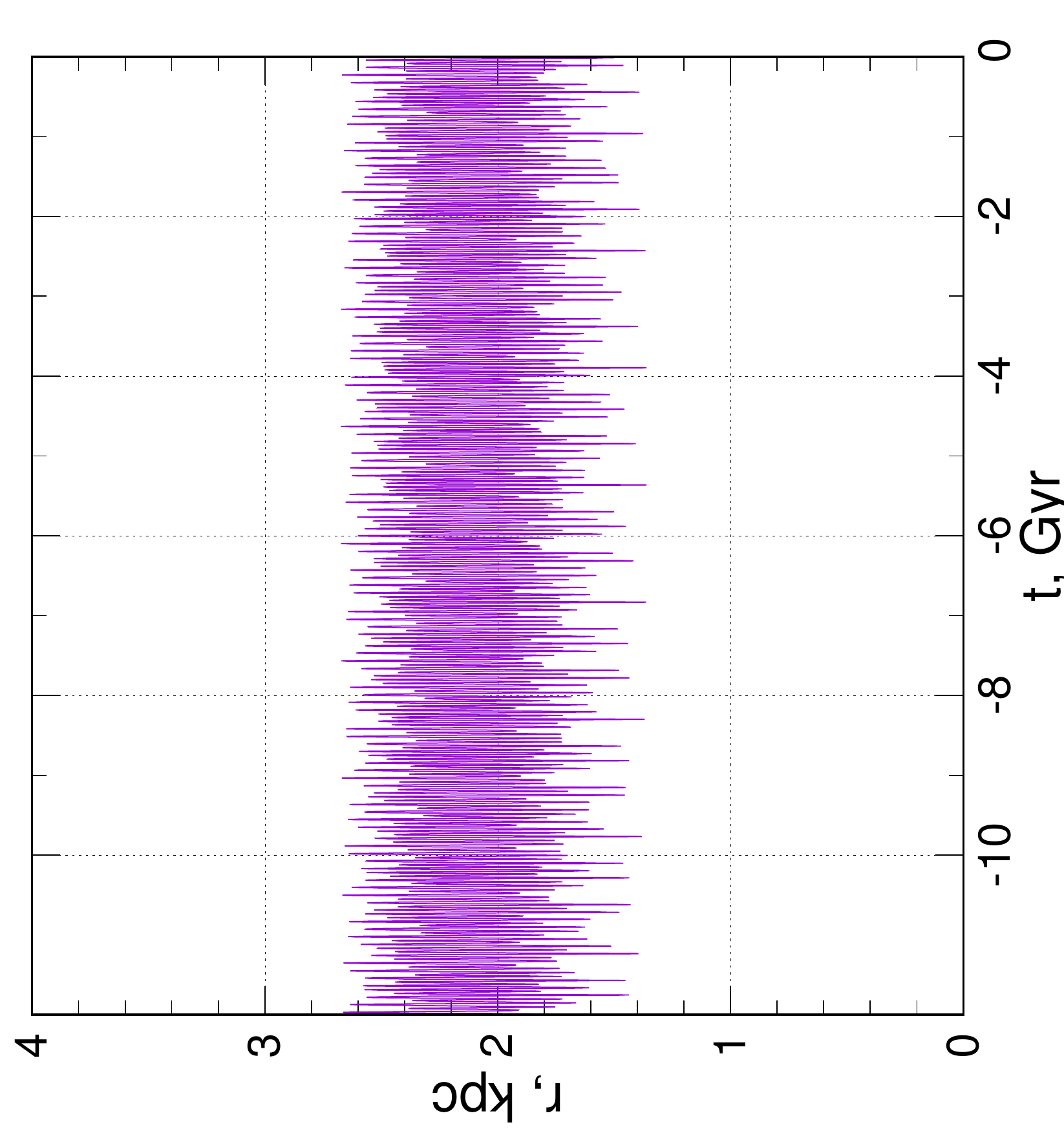}
       \includegraphics[width=0.275\textwidth,angle=-90]{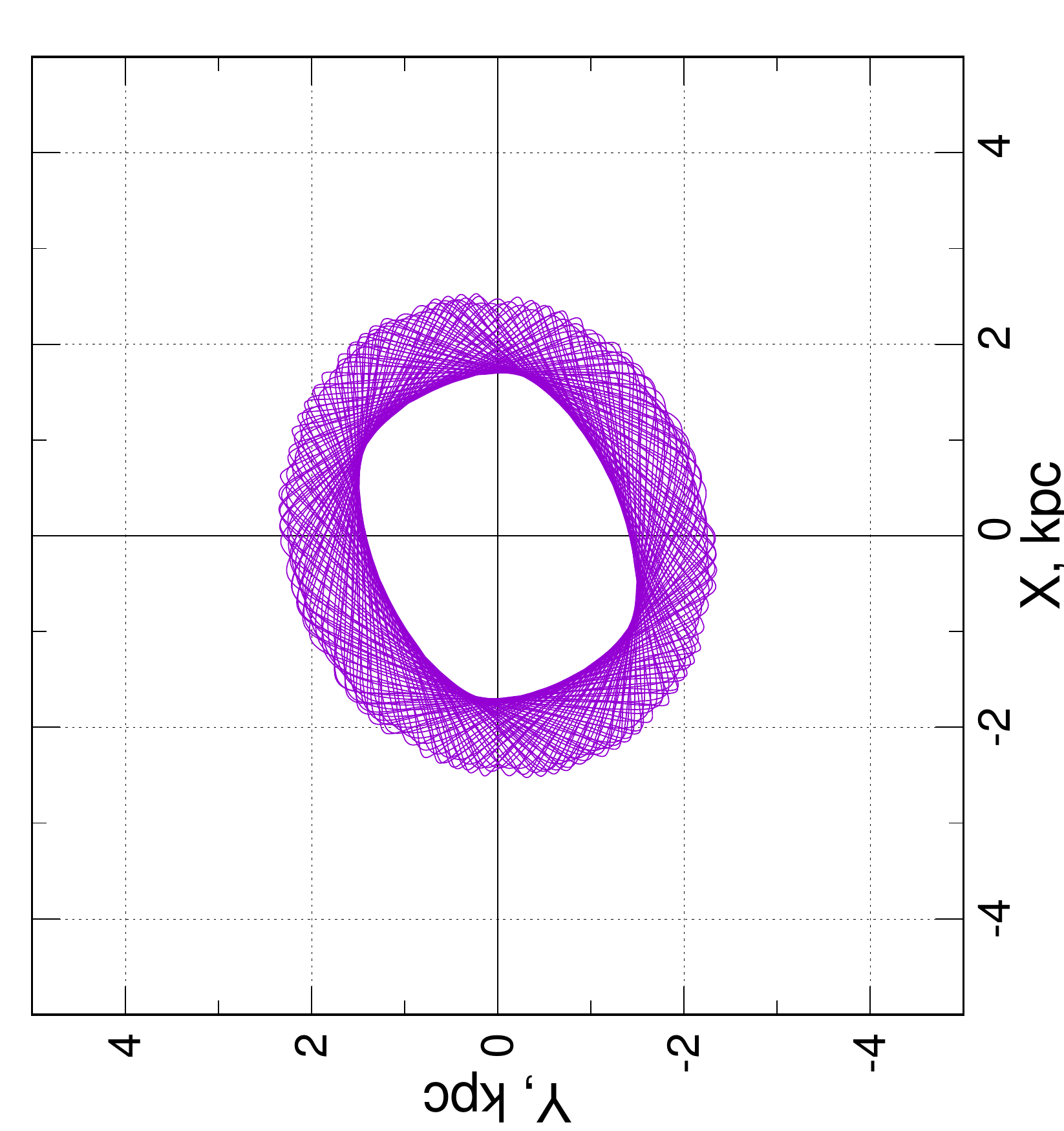}
       \includegraphics[width=0.275\textwidth,angle=-90]{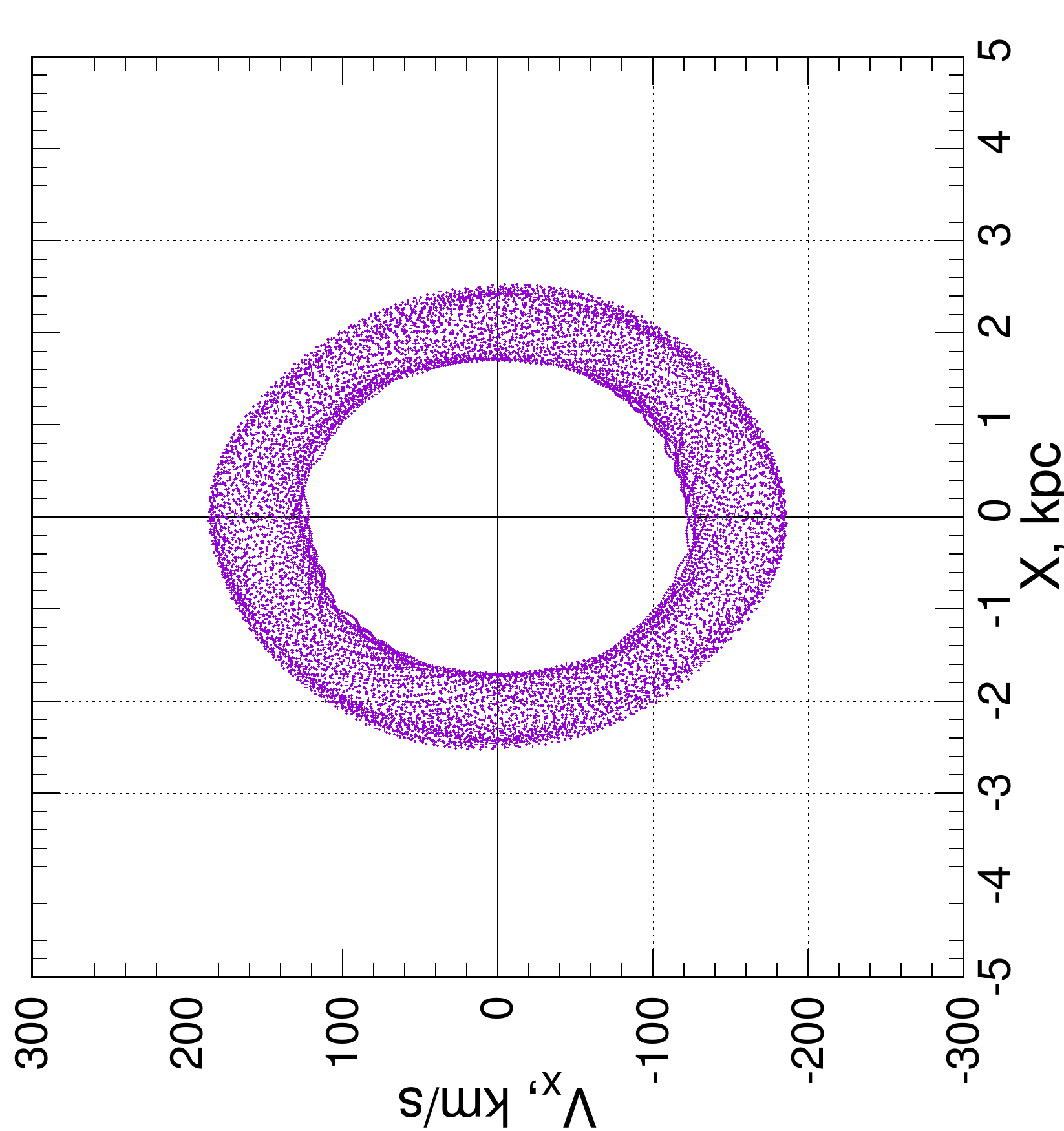}
\caption{\small Dynamics of motion of the globular cluster NGC6266 with unperturbed $(x_0, y_0, z_0)$ (red color) and perturbed $(x_1, y_1, z_1)$ (purple color) initial points, obtained by integrating orbits over a time interval of 12 billion years ago. The left panels show graphs of changes in the radius $r$ of the original and perturbed orbits versus time $t$, the middle panels show $X-Y$ projections of the corresponding orbits, and the right panels show the relationship between the velocity $V_X$ and the coordinate $X$.}
\label{Lyap2}
\end{center}}
\end{figure*}

As an example, Fig.~\ref{Lyap2} shows graphs displaying the dynamics of the motion of the globular cluster NGC6266 over a time interval of 12 billion years for a given $(x_0, y_0, z_0)$ and a perturbed $(x_1, y_1, z_1)$ starting points, shown in red and purple, respectively. From a visual analysis of the graphs, the conclusion follows that the orbital motion of the GC is highly stable over a long time interval, comparable to the age of the Universe. A graphical analysis of the dynamics of other GCs did not reveal chaotic motion in any object, which is consistent with the obtained pattern of convergence of $L(n)$ to zero with increasing $n$ (Fig.~\ref{Lyap}).

\section{CONCLUSIONS}

The orbital dynamics of 45 globular clusters in the central region of the Galaxy was simulated over
a wide range of changes in the mass, length, and angular velocity of rotation of the bar built into an axisymmetric potential. To integrate the orbits, we used the most accurate astrometric data to date from the Gaia satellite (Vasiliev and Baumgardt, 2021), as well as new refined average distances to globular clusters (Vasiliev and Baumgardt, 2021).

The constructed orbits were studied using spectral dynamics methods (Binney and Spergel, 1982),
based on calculating the ratio of dominant frequencies $f_R/f_X$. At $f_R/f_X=2\pm 0.1$ it was considered that the orbit was captured by the bar. As a result, the orbits supporting the bar were identified for each set of its parameters. The bar parameters have been established at which the number of captured orbits of globular clusters is maximum. Examples of orbits with $f_R/f_X\approx 2$ are given.

It was established that the dominant frequency $f_R$ is practically independent of the angular velocity of rotation of the bar $\Omega_b$, and the dominant frequency $f_X$ depends on b according to the law
\begin{equation}
f_X(\Omega_b) \approx f_X(0) \pm K\times\Omega_b,
\end{equation}
where K = 0.1587 kpc s km$^{-1}$ Gyr$^{-1}$, which leads to a systematic shift in the frequency ratio $f_R/f_X$ with changes in $\Omega_b$. This interesting dependence was obtained by us for the first time and requires further study.

As a result of modeling the orbital motion of GCs, the probabilities of their capture by the bar were determined when the parameters of the latter were varied randomly according to a uniform distribution law in the following ranges of values: bar mass $M_{bar}=[330 - 430]\times M_G$, bar length $q_b=[2-5]$ kpc, angular velocity of rotation $\Omega_b =[30-50]$  km s$^{-1}$ kpc$^{-1}$, degree of elongation of the bar $a/c=[2-5]$ in the projection $X-Y$. At the same time, using the Monte Carlo method, the uncertainties in the initial positions and velocities of the GCs, as well as the peculiar velocity of the Sun, were taken into account. As a result, a list of 14 globular clusters with the most significant capture probabilities was obtained, namely: ESO456-SC78, NGC6256, NGC6266, NGC6304, NGC6342, NGC6522, NGC6539, NGC6540, NGC6569, NGC6717, NGC6723, Terzan 3, Terzan 4, Terzan 5, from which the GCs NGC6266, NGC6569, Terzan 5, NGC6522 and NGC6540 are distinguished with the highest probabilities being 0.2, 0.2, 0.24, 0.38 and 0.68, respectively.

Based on the calculation of approximations of the maximum characteristic Lyapunov exponents and
graphical analysis of the dynamic characteristics of orbits over long time intervals, it was concluded that the orbits of all 45 studied globular clusters are regular in the accepted gravitational potential.

\section*{ACKNOWLEDGMENTS}
The authors are grateful to the reviewer for a number of useful comments. Special thanks to A. V. Melnikov for discussing the problem of orbital stability.

\section*{CONFLICT OF INTEREST}
The authors declare no conflict of interest.

\end{document}